\documentclass[11pt,a4paper]{article}
\pdfoutput=1
\usepackage{jheppub}
\usepackage{amsmath,amssymb}
\usepackage{dsfont}
\usepackage{verbatim} 
\usepackage{wrapfig}
\newcommand{\be}{\begin{equation}}
\newcommand{\ee}{\end{equation}}
\newcommand{\slsh}[1]{{#1}\!\!\!\! / \, }
 
\newcommand{\zet}{}
\newcommand{\nn}{\nonumber}
\title{The generalized cusp in ABJ(M) $\mathbf{{\cal N}=6}$ Super Chern-Simons theories}
\author[a]{Luca Griguolo}  
\author[a]{Daniele Marmiroli}
\author[b]{Gabriele Martelloni}
\author[b]{and Domenico Seminara}
\affiliation[a]{Dipartimento di Fisica, Universit\`a di Parma and INFN Gruppo Collegato di Parma, \\ Viale G.P. Usberti 7/A, 43100 Parma, Italy} 
\affiliation[b]{Dipartimento di Fisica e Astronomia, Universit\`a di
Firenze and INFN Sezione di Firenze,\\ Via  G. Sansone 1, 50019 Sesto Fiorentino, Italy} 
\emailAdd{luca.griguolo@fis.unipr.it} 
\emailAdd{daniele.marmiroli@fis.unipr.it} 
\emailAdd{martelloni@fi.infn.it} 
\emailAdd{seminara@fi.infn.it}
\abstract{We construct a generalized cusped Wilson loop operator in ${\cal N}=6$ super Chern-Simons-matter theories which is locally invariant under half of the supercharges. It depends on two parameters and interpolates smoothly between the 1/2 BPS line or circle and a pair of antiparallel lines, representing a natural generalization of the quark-antiquark potential in ABJ(M) theories. For particular choices of the parameters we obtain 1/6 BPS configurations that, mapped on $S^2$  by a conformal transformation, realize a three-dimensional analogue of the wedge DGRT Wilson loop of ${\cal N}=4$. The cusp couples, in addition to the gauge and scalar fields of the theory, also to the fermions in the bi-fundamental representation of the $U(N)\times U(M)$ gauge group and its expectation value is expressed as the holonomy of a suitable 
super-connection. We discuss the definition of these observables in terms of traces and the role of the boundary conditions of fermions along the loop. We perform a complete two-loop analysis, obtaining an explicit result for the generalized cusp at the second non-trivial order, from which we read off the interaction potential between heavy 1/2 BPS particles in the ABJ(M) model. Our results open the possibility to explore in the three-dimensional case the connection between localization properties and integrability, recently advocated in $D=4$.}
\keywords{} 
\begin{document} 
\maketitle

\section{Introduction and results}
The duality between string theory on asymptotically AdS spaces and conformal gauge theories, usually known as the AdS/CFT correspondence, has experienced an important evolution in the last few years. General non-BPS observables, as anomalous dimensions of composite operators and scattering amplitudes, can now be studied at high precision level providing sophisticated tests of the correspondence and sometimes offering non-trivial interpolating functions between weak and strong coupling. In both cases, the underlying integrability properties of the planar theory play a crucial role in the exact quantum evaluation and allow to follow the transition between the opposite regimes. Dramatic progresses have also concerned more traditional investigations, as the study of protected sectors of supersymmetric gauge theories: the introduction of powerful localization techniques makes now possible the exact computation of complicated path-integrals, providing again examples of interpolation between perturbative and asymptotic behaviors. It is tempting to speculate if the two different approaches, integrability and localization, could be somehow connected, at least in the computation of specific observables. 

\noindent
This evocative possibility has been vigorously advocated in \cite{Drukker:2011za} for a general class of Wilson loops in $\mathcal{N}=4$ super Yang-Mills theory and concretely realized in a series of recent papers \cite{Correa:2012at,Correa:2012nk,Drukker:2012de,Correa:2012hh}, where a new set of integral equations of TBA type, describing exactly a generalized cusp anomalous dimension, have been derived and checked against localization and perturbation theory at three loops. The result is striking and it contains, in principle, the all-order expression of the static potential between two heavy charged particles in four-dimensional maximally supersymmetric gauge theory.  

\noindent
Wilson loops are important observables in nonabelian gauge theories: they compute the potential between heavy colored probes, representing an order parameter for confinement \cite{Wilson:1974sk} and encode a large part of information of the high-energy scattering between charged particles
 \cite{Korchemsky:1988si,Korchemsky:1992xv}. In ${\cal N}=4$ super Yang-Mills theory they play a prominent role, being the observables directly related to the fundamental string of the dual theory in $AdS_5 \times S^5$ \cite{Rey:1998ik,Maldacena:1998im}.  They are conjectured to calculate scattering amplitudes exactly \cite{Bern:2005iz,Alday:2007hr} and in particular cases they are BPS operators \cite{Dymarsky:2009si,Cardinali:2012sy}, whose quantum expectation value can be derived for any strength of the coupling constant \cite{Erickson:2000af,Drukker:2000rr,Pestun:2007rz}. In \cite{Drukker:2011za} a two-parameters family of Wilson loops in ${\cal N}=4$ SYM has been proposed and studied both at weak and strong coupling: it consists of two rays in $\mathbb{R}^4$ meeting at a point, with a cusp angle denoted by $\pi-\varphi$. Because the Maldacena-Wilson loops in ${\cal N}=4$ SYM couple to a real scalar field, it is natural to consider different scalars on different rays, connected by an $R$-symmetry rotation of parameter $\theta$. By varying continuously $\varphi$ and $\theta$ and performing suitable conformal transformations, these observables can be related to important physical quantities and to interesting BPS configurations. Mapping the theory to $S^3\times\mathbb{R}$ one obtains a pair of antiparallel lines, separated by angle $\pi-\varphi$ on the sphere, and derives the potential $V(\lambda,\varphi,\theta)$ between two heavy $W$-bosons propagating over a large time $T$:
\begin{equation}
\langle \mathcal{W}_{\rm lines}\rangle\simeq \exp \left(-TV(\lambda,\varphi,\theta)\right).
\end{equation}
The usual potential in the flat space is easily recovered by taking the residue of $V(\lambda, \varphi ,0)$ as $\varphi\to\pi$ \cite{Drukker:2011za}. In the cusped version, the Wilson loop has the leading form
\be
\langle
\mathcal{W}_{\rm cusp}
\rangle\simeq \exp\left(-\Gamma_{cusp}(\lambda, \varphi,\theta)\log\left(\frac{\Lambda}{\epsilon}\right)\right),\ee
and it turns out that $\Gamma_{cusp}(\lambda, \varphi,\theta)=V(\lambda, \varphi,\theta)$, with $\Lambda$ and $\epsilon$ being IR and UV cut-offs respectively \cite{Polyakov:1980ca,Brandt:1981kf}.  The cusped Wilson loops are strictly related to scattering amplitudes in Minkowski space: taking $\varphi$ imaginary and large, $\varphi=i\phi$, produces to the so-called universal cusp anomalous dimension $\gamma_{cusp}(\lambda)$ \cite{Korchemsky:1987wg}
\be \lim_{\phi\to\infty}\Gamma_{cusp}(\lambda, i\varphi,\theta)=\frac{\varphi\,\gamma_{cusp}(\lambda)}{4}.\ee
Remarkably BPS configurations are also included in the family: for $\theta=\pm\varphi$ the cusped Wilson loop is of Zarembo's type \cite{ Zarembo:2002an}, implying the vanishing of  $\Gamma_{cusp}(\lambda, \varphi,\theta)$ in this case. By mapping conformally the rays into $S^2$ we recover instead the DGRT wedge, a well studied 1/4 BPS operator \cite{Drukker:2007qr,Bassetto:2008yf,Young:2008ed} belonging to the general class of loops on $S^3$ introduced in \cite{Drukker:2007qr}. The quantum expectation value of the wedge is computed exactly by pertturbative two-dimensional Yang-Mills theory \cite{Bassetto:1998sr}, a property shared with other DGRT loops on $S^2$  \cite{Pestun:2009nn} and with a certain class of loop correlators \cite{Bassetto:2009rt,Bassetto:2009ms,Giombi:2009ds,Giombi:2012ep}. We remark that path-integral localization properties are essential in order to derive these results. 

\noindent
In the limit of small $\varphi$, $\Gamma_{cusp}(\lambda, \varphi,0)\simeq-B(\lambda)\varphi^2$ computes also the radiation of a particle moving along an arbitrary smooth path \cite{Correa:2012at} and an exact expression can be obtained by exploiting the BPS properties at $\varphi=0$. A simple modification applies as well in expanding around the general BPS points $\theta=\pm\varphi$, thanks to the knowledge of the all-orders expression of the DGRT wedge. Further, in \cite{Drukker:2012de,Correa:2012hh}, a powerful set of TBA type integral equations for the generalized cusp was derived, using integrability: the explicit one-loop perturbative result \cite{Drukker:2012de} and the three-loop expression of the near BPS limit \cite{Correa:2012nk} were recovered as a check of the procedure.

\noindent
In view of these recent and exciting developments, it appears natural to wonder wether similar results could also be obtained for other superconformal gauge theories with integrable structures: the obvious choice is to investigate what happens in ${\cal N}=6$ super Chern-Simons theories with matter, also known as ABJ(M) \cite{Aharony:2008ug,Aharony:2008gk}. Wilson loops in ABJ(M) theory are still a rather unexplored subject: equivalence with scattering amplitudes has been shown at the second order in perturbation theory \cite{Henn:2010ps,Chen:2011vv,Bianchi:2011dg,Bianchi:2011fc} and a quite mysterious functional similarity with their four-dimensional cousins seems to emerge. On the other hand, supersymmetric configurations, supported on straight lines and circles, have been discovered \cite{Drukker:2009hy,Drukker:2008zx,Chen:2008bp,Rey:2008bh} and a localization formula, reducing the computation to an explicit matrix model average, has been derived \cite{Kapustin:2009kz}. Concrete results at weak and strong coupling, using matrix model and topological string techniques, are presented in a beautiful series of papers \cite{Marino:2009jd,Drukker:2010nc,Drukker:2011zy,Marino:2011eh}, where also various aspects of the partition function on $S^3$ have been discussed. Nevertheless many issues should be understood in order to extend the generalized cusp program. First of all it does not even exist a computation of the standard quark-antiquark potential nor of the conventional cusped Wilson loop. Secondly, BPS lines and circles appear in two fashions, distinguished by the degree of preserved supersymmetry: we have the 1/6 BPS operators, originally defined in \cite{Drukker:2008zx,Chen:2008bp} (see also \cite{Rey:2008bh}), that are a straightforward extension of the Maldacena-Wilson loop to the three-dimensional case. The (quadratic) coupling with the scalar fields of the theory is governed by a mass matrix $M^I_J$, preserving an $SU(2)\times SU(2)$ of the original $R$-symmetry, while gauge fields appear in the usual way. Although its simplicity, this kind of loop cannot be considered the dual of the fundamental string living in $AdS_4\times \mathbb{CP}^3$, because supersymmetries do not match \cite{Drukker:2008zx}. The field theoretical partner of the fundamental string is instead the 1/2 BPS loop discovered in \cite{Drukker:2009hy} (see \cite{Lee:2010hk} for a derivation arising from the low-energy dynamics of heavy $W$-bosons): the loop couples, in addition to the gauge and scalar fields of the theory, also to the fermions in the bi-fundamental representation of the $U (N )\times U (M )$ gauge group. These ingredients are combined into a superconnection whose holonomy gives the Wilson loop, which can be defined for any representation of the supergroup $U(N |M )$. Supersymmetry is realized through a highly sophisticated mechanism, as a super-gauge transformation, requiring therefore the full non-linear structure of the path-ordering.  Actually both loops turn out to be in the same cohomology class, differing by a BRST exact term with respect the localization complex: their quantum expectation value should be therefore the same \cite{Drukker:2009hy}. The above equivalence has not been checked at weak-coupling, where perturbative computations have been performed just for the 1/6 BPS circle 
\cite{Drukker:2008zx,Chen:2008bp,Rey:2008bh}, the presence of fermions complicates the calculations and rises delicate issues on the regularization procedure. Crucially there are also no examples of loops with fewer supersymmetries, including the known BPS lines and circle as particular cases: it would be interesting to find configurations of this type that could also help to understand better the mysterious cohomological equivalence. 

\noindent
This paper represents a first step towards a systematic study of generalized cusps in ABJ(M) theories: similar configurations have been discussed, at strong coupling, in \cite{Forini:2012bb}. We hope that our investigations could stimulate the application of integrability and cohomological techniques in the exact evaluation of non-BPS observables, such as the heavy-bosons static potential. Our main concern here is the construction of a generalized cusp using two 1/2 BPS rays, the study of its supersymmetric properties and its quantum behavior at weak-coupling. The additional $R$-symmetry deformation is obtained by preserving different $SU(3)$ subgroups on the two lines: from the bosonic point of view this amounts to deform the mass-matrix $M^I_J$, by rotating two directions of opposite eigenvalues. The fermionic couplings experience a similar deformation and are also explicitly affected by the geometric parameter $\varphi$, because they transform as spinors under spatial rotations. We study the supersymmetry shared by the two rays and we discover that for $\theta=\pm\varphi$ two charges are still globally preserved: in this case the super-gauge transformations, encoding the supersymmetry variation on the two edges of the cusp, become smoothly connected at the meeting point.  A key observation made in the original paper of Drukker and Trancanelli \cite{Drukker:2009hy} was that, for the 1/2 BPS circle, only the trace of the super-holonomy turns out to be supersymmetric invariant and not its super-trace. Conversely the fermionic couplings were assumed to be anti-periodic on the loop: here we examine the same problem on the new BPS configurations. By performing an explicit conformal transformation, we map our cusp on $S^2$, obtaining a wedge: the fermionic couplings, constant on the plane, become space-dependent as an effect of the conformal map, and connected by a non-trivial rotation on the upper point of the wedge. In the BPS case this matrix simply appears as an anti-periodicity,  and therefore it is again the trace that leads to supersymmetric invariance. The loops constructed in this way are a sort of ABJ(M) version of the DGRT wedge \cite{Drukker:2007qr} and preserve 1/6 of the original supersymmetries. We consistently define our generalized cusp as the trace of the super-holonomy and attempt the computation of its quantum expectation value in perturbation theory. We observe two basic differences with the analogous four-dimensional computation performed in \cite{Drukker:2011za}: first of all the effective propagators here, attaching on one side of the cusp only, are not automatically vanishing, as it happens for ${\cal N}=4$ SYM in Feynman gauge, and the fermionic sector gives a divergent contribution at one-loop, that has to be regularized and renormalized. Secondly, because of the presence of a supergroup structure, involving fermions coupled to the external lines, it is not obvious to extend the non-abelian exponentiation theorem to this setting: we could not rely on such powerful device to reduce the amount of computations and to properly isolate the cusp anomalous dimensions. Concerning this second point we make the natural assumption that our cusped loops undergo through a ``double-exponentiation" 
\be
\label{doppia}
\mathcal{W}=\frac{M \exp(V_{N})+ N \exp(V_{M})}{N+M}, 
\ee
where the generalized potentials\footnote{With an abuse of language we have  referred  to $V_{N}$ and $V_{M}$ as the {\it  generalised potentials.} Actually only  the coefficient of the $1/\epsilon$ pole  has this meaning.}
 $V_N$ and $V_M$ are simply related by exchanging $N$ with $M$. A highly non-trivial check of the above assumption is the actual exponentiation of the one-loop term, constraining in particular the structure of the double-poles (in the dimensional regularization parameter $\epsilon=(3-D)/2$) appearing at two-loops. We find a perfect agreement of our results with the double-exponentiation hypothesis, recovering at the second order in perturbation theory the quadratic contributions coming from the first order one. Our final expression for the unrenormalized $V_N$ is
\begin{align}
V_{N}=&\left(\frac{2\pi}{\kappa}\right) N\left(\frac{ \Gamma(\frac{1}{2}-\epsilon)}{4\pi^{3/2-\epsilon}}\right)
(\mu L)^{2\epsilon}\left[\frac{1}{\epsilon}\left(\frac{\cos\frac{\theta}{2}}{\cos\frac{\varphi}{2}}-2\right)-2\frac{\cos\frac{\theta}{2}}{\cos\frac{\varphi}{2}}\log \left(\sec \left(\frac{\varphi }{2}\right)+1\right)\right]+\nonumber\\
&\left(\frac{2\pi}{\kappa}\right)^{\!\!2}N^2\!\left(\frac{\Gamma\!\left(\frac{1}{2}-\epsilon\right)}{4 \pi^{3/2-\epsilon}}\right)^2(\mu L)^{4\epsilon}\left[\frac{1}{\epsilon}\log\left(\cos\frac{\varphi}{2}\right)^2\left(\frac{\cos\frac{\theta}{2}}{\cos\frac{\varphi}{2}}-1\right)+O(1)\right].
\end{align}
Here $k$ is the Chern-Simons level, $L$ is the length of the lines and $\mu$ is the renormalization scale introduced by dimensional regularization. To extract the cusp anomalous dimension we have to carefully subtract the divergences coming from single-leg diagrams: for closed contours in four dimensions these are usually associated, in a generic gauge theory, to a linear divergence proportional to the perimeter loop \cite{Polyakov:1980ca}. In the smooth case once subtracted this perimeter term. the standard lagrangian renormalization makes the quantum expectation value finite \cite{Brandt:1981kf,Exp}. 

\noindent
When open contours are considered the situation changes and some subtleties in the renormalization procedure arise: a systematic analysis of these problems have been performed in eighties \cite{Aoyama:1981ev,Dorn1,Dorn2,Knauss:1984rx,Dorn:1986dt} and (somehow) forgotten. The outcome is essentially contained in the introduction of a further a gauge-dependent renormalization constant, sometimes called $Z_{\rm open}$, taking into account shape-independent extra-divergencies associated to the end-points of the contour. To isolate the true gauge-invariant cusp-divergence these spurious contributions should be subtracted because, in general, they appear for finite lenght of the lines. ${\cal N}=4$ SYM in Feynman gauge represents a lucky situation in which, due to the peculiar combination of the gauge/scalar propagator, these additional effects are not present. We remark that in general $\alpha$-gauge a $Z_{\rm open}$ should be taken into account. We will carefully review all these topics in subsec. \ref{outcome}. 
\noindent
  
In three dimensions the superconformal case, due to the fermionic couplings, inevitably implies the appearence of the spurios single-length contributions: we will carefully discuss the subtraction procedure, examing in details the paradigmatic case of the 1/2 BPS infinite-line, and we hope to clarify the structure of the divergences for these family of loops. We will also comment on the difference between the 1/2 BPS and 1/6 BPS cases, showing that at finite-length the cohomological equivalence is broken by boundary terms, generating the unexpected divergence at quantum level. 

Our final receipt amounts to subtract, in the second order computation,  the one-loop poles associated to single line diagrams, normalizing in this way the final result to the straight line, 1/2 BPS contour
\begin{align}  
\!\!V^{\rm Ren.}_{N}=&\left(\frac{2\pi}{\kappa}\right) N\left(\frac{ \Gamma(\frac{1}{2}-\epsilon)}{4\pi^{3/2-\epsilon}}\right)\!
(\mu L)^{2\epsilon}
\!\!\left[\frac{1}{\epsilon}\left(\frac{\cos\frac{\theta}{2}}{\cos\frac{\varphi}{2}}-1\right)\!\!-2\frac{\cos\frac{\theta}{2}}{\cos\frac{\varphi}{2}}\log \left(\sec \left(\frac{\varphi }{2}\right)\!+\!1\right)\!\!+\log 4\right]\!+\nonumber\\
&+\left(\frac{2\pi}{\kappa}\right)^{\!\!2}N^2\!\left(\frac{\Gamma\!\left(\frac{1}{2}-\epsilon\right)}{4 \pi^{3/2-\epsilon}}\right)^2(\mu L)^{4\epsilon}\left[\frac{1}{\epsilon}\log\left(\cos\frac{\varphi}{2}\right)^2\left(\frac{\cos\frac{\theta}{2}}{\cos\frac{\varphi}{2}}-1\right)+O(1)\right].
\end{align}
From the above expression we can easily recover the quark-antiquark potential\footnote{Actually we have two potentials, $V^{(s)}_N$ and $V^{(s)}_M$, associated respectively to singlets in the $N\times\bar{N}$ and $M\times\bar{M}$ direct product.}, taking the limit $\varphi\to \pi$ and following the prescription of \cite{Drukker:2011za}
\begin{equation}
V^{(s)}_{N}(R)=\frac{N}{k}\frac{1}{R}-\left(\frac{N}{k}\right)^2\frac{1}{R}\log\left(\frac{T}{R}\right).
\end{equation}

We find a logarithmic, non-analytic term in $T/R$ at the second non-trivial order that, as in four dimensions, is expected to disappear when resummation of the perturbative series is performed. In the opposite limit, for large imaginary $\varphi$, we get the universal cusp anomaly (using the four-dimensional definition)
\be
\gamma_{cusp}=\frac{N^2}{k^2},
\ee
reproducing the result obtained directly from the light-like cusp \cite{Henn:2010ps}.

\noindent
The plan of the paper is the following: in Section 2 we review the construction of 1/2 BPS Wilson lines in ABJ models, giving us the possibility to introduce the peculiar structures entailed by maximal supersymmetric loops in ${\cal N}=6$ super Chern-Simons-matter theories. Section 3 is devoted to the explicit  realization of the generalized cusp: we obtain the appropriate bosonic and fermionic couplings and their deformations and discuss how the supersymmetry properties depends on the relevant parameters. The conformal transformation, mapping the cusp on a wedge of $S^2$, is also presented: the periodicity properties of the fermions are derived and BPS observables are obtained taking the trace of the super-holonomy. In Section 4 we start the quantum investigation computing the expectation value at the first order in perturbation theory. The two-loop analysis is contained in Section 5. The final result, obtained by summing up all the contributions and performing the renormalization procedure is presented in Section 6, where the peculiar divergences structure of these observables is carefully discussed. We present a rather detailed review of known facts on the renormalization of closed, open and cusped Wilson loops, that we think will clarify the apparent intricacy of our subtraction procedure and unveil its gauge-independent meaning. Some conclusions and outlooks appear in Section 7. We complete the paper with some appendices, containing our conventions and the technical details of the computations.

\section{1/2 BPS straight-line in ABJ theories}
\label{sec2}
We start by reviewing the construction of the $1/2-$BPS Wilson line  given in \cite{Drukker:2009hy,Lee:2010hk}: the mechanism leading to its gauge  invariance is carefully reconsidered, since it is substantially different from the four dimensional analogue.

The  central idea  of \cite{Drukker:2009hy} is to replace the obvious $U(N)\times U(M)$  gauge  connection with  a super-connection\be
\label{superconnection}
 \mathcal{L}(\tau) \equiv -i \begin{pmatrix}
i\mathcal{A}
&\sqrt{\frac{2\pi}{k}}  |\dot x | \eta_{I}\bar\psi^{I}\\
\sqrt{\frac{2\pi}{k}}   |\dot x | \psi_{I}\bar{\eta}^{I} &
i\hat{\mathcal{A}}
\end{pmatrix} \ \  \ \ \mathrm{with}\ \ \ \  \left\{\begin{matrix} \mathcal{A}\equiv A_{\mu} \dot x^{\mu}-\frac{2 \pi i}{k} |\dot x| M_{J}^{\ \ I} C_{I}\bar C^{J}\\
\\
\hat{\mathcal{A}}\equiv\hat  A_{\mu} \dot x^{\mu}-\frac{2 \pi i}{k} |\dot x| \hat M_{J}^{\ \ I} \bar C^{J} C_{I},
\end{matrix}\ \right.
\ee
belonging to the super-algebra\footnote{In Minkowski space-time, where $\psi$ and $\bar\psi$ are related by complex conjugation, $\mathcal{L}(\tau)$ belongs to $\mathfrak{u}(N|M)$ if $\bar\eta=i (\eta)^{\dagger}$. In Euclidean space, where the reality condition among spinors are lost,  we shall deal with the complexification  of this group $\mathfrak{sl}(N|M)$.} of $U(N|M)$. In \eqref{superconnection} the coordinates $x^{\mu}(\tau)$  
draw the contour along which the loop operator is defined,  while  $M_{J}^{\ \ I}$, $\hat M_{J}^{\ \ I}$, $\eta_{I}^{\alpha}$ and $\bar{\eta}^{I}_{\alpha}$ are  free parameters. The latter two, in particular, are Grassmann even quantities even though they transform in the spinor representation of the Lorentz group.

The dependence of $\mathcal{L}(\tau)$ on the fields is largely dictated by dimensional analysis and transformation properties. Since the {\it classical} dimension of the  scalars in $D=3$ is $1/2$,  they could only appear as bilinears,  transforming  in the adjoint and thus entering in the diagonal blocks together with the gauge fields. Instead  the fermions  have  dimension $1$ and should appear linearly. Since they transform in the bi-fundamentals of the gauge group, they are naturally placed in the off-diagonal entries of the matrix \eqref{superconnection}. 

When the contour $x^{\mu}(\tau)$ is a straight-line $S$, the  invariance  under translations along the direction  defined by $S$ ensures that all the couplings can be chosen to be  independent of $\tau$, {\it i.e.} constant. 
Moreover the requirement of having  an unbroken  $SU(3)$ $R-$symmetry, as that of the dual string configuration, restricts  the couplings $M_{J}^{\ \ I}$,\  $\hat M_{J}^{\ \ I}$,\ $\eta_{I}^{\alpha}$\  and\  $\bar{\eta}^{I}_{\alpha}$ to be   of the 
form
\be
\begin{split}
\label{cc}
\eta_{I}^{\alpha}=n_{I} \eta^{\alpha},\ \ \ \   \bar\eta^{I}_{\alpha}=\bar n^{I} \bar\eta_{\alpha},\ \ \ \ 
M_{J}^{\ \ I}=p_{1}\delta^{I}_{ J}-2 p_{2} n_{J}  \bar n^{I},\ \ \ \ 
\widehat M_{J}^{\ \ I}=q_{1} \delta^{I}_{J}-2 q_{2} n_{J} \bar n^{I}.
\end{split}
\ee
Here $n_{I}$  and $\bar n^{I}$ are two complex conjugated vectors which transform in the fundamental 
and anti-fundamental representation and determine the embedding of 
the $SU(3)$ subgroup in $SU(4)$\footnote{In the internal $R-$symmetry space $n_{I}$ identifies the direction preserved by the action of the  $SU(3)$ subgroup}. By rescaling $\eta^{\alpha}$ and $\bar\eta_{\alpha}$, we  can always choose $n_{I}\bar n^{I}=1$. The parameters $p_{i}$ and $q_{i}$  in the definition of $M$ and $\hat M$ instead  control the eigenvalues of the two matrices.

The  free  parameters appearing in \eqref{cc} can be then  constrained by imposing that the resulting Wilson loop is globally supersymmetric. This issue is subtle: the usual requirement $\delta_{\rm susy}\mathcal{L}(\tau)=0$ does not  yield  any 1/2 BPS solution indeed. We  just obtain   loop operators which are merely  bosonic ($\eta=\bar\eta=0$) and   at most $1/6$ BPS \cite{Drukker:2008zx,Rey:2008bh}. In order to obtain 1/2 BPS solution, we must replace  $\delta_{\rm susy}\mathcal{L}(\tau)=0$  with  the weaker condition \cite{Drukker:2009hy,Lee:2010hk}
\be
\label{var1}
\delta_{\rm susy}\mathcal{L}(\tau)=\mathfrak{D}_{\tau} G\equiv\partial_{\tau} G+ i\{ \mathcal{L},G],
\ee
where  the r.h.s. is the super-covariant derivative  constructed out of the connection  
$\mathcal{L}(\tau)$  acting on a super-matrix $G$ in $\mathfrak{u}(N|M)$. The requirement \eqref{var1} 
assures that the action of the relevant  supersymmetry charges translates into an infinitesimal
 super-{\it gauge} transformation for $\mathcal{L}(\tau)$ and thus   the ``{\it traced}'' loop operator is invariant. 

Now we shall recapitulate  the analysis  of \cite{Drukker:2009hy} leading to fix the  free parameters in \eqref{cc}.
However, for future convenience, we shall present the result  in a  {\it covariant} notation {\it i.e.} without referring to a specific form of the straight line.

We start by considering the structure of the infinitesimal gauge parameter in \eqref{var1}.
Since the supersymmetry transformation of the bosonic fields does not contain any derivative of the fields,  the super-matrix $G$ in \eqref{var1}
must be anti-diagonal 
\be
\label{susy2}
G=\begin{pmatrix}
0 & g_{1}\\
\bar g_{2}  & 0
\end{pmatrix}\ \ \Rightarrow \  \  \mathfrak{D}_{\tau} G=\begin{pmatrix}
\sqrt{\frac{2\pi}{k}}  |\dot x | (\eta_{I}\bar\psi^{I} \bar g_{2}+g_{1}\psi_{I}\bar \eta^{I}) &\mathcal {D}_{\tau} g_{1}\\
\mathcal{D}_{\tau }\bar g_{2}  & \sqrt{\frac{2\pi}{k}}  |\dot x | (\bar g_{2}\eta_{I}\bar\psi^{I} +\psi_{I}\bar \eta^{I} g_{1}) 
\end{pmatrix} .
\ee
Here $\mathcal{D}_{\tau}$ is the covariant derivative constructed out of  the {\it dressed} bosonic connections $\mathcal{A}$ and $\hat{\mathcal{A}}$ and given by
\be
\begin{aligned}
\mathcal{D}_\tau g_{1}&= \partial_\tau  g_{1} + i (\mathcal{A}\,  g_{1} - g_{1}\, \hat {\mathcal{A}})\,, \ \ \ \ 
\mathcal{D}_\tau \bar g_{2} &= \partial_\tau \bar g_{2} 
- i (\bar g_{2}\,\mathcal{A}-   \hat{\mathcal{A}}\, {\bar g}_{2}) .\,
\end{aligned}
\ee
 The condition \eqref{var1} for the anti-diagonal entries  first constrains   the form 
 of the spinor $\eta$ and $\bar\eta$  to obey the two conditions
\be
\label{orto1}
{(\dot{x}^{\mu}\gamma_{\mu})_{\alpha}^{\ \ \beta}}=\frac{1}{(\eta\bar\eta)} |\dot x|(\eta^{\beta} \bar\eta_{\alpha}+
\eta_{\alpha} \bar\eta^{\beta})\ \ \ \ \ \   (\eta^{\beta} \bar\eta_{\alpha}-
\eta_{\alpha} \bar\eta^{\beta})=(\eta\bar\eta)\delta^{\beta}_{\alpha},
\ee
which assure that the covariant derivatives appearing  in the supersymmetry transformations of  $\psi\bar\eta$ and $\eta\bar\psi$  are  only evaluated along the circuit.  The value of the parameters $p_{i}$ and $q_{i}$, appearing in the matrices $M$ and $\hat M$, is equal to $1$ for the same reason. 

The requirement \eqref{var1} for the diagonal entries  does not yield, instead, new conditions, simply fixing the normalization 
\be
\label{norm}
\eta\bar\eta=2 i.
\ee
In particular the vector $n_{I}$ continues to be unconstrained. 

The origin of the superconnection was also investigated from the point of view of the low-energy dynamics of heavy {\it ``W-bosons''}  in \cite{Lee:2010hk}. It was shown that when the theory is higgsed preserving half of the total supersymmetry, the corresponding low-energy Lagrangian enjoys a larger gauge invariance, given by the supergroup $U(N|M)$. The light fermions do not decouple from the dynamics, at variance with the case of ${\cal N}=4$ SYM, and their interactions with heavy $W$-bosons are described by $\eta_{I}^{\alpha},\bar{\eta}^{I}_{\alpha}$. The role of the mass-matrix is instead played by $M_{J}^{\ \ I},\widehat M_{J}^{\ \ I}$. This result unveils the physical nature of the potential related to the rectangular Wilson loops, constructed with 1/2 BPS lines in ABJ(M) theories.

Armed with the explicit form for the couplings, we can find  twelve supercharges \cite{Drukker:2009hy}   whose action on $\mathcal{L}(\tau)$ 
can be cast into the form \eqref{var1}. There are six supercharges of the Poincar\`e type\footnote{Recall that the counting is performed in terms of complex supercharges in Euclidean space-time,
while we use {\it real} supercharges in Minkowski signature.},
 \be
 \label{susypar}
 \bar\theta^{IJ\beta}
= (\bar\eta^{I\beta} \bar v^{J}-\bar\eta^{J\beta} \bar v^{I})
-i\epsilon^{IJKL} \eta_{K}^{\beta} u_{L}=
(\bar n^{I} \bar v^{J}-\bar n^{J} \bar v^{I})\bar\eta^{\beta}
-i\epsilon^{IJKL} n_{K} u_{L}\eta^{\beta},
 \ee 
parametrized by  two  vectors $u^{I}$ and $\bar v^{I}$ that satisfy $(n_{I}\bar v^{I})=(\bar n^{I} u_{I})=0$.  We remark that these vectors  are really  independent in Euclidean signature, while  $\bar{v}^{I}=u_{I}^{\dagger}$ as a result of  the reality conditions  present in  the Minkowski case.
 Next to the above $\bar\theta^{IJ}$  we  can also identify six super-conformal   charges\footnote{We have parametrized a generic supercharge as follows
\[\bar\Theta^{IJ}=\bar\theta^{IJ}+x^{\mu}\gamma_{\mu}\bar\vartheta^{IJ},
\]
where $\bar\theta^{IJ}$ generates the Poincar\`e supersymmetries, while $\bar\vartheta^{IJ}$ yields the conformal ones.
} $\bar\vartheta^{IJ\beta}$, whose structure is again given by an expansion of  the form \eqref{susypar}. 
The origin of this second set of supersymmetries is easily understood: they are obtained by combining
the  Poincar\`e supercharges \eqref{susypar} with a special conformal transformation in the direction associated to the straight-line.

The  analysis presented in \cite{Drukker:2009hy,Lee:2010hk} also provides the explicit form of the gauge function  
in terms of the  scalar fields, the spinor couplings and the supersymmetry parameters $\bar\theta^{IJ}$.
In our notation,  they take the form 
\be
\label{gaugefunc}
g_{1}=  2\sqrt{\frac{2\pi}{k}}  n_{K}(\eta\bar\theta^{KL})C_{L}\ \  \ \ \ \  \mathrm{and} \ \ \ \ \ \ \ 
\bar g_{2}=-\sqrt{\frac{2\pi}{k}}   \varepsilon_{IJKL} \bar n^{K}(\bar\theta^{IJ}\bar\eta) \bar C^{L}.
\ee 
Now we come back to  analyze the issue  of  supersymmetry invariance  for a generic  Wilson loop defined  by \eqref{superconnection} when its variation can be cast into the form \eqref{var1}.   The finite transformation  of the untraced operator 
\be
W(\tau_{1},\tau_{0})=\mathrm{P}\exp\left(-i \int_{\tau_{0}}^{\tau_{1}} d\tau \mathcal{L}(\tau)\right),
\ee
under the  gauge transformation  generated by $G(\tau)$ in \eqref{susy2}, can be written as \cite{Lee:2010hk}
\be
 W(\tau_{1},\tau_{0})\mapsto  U^{-1} (\tau_{1}) W(\tau_{1},\tau_{0}) U(\tau_{0}),
\ee 
where $U(\tau)=e^{i G(\tau)}$.  For a closed path $\gamma$ $[\gamma(\tau_{0})=\gamma(\tau_{1})]$, 
we must carefully consider the boundary conditions  obeyed by the gauge functions $g_{1}$ and 
$\bar g_{2}$ in order to define the gauge invariant operator. If they are {\it periodic}, {\it i.e.} $g_{1}(\tau_{0})=g_{1}(\tau_{1})$ and $\bar g_{2}(\tau_{0})=\bar g_{2}(\tau_{1})$, we find that $U(\tau_{0})=U(\tau_{1})$ and a gauge invariant operator  is obtained  by taking
the usual super-trace
\be
\mathcal{W}=\mathrm{Str}(W).
\ee
Actually it is the super-trace to be invariant under similitude transformations.  However we can have  different situations: in \cite{Drukker:2009hy} it was examined another 1/2-BPS loop, the circle, and pointed out that the function $g_{1}$ and $\bar g_{2}$ are anti-periodic in this case. Consequently the untraced operator, because $U(\tau_{1})=U^{-1}(\tau_{0})$, transforms as follows
\be
 W(\tau_{1},\tau_{0})\mapsto  U (\tau_{0}) W(\tau_{1},\tau_{0}) U(\tau_{0}).
\ee
To construct a supersymmetric operator, we first observe that
\be\mbox{\scriptsize $\begin{pmatrix} \mathds{1} & 0\\ 0 &-\mathds{1}\end{pmatrix}$}U(\tau_{0}) \mbox{\scriptsize$\begin{pmatrix} \mathds{1} & 0\\ 0 &-\mathds{1}\end{pmatrix}$}=U^{-1}(\tau_{0})
\ee
  for a gauge transformation generated by the matrix $G$ in \eqref{susy2}.
Then   the operator 
\be
\mathcal{W}=\mathrm{Str}\left[\mbox{\footnotesize $\begin{pmatrix} \mathds{1} & 0\\ 0 &-\mathds{1}\end{pmatrix}$}\mbox{$\displaystyle\mathrm{P}\!\exp\left(-i \int_{\tau_{0}}^{\tau_{1}} d\tau \mathcal{L}(\tau)\right)$}\right]\equiv\mathrm{Tr}\left[\mbox{$\displaystyle\mathrm{P}\!\exp\left(-i \int_{\tau_{0}}^{\tau_{1}} d\tau \mathcal{L}(\tau)\right)$}\right]
\ee
turns out to be  invariant.  In the case of a straight line the situation  is more intricate, since we deal with an open infinite circuit. The invariance under supersymmetry is recovered by choosing a set of suitable boundary conditions for the fields, in particular for the scalars appearing in the definition of $g_{1}$ and $\bar 
g_{2}$. The naive statement that they must vanish when $\tau=\pm \infty$ seems to leave open a double
possibility for defining a supersymmetric operator
\be
\label{Wpm}
\begin{split}
\mathcal{W}_{-}=&\frac{1}{N-M}
\mathrm{Str}\left[\mbox{$\displaystyle\mathrm{P}\!\exp\left(-i \int_{\tau_{0}}^{\tau_{1}} d\tau \mathcal{L}(\tau)\right)$}\right]\   {\mathrm{or}}\\\mathcal{W_{+}}=&\frac{1}{N+M}
\mathrm{Tr}\left[\mbox{$\displaystyle\mathrm{P}\!\exp\left(-i \int_{\tau_{0}}^{\tau_{1}} d\tau \mathcal{L}(\tau)\right)$}\right],
\end{split}
\ee
since $U(\pm\infty)=\mathds{1}$ and $W(\tau_{1},\tau_{0})$ itself is invariant. We shall consider in our explicit quantum computation the second possibility: as we will see, the trace is the correct option to generate BPS observables from closed contours, connected to ours through conformal transformations. It also seems to provide a result consistent with the interpretation of the Wilson loop in terms of quark-antiquark potentials.

\section{The generalized cusp}
We discuss here in detail  the Wilson loop observables we will study in the rest of the paper. After constructing the bosonic and fermionic couplings for the generalized cusp, we study the possibility to find novel BPS configurations. We determine the BPS conditions for the cusp parameters and derive the explict form of the related supercharges. Finally we map our new configurations on the sphere $S^2$, by means of conformal transformations, and we obtain a non-trivial BPS deformation of the BPS circle constructed in \cite{Drukker:2009hy}.

\subsection{Bosonic and fermionic couplings}
\label{subseccoupl}
\begin{figure}[ht]
\centering
	\includegraphics[width=.6 \textwidth]{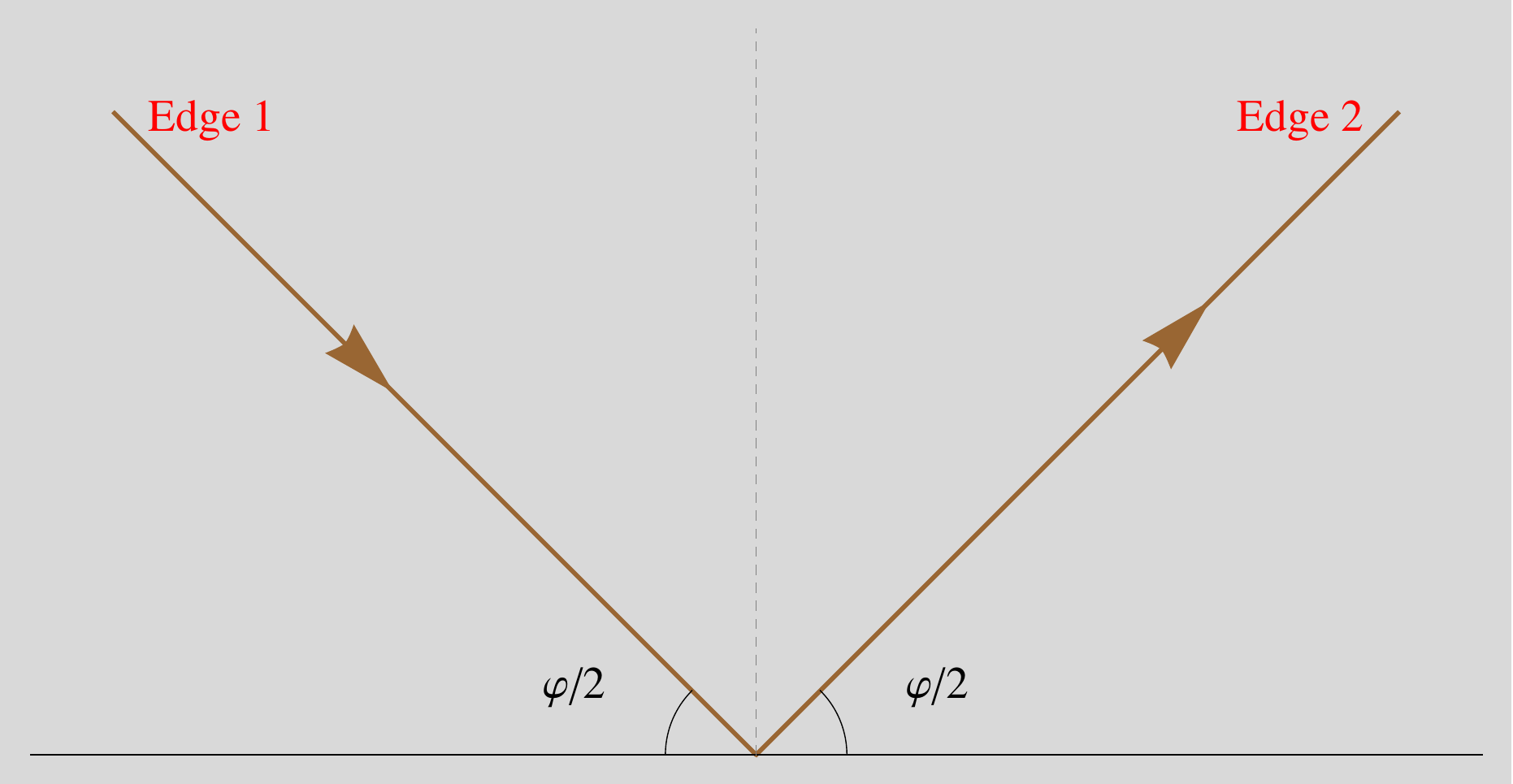}
\caption{\label{cusp1} Eq. \eqref{cusp} represents a planar cusp, whose angular extension is given by $\pi-\varphi$.}
\end{figure}

We start by considering  the  theory on the Euclidean space-time.  We shall consider two rays  in the plane $(1,2)$ intersecting at the origin as  illustrated  in fig.  \ref{cusp1}. The angle between the rays is $\pi-\varphi$, such that for $\varphi= 0$ they form a continuous straight line.
The path in fig. \ref{cusp1} is given by
\be
\label{cusp}
 x^{0}=0 \ \ \ \ \ x^{1}=s\cos\frac{\varphi}{2} \ \ \ \ \  x^{2}=|s|\sin\frac{\varphi}{2}\ \ \ \ \  -\infty\le s\le \infty.
\ee
 The fermionic couplings on each straight-line  possess the  factorized structure discussed in the previous section, {\it i.e.}
\be
\label{cuspcoupling}
\eta_{i M}^{\alpha}=n_{i M} \eta^{\alpha}\ \ \ \ \ \mathrm{and}\ \ \ \ \  \bar\eta^{M}_{i\alpha}=\bar n_{i}^{M} \bar\eta_{i\alpha}.
\ee
The  additional  index $i=1,2$ in \eqref{cuspcoupling} specifies which edge of the cusp  we are considering.
For $i=\mathrm{1}$,  $\eta_{1\alpha}$ is constructed as the  eigenspinor of eigenvalue $1$ of the matrix 
$\frac{\dot{x}_{1}^{\mu}}{|\dot x_{1}|}\gamma_{\mu}$:
\be
\frac{\dot{x}_{1}^{\mu}}{|\dot x_{1}|}\gamma_{\mu}\bar\eta_{1}=
\left(\cos\frac{\varphi}{2}\gamma_{1}-\sin\frac{\varphi}{2}\gamma_{2}\right)\bar\eta_{1}=\bar\eta_{1}\ \ \ \ \Rightarrow\ \ \ \ 
\bar\eta_{1\alpha} =i\begin{pmatrix} e^{i\frac{\varphi}{4}}\\ e^{-i\frac{\varphi}{4}}\end{pmatrix}
\ee
as a result\footnote{We have dropped a global phase in $\bar\eta_{1}$ since it can be eliminate through a global $U(1)$ redefinition of the matter fields.} of the  two constraints \eqref{orto1}.
Similarly
the spinor $\eta_{1}^{\alpha}$ obeys the hermitian conjugate of the above equation and thus
$\eta_{1}^{\alpha}\propto (\bar\eta_{1\alpha})^{\dagger}$. The  condition \eqref{norm}
fixes the relative normalization and we find
\be
\eta_{1}^{\alpha}= ( e^{-i\frac{\varphi}{4}}\ \ \  e^{i\frac{\varphi}{4}}).
\ee
On the other hand
the $R-$symmetry part of the couplings   is arbitrary and in fact $n_{1M}$ and $\bar n_{1}^{M}$ are totally unconstrained. For future convenience we  choose 
\be
n_{1M}=\mbox{\small$ \left(\cos\frac{\theta}{4}\ \ \sin\frac{\theta}{4}\ \ 0\ \ 0\right)$}
\ \  \  \ \mathrm{and}
\ \ \ \  
\bar n_{1}^{M}=\mbox{\footnotesize $
\begin{pmatrix}\cos\frac{\theta}{4}\\ \sin\frac{\theta}{4}\\ 0\\ 0\end{pmatrix}$}.
\ee
On the second edge, again as a result of  \eqref{orto1},   $\bar\eta_{2}$ must be  the 
eigenspinor of eigenvalue $1$ of $\frac{\dot{x}_{2}^{\mu}}{|\dot x_{2}|}\gamma_{\mu}$ and following the same route we get
\be
\bar\eta_{2\alpha} =i e^{i\delta}\begin{pmatrix} e^{-i\frac{\varphi}{4}}\\ e^{i\frac{\varphi}{4}}\end{pmatrix}\,\,\,\,\,\,\,\,\,\,\,
\eta_{2}^{\alpha}=e^{-i\delta} ( e^{i\frac{\varphi}{4}}\ \ \  e^{-i\frac{\varphi}{4}}).
\ee
The arbitrary phase $\delta$ cannot be reabsorbed into a redefinition of the fields without altering the structure of the fermionic couplings on the first edge. For the $R-$symmetry sector in \eqref{cuspcoupling}
we instead set 
\be
n_{2M}=\mbox{\small$ \left(\cos\frac{\theta}{4}\ \ -\sin\frac{\theta}{4}\ \ 0\ \ 0\right)$}
\ \  \  \ \mathrm{and}
\ \ \ \  
\bar n_{2}^{M}=\mbox{\footnotesize $
\begin{pmatrix}\cos\frac{\theta}{4}\\ -\sin\frac{\theta}{4}\\ 0\\ 0\end{pmatrix}$}.
\ee
The two matrices which couple the scalars are then determined through the relations \eqref{cc}, which give
$M$ and $\widetilde{M}$ in terms of $n$ and $\bar n$. On the two edges we have respectively
\be
M_{1J}^{\ \ I}=
\widehat M_{1J}^{\ \ I}=\mbox{\small $\left(
\begin{array}{cccc}
 -\cos \frac{\theta }{2}& -\sin \frac{\theta }{2} & 0 & 0 \\
 -\sin \frac{\theta }{2}& \cos\frac{\theta }{2} & 0 & 0 \\
 0 & 0 & 1 & 0 \\
 0 & 0 & 0 & 1
\end{array}
\right)$}\ \ \ \ \mathrm{and}\ \ \ \  M_{2J}^{\ \ I}=
\widehat M_{2J}^{\ \ I}=\mbox{\small $\left(
\begin{array}{cccc}
 -\cos \frac{\theta }{2} & \sin \frac{\theta }{2} & 0 & 0 \\
 \sin \frac{\theta }{2} & \cos\frac{\theta }{2} & 0 & 0 \\
 0 & 0 & 1 & 0 \\
 0 & 0 & 0 & 1
\end{array}
\right)$}.
\ee
\subsection{Intermediate BPS configurations}
\label{subsecBPS}
In the following we would like to explore if there is a choice of the  $(\varphi,\theta,\delta)$ such 
that the generalized cusp turns out to be  BPS\footnote{Of course we have an obvious one: $\varphi=\theta=\delta=0$.}.
These configurations may provide useful checks for the perturbative computations, but they can also provide
a tool for addressing the issue of nonperturbative computations \cite{Correa:2012at}.

Let us consider one of  the  Poincar\`e  supersymmetries  preserved by the first edge of the cusp in fig. \ref{cusp1}.
 As discussed in sec. \ref{sec2},  it admits the following expansion
 \be
 \label{charge1}
 \bar\theta_{1}^{IJ\beta}= (\bar\eta_{1}^{I\beta} \bar v^{J}_{1}-\bar\eta_{1}^{J\beta} \bar v^{I}_{1})
-i\epsilon^{IJKL} \eta_{1K}^{\beta} u_{1L}=
(\bar n^{I}_{1} \bar v^{J}_{1}-\bar n^{J}_{1} \bar v^{I}_{1})\bar\eta_{1}^{\beta}
-i\epsilon^{IJKL} n_{1K} u_{1L}\eta_{1}^{\beta},
\ee
 where $ \eta_{1K}$   $\bar\eta_{1}^{I}$ are the spinor couplings on the first line. The choice of the two vectors $u_{1}$ and $\bar v_{1}$ selects the  charge that  we are considering. We observe that if \eqref{charge1} defines 
a supercharge shared  with the second edge it must admit a similar  expansion in terms of the spinor couplings of  the second line. Expanding $\theta^{IJ}_{1}$ in the basis provided by $\eta_{2}$ and $\bar\eta_{2}$ , we obtain the following system of equation    
\begin{subequations}
\label{cond1}
\begin{align}
-i\epsilon^{IJKL} n_{2K} u_{2L}=&-i e^{i\delta}\epsilon^{IJKL} n_{1K} u_{1L}\cos\frac{\varphi}{2}+(\bar n^{I}_{1} \bar v^{J}_{1}-\bar n^{J}_{1} \bar v^{I}_{1})e^{i\delta} \sin\frac{\varphi}{2}\label{cond1a}\\
(\bar n^{I}_{2} \bar v^{J}_{2}-\bar n^{J}_{2} \bar v^{I}_{2})=& ~i e^{-i\delta}\epsilon^{IJKL} n_{1K} u_{1L}\sin\frac{\varphi}{2}+(\bar n^{I}_{1} \bar v^{J}_{1}-\bar n^{J}_{1} \bar v^{I}_{1})e^{-i\delta} \cos\frac{\varphi}{2}.\label{cond1b}
\end{align}
\end{subequations}
When  this set of equations can be consistently solved both for $u_{2}$ and $\bar v_{2}$, we have found a 
candidate BPS configuration.   To begin with, we shall multiply \eqref{cond1a} by  $ n_{2J}$. The resulting
condition does not contain $u_{2}$ and $\bar v_{2}$: it is actually a constraint on the super-charge $\bar\theta_{1}^{IJ}$
\be
\label{consist1}
\epsilon^{IJKL}n_{2 J} n_{1K} u_{1L}\cos\frac{\varphi}{2}+i \left(\bar n^{I}_{1} (n_{2J}\bar v^{J}_{1})-\cos\frac{\theta}{2} \bar v^{I}_{1}\right) \sin\frac{\varphi}{2}=0.
\ee
If we project  \eqref{consist1} onto the direction $n_{1I}$, we have immediately
\be
(n_{2 J} \bar v_{1}^{J}) \sin\frac{\varphi}{2} e^{i\delta}=0\ \ \ \ \Rightarrow\ \ \ \  (n_{2 J} \bar v_{1}^{J}) =0.
\ee
and consequently from \eqref{consist1}
\be
\label{v1}
\bar v^{I}_{1} =-i\frac{\cot\frac{\varphi}{2}}{\cos\frac{\theta}{2}}
\epsilon^{IJKL}n_{2 J} n_{1K} u_{1L}.
\ee
Next we multiply  \eqref{cond1b} by $\epsilon_{IJMN} \bar n_{2}^{M}$. Again the dependence on $u_{2}$ and $v_{2}$ drops out and we end up with the following constraint
\be
0= i\left (\cos\frac{\theta}{2} u_{1N}-n_{1N} (\bar n_{1}^{M}u_{1M})\right)\sin\frac{\varphi}{2}+\epsilon_{IJMN}\bar n^{I}_{1} \bar v^{J}_{1}\bar n_{2}^{M}\cos\frac{\varphi}{2},
\ee
which is equivalent to 
\be
\label{u1}
u_{1 L}=i\frac{\cot\frac{\varphi}{2}}{\cos\frac{\theta}{2}} \epsilon_{RSML}\bar n^{R}_{1} \bar v^{S}_{1}\bar n_{2}^{M}.
\ee
The relations \eqref{v1} and \eqref{u1} are consistent  if and only if
\be
\bar v^{I}_{1} =-i\frac{\cot\frac{\varphi}{2}}{\cos\frac{\theta}{2}}
\epsilon^{IJKL}n_{2 J} n_{1K} u_{1L}=\frac{\cot^{2}\frac{\varphi}{2}}{\cos^{2}\frac{\theta}{2}}
\epsilon^{IJKL}n_{2 J} n_{1K} \epsilon_{RSML}\bar n^{R}_{1} \bar v^{S}_{1}\bar n_{2}^{M}=
\frac{\cot^{2}\frac{\varphi}{2}}{\cot^{2}\frac{\theta}{2}} v_{1}^{I},
\ee
namely $\cot^{2}\frac{\varphi}{2}=\cot^{2}\frac{\theta}{2}.$  
Therefore for $\theta=\pm\varphi$ we expect that  the loop operator defined in the previous section is  BPS.
In fact for this value of the parameters we can find an explicit and simple solution for $u_{2}$ and $\bar v_{2}$
\begin{align}
u_{2I}=&e^{i\delta } u_{1I}\ \ \mathrm{and} \ \  \bar v^{I}_{2}= e^{-i\delta} v^{I}_{1}.
\end{align}
We remark we still need another property to fully confirm the presence of a BPS configuration at $\theta=\pm\varphi$: we should prove that the gauge functions $g_{1}$ and $\bar g_{2}$ on the two edges define a globally well-defined gauge transformation, which is continuous when we cross the cusp.  The values of  $g_{1}$ 
on the two edges are given by
\begin{equation}
\mbox{\footnotesize \sc [First Edge]}: 4i\sqrt{\frac{2\pi}{k}}  \bar v^{L}_{1}C_{L} \ \ \  \ \ \ \ \ \ \ \ \ \ \ 
\mbox{\footnotesize  \sc [Second Edge]}:4i\sqrt{\frac{2\pi}{k}}\bar v^{L}_{1}e^{-i\delta}C_{L} ,\
\end{equation}
while for $\bar g_{2}$ we find
\be
\mbox{\footnotesize \sc [First Edge]}: 4\sqrt{\frac{2\pi}{k}}  u_{1L}\bar C^{L} \ \ \  \ \ \ \ \ \ \
\mbox{\footnotesize  \sc [Second Edge]}: 4\sqrt{\frac{2\pi}{k}} e^{i\delta} u_{1L}\bar C^{L}.
 \ee
 Only  for $\delta=0$ the two gauge function are continuous through the cusp.  Summarizing, 
 for the $\theta=\pm \varphi$ and $\delta=0$ the generalized cusp of fig.\ref{cusp1}
 is BPS. The preserved Poincar\`e supercharges in terms of the quantity of the first line  can be then 
 written in the following two equivalent ways
  \begin{align}
  \label{34}
 \bar\theta_{1}^{IJ\beta}=&
-i\left(\frac{\bar n^{I}_{1} }{\sin\frac{\varphi}{2}}
\epsilon^{JMKL}n_{2 M} n_{1K} u_{1L}-\frac{\bar n^{J}_{1} }{\sin\frac{\varphi}{2}}
\epsilon^{IMKL}n_{2 M} n_{1K} u_{1L}\right)\bar\eta_{1}^{\beta}
-i\epsilon^{IJKL} n_{1K} u_{1L}\eta_{1}^{\beta}=\nonumber\\
 =& 
(\bar n^{I}_{1} \bar v^{J}_{1}-\bar n^{J}_{1} \bar v^{I}_{1})\bar\eta_{1}^{\beta}
-\frac{1}{\sin\frac{\varphi}{2}}\left(\left(\bar n^{I}_{2}- \cos\frac{\varphi}{2} \bar n^{I}_{1}\right) 
\bar v^{I}_{1}-\left(\bar n^{J}_{2}- \cos\frac{\varphi}{2} \bar n^{J}_{1}\right) 
\bar v^{I}_{1}\right)\eta_{1}^{\beta}.
\end{align}
The vector  $ u^{I}_{1}$ in the first line  of \eqref{34} and the vector $\bar v^{I}_{1}$ in the second one
must obey  $ (n_{2 J} \bar v_{1}^{J})=(n_{1 J} \bar v_{1}^{J})=0$ and $ (\bar n_{2 J} u_{1}^{J})=(\bar n_{1 J} u_{1}^{J})=0$ respectively. Thus we have   two shared  Poincar\`e supercharges. 

A remark on the conformal supercharges $\bar\vartheta^{IJ}$ is now in order: for each edge of the cusp they admit the same expansion \eqref{susypar} which was obtained for the Poincar\`e ones. The above analysis implies therefore that there are two shared superconformal charges as well.

 \subsection{Mapping  the cusp to the spherical wedge}
 \label{subsecwedge}
 
Recently  \cite{Correa:2012at}  it was noticed that the DGRT spherical wedge \cite{Drukker:2007qr}, which is a BPS loop
 operator, can be used to extract nonperturbative information about the generalized cusp
 \begin{wrapfigure}[14]{l}{68mm}
\vskip -.9cm
    \includegraphics[width=.40\textwidth, height=.35 \textwidth]{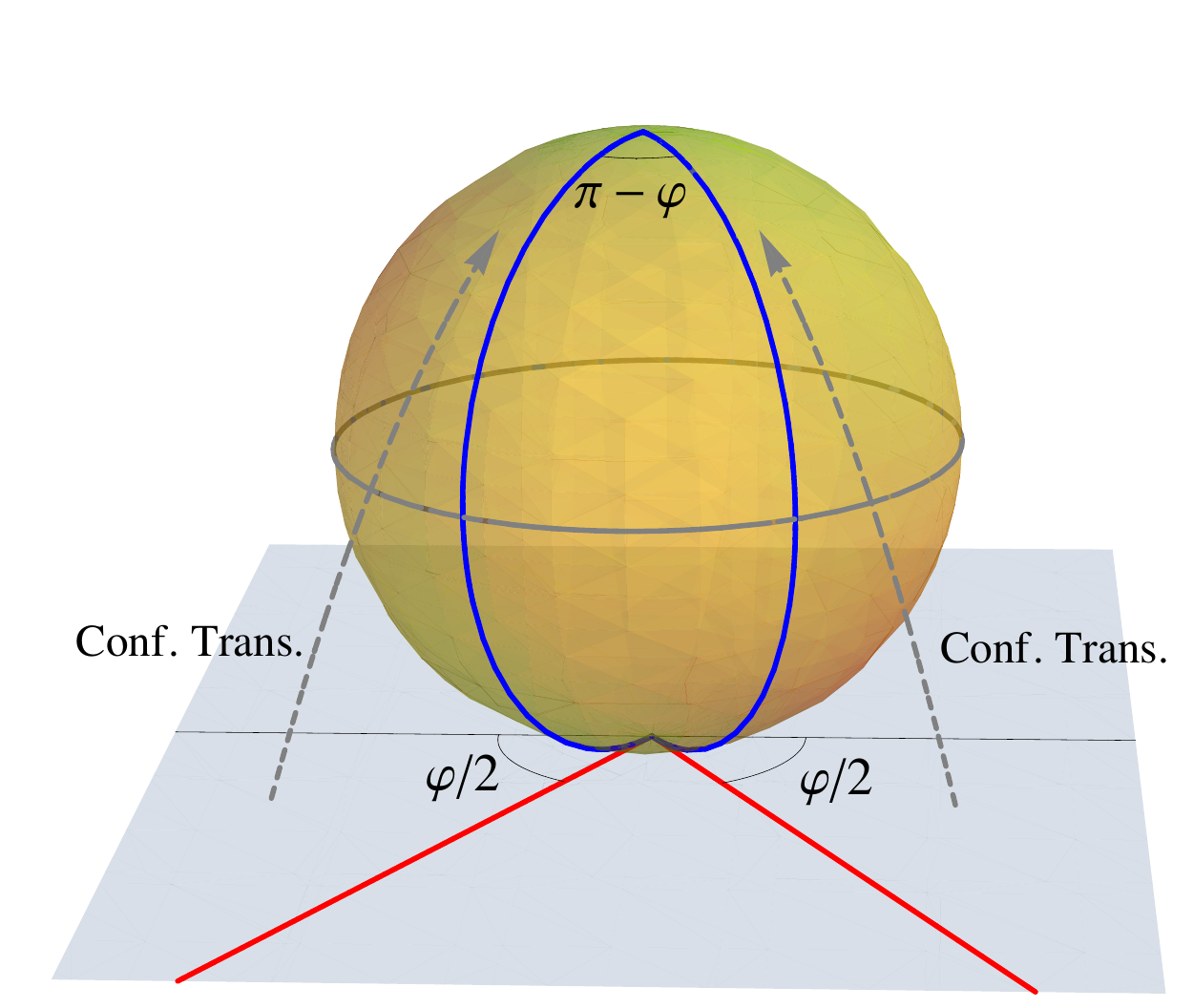}
\vskip -.3cm
\caption{\label{cusptowedge}The cusp in the plane $x^{0}=1/2$ is mapped into the spherical wedge under the conformal transformation generated by the vector $(1,0,0)$.}
\end{wrapfigure}
 in $\mathcal{N}=4$ SYM, since its value is known at all order in the coupling constant.  It was argued in \cite{Drukker:2007qr} that the exact quantum result is obtained from the ordinary circular Wilson loop computation, with $\lambda$  replaced by
 \be
\lambda \mapsto \tilde\lambda=\lambda\frac{4 A_{1} A_{2}}{A^{2}},
\ee
 where $A_{1}$ and $A_{2}$ are the areas of the two sides of the contour and $A = A_{1} + A_{2}$ is the total
area of the two-sphere. Since the DGRT spherical wedge can be related to the BPS  configuration of the generalized cusp in $\mathcal{N}=4$ through a special conformal transformation, it is tempting to 
investigate what is the image of the operator defined in subsect. \ref{subseccoupl} when we map the plane
supporting it  onto a sphere $S^{2}$.
 
Our starting point is not the  cusp in the plane $(1,2)$, as  in subsect. \ref{subseccoupl}, but for simplicity one located   in the plane $x^{0}=1/2$, {\it i.e.}
\be
\label{cusp3}
x^{0}=1/2 
\ \ \ \ \ 
x^{1}=s\cos\frac{\varphi}{2} \ \ \ \ \  x^{2}=|s|\sin\frac{\varphi}{2} \ \ \  \  -\infty\le s,\le \infty.
\ee 
The  plane  $x^{0}=1/2$ can be mapped into the usual unit sphere  centered in the origin through the special conformal transformation generated by the vector  $b^{\mu}=(1,0,0)$
\be
\label{conf2}
 y^{0}=\frac{x^{0}- x^{\mu} x_{\mu}}{1-2 x^{0}+ x^{\mu} x_{\mu}},\ \ \ \ \ 
 y^{1}=\frac{x^{1}}{1-2x^{0}+ x^{\mu} x_{\mu}},\ \ \ \ \ y^{2}=\frac{x^{2}}{1-2x^{0}+ x^{\mu} x_{\mu}}.
\ee
Then  the image of the original contour is the wedge  illustrated in fig. \ref{cusptowedge}.  The new path 
in terms of $s$ is given by 
%
\be
\label{wedges}
\!\!
y^{0}=\frac{\frac{1}{4}-s^{2}}{\frac{1}{4}+s^{2}},\ \ y^{1}=\frac{ s\cos\frac{\varphi}{2}}{\frac{1}{4}+s^{2}},\ \ 
y^{2}=\frac{ |s|\sin\frac{\varphi}{2}}{\frac{1}{4}+s^{2}}.
\ee
When $s$ ranges from $-\infty$ to $\infty$ we move from  the south  of the sphere $(-1,0,0)$ to the north pole $(1,0,0)$ [$s=0$] and back to the south pole. The usual parametrization in polar coordinates is recovered by 
performing the substitution  $s=\frac{1}{2}\tan\frac{t}{2}$ in \eqref{wedges}
\be
\!\!
y^{0}\!=\!\cos t,\  y^{1}\!=\sin t\cos\frac{\varphi}{2},\ 
y^{2}=|\sin t|\sin\frac{\varphi}{2}
\ee
where $ t\in[-\pi,\pi]$.
The effect of the change of coordinates \eqref{conf2} on the fermionic couplings  $\eta_{I}$ and $\bar\eta^{I}$ is more  interesting and straightforward  to  determine once we recall the result of a finite special conformal transformation on a spinor field\footnote{In three dimensions, the finite form of a special conformal transformation on a spinor field is
\[
\psi^{\prime }( y)=(1-2 (b\cdot x) +b^{2} x^{2}) ^{1/2}
 (\mathds{1}-b^{\mu}\gamma_{\mu} x^{\alpha}\gamma_{\alpha})\psi(x).
 \] } (see {\it e.g.} \cite{Ho}). Comparing the anti-diagonal term of the matrix \eqref{superconnection} in the two different coordinates, we find for instance
 \be
 |\dot y| \eta^{\prime}_{I}\bar\psi^{\prime I}( y)=|\dot{x}|(1-2 (b\cdot x) +b^{2} x^{2}) ^{-1/2}
 \eta^{\prime }_{I}  (\mathds{1}-b^{\mu}\gamma_{\mu} x^{\alpha}\gamma_{\alpha})\bar\psi^{I }(x)=
 |\dot{x}|\eta_{I}\bar\psi^{I}(x),
 \ee
 namely
\be
\label{transfeta}
  \eta^{\prime }_{I}  =
 \eta_{I} \frac{(\mathds{1}- x^{\alpha}\gamma_{\alpha}b^{\mu}\gamma_{\mu})}{\sqrt{1-2 (b\cdot x) +b^{2} x^{2}}}.
\ee
In other words, in the case of  the spinor couplings, the  effect of mapping the cusp into spherical  wedge  translates into a local rotation defined by the matrix appearing in \eqref{transfeta}.  We have a different rotation on each edge
\begin{subequations}
 \paragraph{\sc Edge I:}  $(s<0)$
 \be
  \label{rotI}
R_{I}=\left(
\begin{array}{cc}
 \frac{1}{\sqrt{4 s^2+1}} & \frac{2 e^{\frac{i \varphi }{2}} s}{\sqrt{4 s^2+1}} \\
 -\frac{2 e^{-\frac{i \varphi }{2}} s}{\sqrt{4 s^2+1}} & \frac{1}{\sqrt{4 s^2+1}}
\end{array}
\right)=\left(
\begin{array}{cc}
 \cos \frac{t}{2} & e^{\frac{i \varphi }{2}} \sin\frac{t}{2} \\
 -e^{-\frac{i \varphi }{2}} \sin \frac{t}{2} & \cos \frac{t}{2}
\end{array}
\right)
\ee
 \paragraph{\sc Edge II:}  $(s>0)$
 \be
 \label{rotII}
R_{II}=\left(
\begin{array}{cc}
 \frac{1}{\sqrt{4 s^2+1}} & \frac{2 e^{-\frac{i \varphi }{2}} s}{\sqrt{4 s^2+1}} \\
 -\frac{2 e^{\frac{i \varphi }{2}} s}{\sqrt{4 s^2+1}} & \frac{1}{\sqrt{4 s^2+1}}
\end{array}
\right)=\left(
\begin{array}{cc}
 \cos \frac{t}{2} & e^{-\frac{i \varphi }{2}} \sin \frac{t}{2} \\
 -e^{\frac{i \varphi }{2}} \sin\frac{t}{2}& \cos\frac{t}{2}
\end{array}
\right).
 \ee
 \end{subequations}
 
 We have expressed the rotation matrices   in terms of both the original parameter $s$ and  the parameter 
 $t=2\arctan(2 s)$.  Now  the fermionic couplings on the two sides of the wedge  are obtained by rotating the old ones  by means of the two matrices  $R_{I}$ and $R_{II}$
 \be
 \upsilon_{1}^{\alpha}=(\eta_{1}R_{I})^{\alpha}=\mbox{\small$\left(e^{-i\frac{\varphi}{4}}\left(\cos \frac{t}{2}-\sin\frac{t}{2}\right)  \  \ e^{i\frac{\varphi}{4}}\left(\cos \frac{t}{2}+\sin\frac{t}{2}\right)\right)$}
 \ee
  \be
 \upsilon_{2}^{\alpha}=(\eta_{2}R_{II})^{\alpha}=\mbox{\small$\left(e^{i\frac{\varphi}{4}}\left(\cos \frac{t}{2}-\sin\frac{t}{2}\right)  \  \ e^{-i\frac{\varphi}{4}}\left(\cos \frac{t}{2}+\sin\frac{t}{2}\right)\right)$}
 \ee
 and obviously $\bar\upsilon_{1,2\alpha}=i (\upsilon_{1,2}^{\alpha})^{\dagger}$.  The  matrices $M$ and 
 $\widehat M$ which  couple the scalars to the loop are instead  unaffected by the  special conformal transformation. 
 
 Next we consider the effect of the change of variables \eqref{conf2} on  the  preserved  super-charges of subsect. \ref{subsecBPS}.  The super-conformal Killing spinors  $(\bar{\theta}^{IJ},\bar\vartheta^{IJ})$   of the cusp    are  transformed into 
 \be
(\bar{\theta}^{IJ},\bar\vartheta^{IJ})\mapsto 
(\bar{\theta}^{IJ},\bar\vartheta^{'IJ})=(\bar{\theta}^{IJ},\bar\vartheta^{IJ}+\bar\theta^{IJ}(b^{\mu}\gamma_{\mu}))=
(\bar{\theta}^{IJ},\bar\vartheta^{IJ}+\bar\theta^{IJ}\gamma_{0}).
\ee
The loop operator defined by this spherical wedge is preserved by the  conformal Killing spinors with a structure given by
$\bar\Theta^{IJ}=\bar\theta^{IJ} (1+y^{\mu}\gamma_{0}\gamma_{\mu})+y^{\mu}\bar\vartheta^{IJ}\gamma_{\mu}. $ 

In doing the conformal transformation we have effectively compactified the contour and we have to understand what happens to the gauge functions at the south pole: continuity of the gauge transformations at north pole is instead inherited by the BPS properties of the open cusp. It  is a straightforward exercise to compute the spinor contractions which are relevant in determining the gauge function $g_{1}$ and $\bar g_{2}$ at the points $t=-\pi$ and  $t=\pi$:
\be
\label{ghfun}
\begin{array}{llll}
\left. g_{1}\right|_{t=-\pi}=   4 i\sqrt{\frac{2\pi}{k}} (\bar{\tilde v}^{L}_{1}C_{L})
& & &
\left. g_{1}\right|_{t=\pi}= 
  -4 i\sqrt{\frac{2\pi}{k}} (\bar{\tilde v}^{L}_{2}C_{L})\\
  & & &\\
\left.\bar g_{2}\right|_{t=-\pi}=
4 \sqrt{\frac{2\pi}{k}}   (\tilde u_{1L}\bar C^{L})
& & &
\left.\bar g_{2}\right|_{t=\pi}=
-4 \sqrt{\frac{2\pi}{k}}   (\tilde u_{2L}\bar C^{L})
\end{array}
\ee 
where we used that $\vartheta^{IJ}$ admits  the same expansion  of the Poincar\`e supercharges in terms of two vectors $\tilde u_{i}$ and  $\bar{\tilde v}_{i}$. These last two vectors will obey the same constraint of $u_{i}$ and $\bar v_{i}$ and in particular for the shared supercharges $\bar{\tilde v}^{I}_{1}=\bar{\tilde v}^{I}_{2}$ and 
$\tilde u^{I}_{1}=\tilde u^{I}_{2}$. We see the gauge functions \eqref{ghfun} are anti-periodic and consequently, to have a BPS loop,  we have to take the trace to obtain a {\it supersymmetric} wedge on $S^{2}$. This is consistent with the result of \cite{Drukker:2009hy}, our wedge being a non-trivial BPS deformation of the BPS circle. It is interesting that within our construction the antiperiodicity of gauge functions appears as an effect of the conformal mapping, rather that being assumed from the beginning. The presence of such supersymmetric configurations suggests also that it should be possible to construct a general class of BPS loops on $S^2$, representing the ABJ analogue of the DGRT loops of ${\cal N}=4$. The explicit construction and the quantum analysis of this new family, as well as of the analogue of Zarembo's loops in superconformal Chern-Simons theories, will be the subjects of a separate publication \cite{CGMS}.

\section{Quantum results}
We shall compute the expectation value of the generalized {\it cusp } operator up to the  second order in the coupling constant $\left(\frac{2\pi}{\kappa}\right)$. So far there are very few results about the perturbative properties of supersymmetric Wilson loops in ABJ(M) theories and they are all strictly confined to the 1/6 BPS {\it bosonic} case \cite{Drukker:2008zx,Chen:2008bp,Rey:2008bh}. We remark that even the matrix model \cite{Drukker:2009hy} - believed  to capture the exact result for the 1/2 BPS circle -  has not been verified by explicit Feynman diagrams computations.

The quantum holonomy of 
the super-connection ${\cal L}$ 
 in a representation ${\cal R}$ of the supergroup $U(N|M)$ is by definition

\be
\label{eq:loopexpectationvalue}
\left\langle \mathcal{W}_{\cal R} \right\rangle= \frac{1}{{\rm dim}_{\cal R}}\int {\cal D}[A,\hat{A},C,\bar{C},\psi,\bar{\psi}]~
{\rm e}^{-S_{\rm ABJ}}~{\rm Tr}_{\cal R} \left[
  {\rm P} \exp \left(- i\int_{\Gamma} d\tau\, {\cal L}(\tau) \right) \right],
\ee
where $S_{ABJ}$ is the Euclidean action for $ABJ(M)$ theories (the part relevant for our computation is spelled out in app. \ref{ABJ}).  In the following $\mathcal{R}$ will be chosen to be the fundamental representation.

 In order to evaluate $\left\langle \mathcal{W}_{\cal R} \right\rangle$  we  shall first focus our attention  on the upper left $N\times N$ block of the super-matrix appearing  in \eqref{eq:loopexpectationvalue}.
 For this sub-sector the trace is obviously taken in the fundamental representation ${\bf N}$ of the first gauge group. The result for the lower 
 diagonal  block can be then recovered from this analysis by exchanging $N$ with $M$. Our perturbative computation  requires to expand the path-exponential   in \eqref{eq:loopexpectationvalue} up to  the fourth order. The  terms in this expansion relevant for the upper block include both  {\it bosonic} and {\it fermionic} monomials:
\begin{align}
\label{expaloop}
\mathbb{W}_{\mathbf{N}}&={\rm Tr}_{\mathbf{N}}\left[1+i\int_\Gamma d\tau_1{\cal A}_1-\int_{\Gamma}d\tau_{\mbox{\tiny $\displaystyle1\!\!>\!\! 2$}}\Biggl({\cal A}_1{\cal A}_2-(\eta\bar{\psi})_1(\psi\bar{\eta})_2 \Biggr) \right.\nonumber\\
&-i\int_{\Gamma}d\tau_{\mbox{\tiny $\displaystyle1\!\!>\!\! 2\!\!>\!\!3$}}\Biggl( {\cal A}_1{\cal A}_2{\cal A}_3+\frac{2\pi}{k}[(\eta\bar{\psi})_1(\psi\bar{\eta})_2{\cal A}_3 
+(\eta\bar{\psi})_1\hat{\cal A}_2 (\psi\bar{\eta})_3+{\cal A}_1(\eta\bar{\psi})_2(\psi\bar{\eta})_3]\Biggr)\nonumber\\
&\left.+\int_{\Gamma}d\tau_{\mbox{\tiny $\displaystyle1\!\!>\!\! 2\!\!>3\!\!>4$}}\left(\left(\frac{2\pi}{\kappa}\right)^{2}(\eta\bar{\psi})_1(\psi\bar{\eta})_2(\eta\bar{\psi})_3(\psi\bar{\eta})_4+\mathcal{A}_{1}\mathcal{A}_{2}\mathcal{A}_{3}\mathcal{A}_{4}-\right.\right.\\ 
&-\left(\frac{2\pi}{\kappa}\right)\mathcal{A}_{1}\mathcal{A}_{2}(\eta\bar{\psi})_3(\psi\bar{\eta})_4-\left(\frac{2\pi}{\kappa}\right)\mathcal{A}_{1}(\eta\bar{\psi})_2\hat{\mathcal{A}}_{3}(\psi\bar{\eta})_4-\left(\frac{2\pi}{\kappa}\right)(\eta\bar{\psi})_1\hat {\mathcal{A}_{2}}\hat{\mathcal{A}_{3}}(\psi\bar{\eta})_4-\nonumber\\
&\left.\left.
-\left(\frac{2\pi}{\kappa}\right)\mathcal{A}_{1}(\eta\bar{\psi})_2(\psi\bar{\eta})_3\mathcal{A}_{4}-\left(\frac{2\pi}{\kappa}\right)(\eta\bar{\psi})_1\hat{\mathcal{A}}_{2}(\psi\bar{\eta})_3\mathcal{A}_{4}-\left(\frac{2\pi}{\kappa}\right)(\eta\bar{\psi})_1(\psi\bar{\eta})_2\mathcal{A}_{3}\mathcal{A}_{4}
\right)\right].\nonumber
\end{align}
In \eqref{expaloop} we have introduced a shorthand notation for the circuit parameter dependence of the fields, namely ${\cal A}_i = {\cal A}(x_i)$ with $x_i = x(\tau_i)$. We have also suppressed the spinor and $SU(4)_R$ indices and choosen the parametrization with $|\dot{x}|=1$. The expression above is not symmetric in the exchange of the two gauge groups: the 
 symmetry between them will be recovered when considering also the contribution coming from the lower right $M\times M$ block.

\subsection{One-loop analysis}
\label{Leadord}
\begin{figure}[ht]
\centering{
    \includegraphics[width=.97\textwidth, height=.15\textwidth]{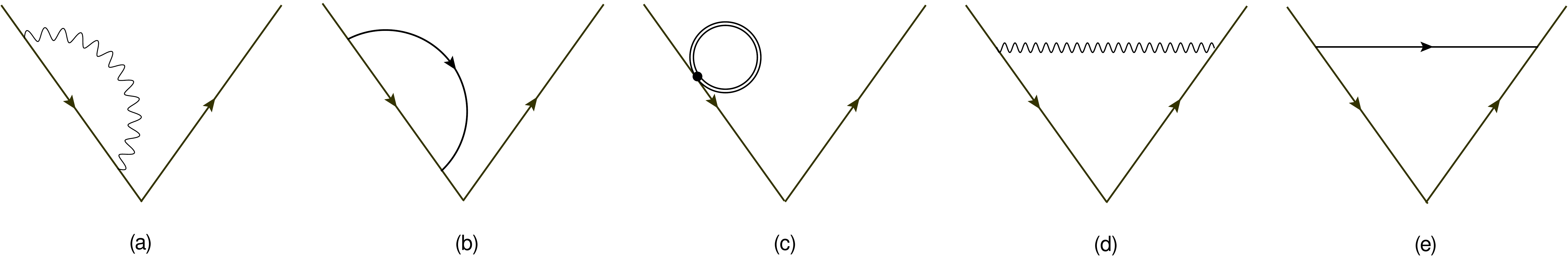}}
\vskip -.3cm
\caption{\label{fig:oneloopgraphs}At one-loop order there are only two classes of diagrams: single-edge diagrams [(a), (b) and (c)] and  exchange diagrams [(d) and (e)]. The scalar field enters the loop operator only through the composite bilinear $M^I_J C_I \bar{C}^J$, and its conjugate $\widehat{M}^I_J \bar{C}^J C_I$, hence the exchange of a single scalar is not permitted.
}
\end{figure}

The first non-trivial contributions are proportional to $\left(\frac{2\pi}{\kappa}\right)$ and involve both bosonic and fermionic diagrams. They are listed in fig. \ref{fig:oneloopgraphs}.  Differently from what occurs for  the  ${\cal N}=4$ generalized cusp, the  diagrams  which involve only one edge do not vanish when we add them up. The situation is a little more intricate and it is actually convenient to deal with them separately, also in view of the two-loop computation.

The evaluation of the  diagrams in fig. \ref{fig:oneloopgraphs} obviously encounter UV divergences which originate  from the part of the integration region where the propagator endpoints  coincide. To tame these 
divergences  we will extensively  use dimensional regularization. However regularizing Chern-Simons-matter theories going
off-dimensions raises some concerns because of the presence of the anti-symmetric $\epsilon^{\mu\nu\rho}$ tensor.
 We will follow the DRED scheme, shifting the dimension to $d=3-2\epsilon$ while keeping the Dirac algebra and $\epsilon^{\mu\nu\rho}$
tensor strictly  in 3 dimensions. Note that this breaks the conformal invariance introducing a mass scale $\mu^{2\epsilon}$ that keeps the action dimensionless. We will also need an explicit IR regulator $L$, representing the finite length of the two rays forming the cusp: because of the underlying conformal invariance we expect that it could be always scaled away, combining into the $(\mu L)^{2\epsilon}$ to some powers that weights the relevant Feynman integrals.

We start by considering the bosonic diagrams:  the scalar tadpole of fig. \ref{fig:oneloopgraphs}.(c), that 
originates from the first term in the expansion  \eqref{expaloop},  vanishes in our regularization procedure and we can safely forget its existence in the following. The  bosonic contributions in  fig. \ref{fig:oneloopgraphs}.(a) and \ref{fig:oneloopgraphs}.(d)  stems from the term with two gauge fields in \eqref{expaloop} and they can be cast  into a single path-ordered integral,
\be
\label{eq:gluon1loop}
 \mathfrak{B}^{(1)}=
 {\rm i} N^2\left( \frac{\mu^{2\epsilon}}{\kappa}\right)
\frac{\Gamma(\frac{3}{2}-\epsilon)}{\pi^{\frac{1}{2}-\epsilon}}\,\int_{\Gamma}d\tau_{\mbox{\tiny $\displaystyle1\!\!>\!\! 2$}}\, 
\frac{\epsilon_{\mu\nu\rho} \dot{x}_1^\mu \dot{x}_2^\nu
(\dot{x}_1 -\dot{x}_2 )^\rho}{|x_1-x_2|^{3-2\epsilon}}=0,
\ee
whose integrand vanishes for any planar loop due to the antisymmetry of the $\epsilon-$tensor\footnote{In general one should also take into account the possibility to have framing contribution \cite{Witten:1988hf}. We assume here that our computation can be consistently done at zero framing.}. We have used here the explicit form of the Chern-Simons propagator in position space, presented in App. \ref{FRSS}. We remark that the same result would have been obtained if we have used 1/6 BPS lines, in spite of the different structure of the mass-matrix $M^I_J$.

\noindent
Next we discuss the fermionic diagrams  in fig. \ref{fig:oneloopgraphs}.  They represent the true  novelty of the present  calculation and originate from taking the vacuum expectation value of the fermionic bilinear in the first line of (\ref{expaloop})
\be
\label{K1}
 \mathfrak{F}^{(1)}=\left(\frac{2\pi}{\kappa} \right)\mu^{2\epsilon}\int_{\tau_{2}<\tau_{1}}\!\!\!\!\! d\tau_{1}d\tau_{2}~\langle \mathrm{Tr}_{\bf N}\left[(\eta\bar\psi)_{1} (\psi\bar\eta)_{2} \right]\rangle=\left(\frac{2\pi}{\kappa} \right)\mu^{2\epsilon}({\cal F}_{b}+{\cal F}_{e}),
\ee
where we have denoted with ${\cal F}_{b}$ and ${\cal F}_{e}$ the contributions corresponding to  the graphs  \ref{fig:oneloopgraphs}.(b) and  \ref{fig:oneloopgraphs}.(e) respectively. At the lowest order, the vacuum expectation value in \eqref{K1} is simply obtained by contracting the fermion propagator \eqref{fermionpropagator} with the
spinors $\eta$ and $\bar\eta$. We find 
\be
\label{contra}
\langle \mathrm{Tr}_{\bf N}\left[(\eta\bar\psi)_{1} (\psi\bar\eta)_{2} \right]\rangle=i M N \frac{\Gamma(1/2-\epsilon)}{4\pi^{3/2-\epsilon}}
\eta_{L1}\gamma^{\mu}\bar\eta_{2}^{L}\partial_{x^\mu_{1}}\left(\frac{1}{(x_{12}^{2})^{1/2-\epsilon}}\right),\ee
where  $x_{ij}=x_i-x_j$. The fermion bilinear $\eta_{L1}\gamma_{\mu}\bar\eta_{2}^{L}$ can be readily evaluated for a general contour (and for general parametrization), thanks to the factorized form \eqref{cc} of the spinor couplings and to the identity (\ref{rg6}). We have
\be
\label{rg6a}
(\eta_{L1}\gamma^{\mu}\bar\eta_{2}^L)=-2\frac{(n_{1}\cdot\bar n_{2})}{(\eta_{2}\bar\eta_{1})}\left[\frac{\dot{x_{1}}^{\mu}}{|\dot x_{1}|}+\frac{\dot{x_{2}}^{\mu}}{|\dot{x}_{2}|}+\frac{1}{2}\frac{\dot{x_{1}}^{\lambda}}{|\dot{x}_{1}|}
\frac{\dot{x_{2}}^{\nu}}{|\dot x_{2}|}{\rm i}\epsilon^{\lambda\mu\nu} \right].
\ee
Since $\dot{x}_{1}$, $\dot{x}_{2}$ and $x_{12}=-x_{21}$ lay on the same plane, we can drop the wedge product in \eqref{rg6a} and we obtain the  following  result
\be
\label{propla}
\langle \mathrm{Tr}_{\bf N}\left[(\eta\bar\psi)_{1} (\psi\bar\eta)_{2} \right]\rangle=-\frac{i M N \Gamma(1/2-\epsilon)}{2\pi^{3/2-\epsilon} (\eta_{2}\bar\eta_{1})}(n_{L1}\bar n^{L}_{2})
\left[\frac{\dot x_{1}^\mu}{|\dot{x_{1}} | }+\frac{\dot{x_{2}}^{\mu}}{|\dot x_{2}|}\right]\partial_{x_{1}^{\mu}}\left(\frac{1}{(x_{12}^{2})^{1/2-\epsilon}}\right),
\ee
which holds  for any planar circuit.
Let us specialize (\ref{K1}) to the diagram of \ref{fig:oneloopgraphs}(e): within our  parametrization of the circuit,  we can rearrange the effective propagator exchanged through the rays as the difference of two total derivatives
\be
\left[\frac{\dot x_{1}^\mu}{|\dot{x_{1}}|}+\frac{\dot{x_{2}}^{\mu}}{|\dot x_{2}|}\right]\partial_{x_{1}^{\mu}}\left(\frac{1}{(x_{12}^{2})^{1/2-\epsilon}}\right)=\left(\frac{d}{d \tau_{1}}-\frac{d}{d \tau_{2}}\right)\frac{1}{(\tau_{1}^{2}+\tau_{2}^{2}-2 \tau_{1} \tau_{2}\cos\varphi)^{1/2-\epsilon}}.
\ee
The integration over the two edges can be done in a rather trivial way
\be
\label{CD0}
\begin{split}
&\int_{0}^{L} d\tau_{1}\int_{-L}^{0} d\tau_{2}\left(\frac{d}{d \tau_{1}}-\frac{d}{d \tau_{2}}\right)\frac{1}{(\tau_{1}^{2}+\tau_{2}^{2}-2 \tau_{1} \tau_{2}\cos\varphi)^{1/2-\epsilon}}=\\
&=-\frac{L^{2\epsilon}}{\epsilon}+2 L^{2\epsilon}\int_{0}^{1} d \tau \frac{1}{(\tau^{2}+2  \tau  \cos\varphi+1)^{1/2-\epsilon}}.
\end{split}
\ee
The remaining integral in \eqref{CD0} is finite as $\epsilon\to0$ and it can be evaluated in terms of hypergeometric functions. However its  exact value for arbitrary $\epsilon$ will not  be relevant for us and we shall only give its expansion around $\epsilon=0$ at the lowest order
\be
\int_{0}^{1} d \tau \frac{1}{(\tau^{2}+2  \tau  \cos\varphi+1)^{1/2-\epsilon}}=\log \left(\sec \left(\frac{\varphi }{2}\right)+1\right)+O(\epsilon).
\ee
We end up with
\be
{\cal F}_{e}=\frac{i M N \Gamma(1/2-\epsilon)}{2\pi^{3/2-\epsilon} (\nu_{1}\bar\nu_{2})}(n_{L2}\bar n^{L}_{1})L^{2\epsilon}\left[\frac{1}{\epsilon}-2\log \left(\sec \left(\frac{\varphi }{2}\right)+1\right)\right]
\ee
and since $(n_{L2}\bar n^{L}_{1})=\cos\frac{\theta}{2}$ and $(\nu_{1}\bar\nu_{2})=2i \cos\frac{\varphi}{2}$ we get
\be
{\cal F}_{e}=M N\left(\frac{ \Gamma(1/2-\epsilon)}{4\pi^{3/2-\epsilon}}\right)
\frac{\cos\frac{\theta}{2}}{\cos\frac{\varphi}{2}}L^{2\epsilon}\left[\frac{1}{\epsilon}-2\log \left(\sec \left(\frac{\varphi }{2}\right)+1\right)\right].
\ee
Next we must consider the case where the fermionic propagator connects two points on the same edge of the cusp, {\it i.e.} the diagrams (b) in fig. \ref{fig:oneloopgraphs}.  We have two  mirror graphs: one for each edge. The result of the first one  is provided by 
\be
\label{sed1l}
-\frac{ M N \Gamma(1/2-\epsilon)}{4\pi^{3/2-\epsilon}}\int_{-L}^{0}\!\!\! d\tau_{1} \int_{-L}^{\tau_{1}} \!\! d\tau_{2}\left(\frac{d}{d \tau_{1}}-\frac{d}{d \tau_{2}}\right)\frac{1}{(\tau_{1}-\tau_{2})^{1-2\epsilon}}=-\frac{ M N \Gamma(1/2-\epsilon)}{4\pi^{3/2-\epsilon}} \frac{L^{2\epsilon}}{\epsilon},
\ee
while the contribution of the second one simply doubles  \eqref{sed1l}  and it yields
\be
{\cal F}_b=-2\frac{ M N \Gamma(1/2-\epsilon)}{4\pi^{3/2-\epsilon}} \frac{L^{2\epsilon}}{\epsilon}.
\ee
Therefore the   complete  one loop  result for the upper left block  can be written as
\be\label{unsub1loop}
\mathfrak{F}^{(1)}=\left(\frac{2\pi}{\kappa}\right)M N\left(\frac{ \Gamma(1/2-\epsilon)}{4\pi^{3/2-\epsilon}}\right)
(\mu L)^{2\epsilon}\left[\frac{1}{\epsilon}\left(\frac{\cos\frac{\theta}{2}}{\cos\frac{\varphi}{2}}-2\right)-2\frac{\cos\frac{\theta}{2}}{\cos\frac{\varphi}{2}}\log \left(\sec \left(\frac{\varphi }{2}\right)+1\right)\right].
\ee
This result may appear  surprising at a first sight: while we could have expected the divergence from the cusp diagram (e), we have also a non-trivial contribution from the propagators living on a single edge (b). In ${\cal N}=4$ SYM theory the analogous contributions, coming from the combined gauge-scalar propagator, are identically zero in Feynman gauge, and their potential divergence never enters into the game. Moreover in the limit $\varphi=\theta=0$ a non-vanishing and divergent result persists, contradicting the naive expectation that the BPS infinite line is trivial. 
To understand the  result \eqref{unsub1loop} and to extract from it the truly gauge-invariant cusp divergence, we have to recall some basics about the renormalization of (cusped) Wilson loops in gauge theories and to adapt the general procedure to our somehow exotic operators: this will be done in the next section, after having completed the two-loop computation.

The full one-loop expression is recovered by considering also the part coming from the lower $M\times M$ block of the super-holonomy: it turns out to be the same, because of the symmetry between $N$ and $M$ at this order. The trace is simply obtained by adding this second contribution.

\section{Two-loop analysis}
We shall compute here the second order contribution to the expectation value of the cusped Wilson loop: we separate the computations of purely bosonic diagrams from fermionic ones, to appreciate technical and conceptual differences.
\subsection{Bosonic diagrams}
\begin{figure}[ht]
\centering{
    \includegraphics[width=.97\textwidth, height=.14\textwidth]{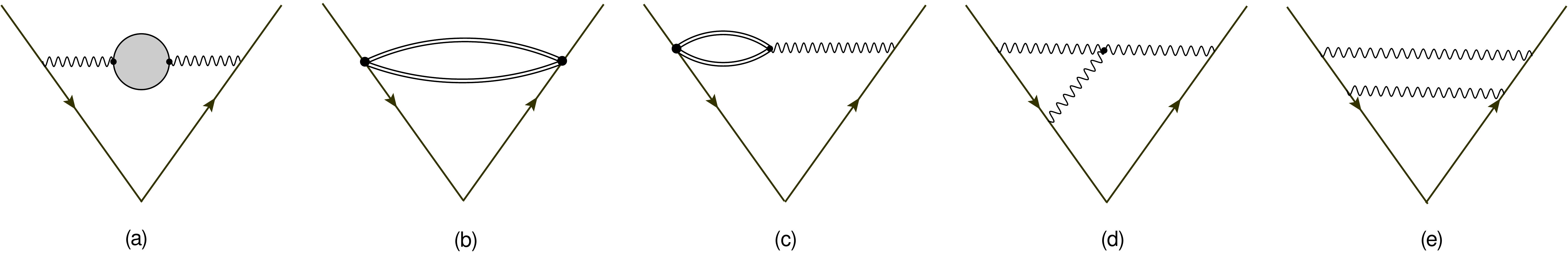}}
\vskip -.3cm
\caption{\footnotesize\label{BosonicDiagrams}\hskip -0.1cm  Two loops bosonic diagrams:  (a) One-loop corrected  gauge propagators; (b)  Correlators of two composite scalar operators;  (c)  Correlators gauge field composite scalar operator; 
(d) Chern-Simons vertex diagrams; (e) Gluon double exchange diagrams.}
\end{figure}

When expanding the Wilson loop operator  at the second order  in the coupling constant, we encounter  the four families of merely bosonic contributions depicted in fig. \ref{BosonicDiagrams}. We consider first the diagrams containing  the  one-loop corrected   gluon propagators (fig.~\ref{BosonicDiagrams}.(a)). As we did in the one-loop analysis, 
 we shall focus our  attention  on the upper diagonal block of the super-matrix, {\it i.e.} on the $U(N)$ sector. With the help of \eqref{oneloopgauge}, where the one-loop propagator is presented, we can immediately write
\be
\label{gaugetwo-loop}
\begin{split}
\left[\ref{BosonicDiagrams}.(a)\right]_{\rm up}\!=\!-M N^{2}\left(\frac{2\pi}{\kappa}\right)^{2}\frac{\Gamma^{2}\left(\frac{1}{2}-\epsilon\right)}{4 \pi^{3-2\epsilon}}\int_{\Gamma} d\tau_{\mbox{\tiny $\displaystyle1\!\!>\!\! 2$}} \! \left[\frac{(\dot{x}_{1}\cdot \dot{x}_{2})}{((x-y)^{2})^{{1}-2\epsilon}} -\partial_{\tau_{1}}\partial_{\tau_{2}}\!\frac{((x_{1}-x_{2})^{2})^{\epsilon}}{4\epsilon(1+2 \epsilon)}\right]\!.
\end{split}
\ee
A similar structure is obtained when considering the correlator of  two scalar composite operators $M_{I}^{\ J} C_{J}\bar C^{I}$ in the diagram \ref{BosonicDiagrams}.(b):
\be
\label{scalartwo-loop}
\begin{split}
\left[\ref{BosonicDiagrams}.(b)\right]_{\rm up}=M N^{2}\left(\frac{2\pi}{\kappa}\right)^{2}\frac{\Gamma^{2}\left(\frac{1}{2}-\epsilon\right)}{16 \pi^{3-2\epsilon}}\int_{\Gamma} d\tau_{\mbox{\tiny $\displaystyle1\!\!>\!\! 2$}} \  \frac{|\dot x_{1}||\dot x_{2}|{\rm Tr}(M_{1} M_{2})}{((x-y)^{2})^{{1}-2\epsilon}} .
\end{split}
\ee
The integrals  \eqref{gaugetwo-loop} and  \eqref{scalartwo-loop}   can be naturally combined together  to give
\be
\label{totBos}
\!
-M N^{2}\left(\frac{2\pi}{\kappa}\right)^{\!\!2}\!\frac{\Gamma^{2}\!\left(\frac{1}{2}-\epsilon\right)}{4 \pi^{3-2\epsilon}}\!\int_{\Gamma} \!d\tau_{\mbox{\tiny $\displaystyle1\!\!>\!\! 2$}} \! \left[\frac{(\dot{x}_{1}\!\cdot \!\dot{x}_{2})-\frac{1}{4}|\dot x_{1}||\dot x_{2}|{\rm Tr}(M_{1} M_{2}) }{((x-y)^{2})^{{1}-2\epsilon}} -
\partial_{\tau_{1}}\partial_{\tau_{2}}\frac{((x_{1}-x_{2})^{2})^{\epsilon}}{4\epsilon(1+2 \epsilon)}\right]\!\!.
\ee
The result \eqref{totBos} deserves some comments. The last term in \eqref{totBos} is a total derivative  and it
would correspond to a gauge transformation -albeit a singular one. In dimensional regularization it yields a 
$(\theta,\varphi)$ independent pole in $\epsilon$ plus finite terms, thus its contribution to the  divergent  part of the cusp  becomes ineffective when we impose the renormalization condition discussed in subsec. \ref{Leadord}.  The other contribution in \eqref{totBos}, as firstly noted in \cite{Drukker:2008zx},  possesses an unforeseen four-dimensional structure. When the two endpoints  lie on the same edge  it is proportional to  the  the tree-level effective propagator in $\mathcal{N}=4$ since $\mathrm{Tr}(M_{1} M_{2})=4$ and thus it vanishes. If they lie  instead  on opposite edges we get the following result
\be
\label{Bos2loop}
\mathfrak{B}^{(2)}=
-M N^{2}\left(\frac{2\pi}{\kappa}\right)^{\!\!2}\!\frac{\Gamma^{2}\!\left(\frac{1}{2}-\epsilon\right)}{4 \pi^{3-2\epsilon}}\left(\cos\varphi-\cos^{2}\frac{\theta}{2}\right )\!\int_{0}^{L} \!\!\!d\tau_{1}  \int_{-L}^{0} \!\!\!d\tau_{2}  \frac{1}{((x-y)^{2})^{{1}-2\epsilon}} ,
\ee
where the integral governing the divergence is the same of the four dimensional case when we replace $2\epsilon$ with $\epsilon$.

Next we examine the graphs \ref{BosonicDiagrams}.(c), \ref{BosonicDiagrams}.(d) and \ref{BosonicDiagrams}.(e). The last one is identically zero for the same reasons of the one-loop
single exchange \ref{fig:oneloopgraphs}.(a). The diagram \ref{BosonicDiagrams}.(c) for the case of 
planar loop was discussed in \cite{Drukker:2008zx} where it was found to vanish.  The same fate 
is shared by \ref{BosonicDiagrams}.(d) as pointed out in \cite{Henn:2010ps}. The only 
contribution originating from the bosonic diagrams is therefore provided by \eqref{Bos2loop}.

 \subsection{Fermionic diagrams}

\begin{wrapfigure}[7]{l}{65mm}
\centering{
    \includegraphics[width=.23\textwidth, height=.13\textwidth]{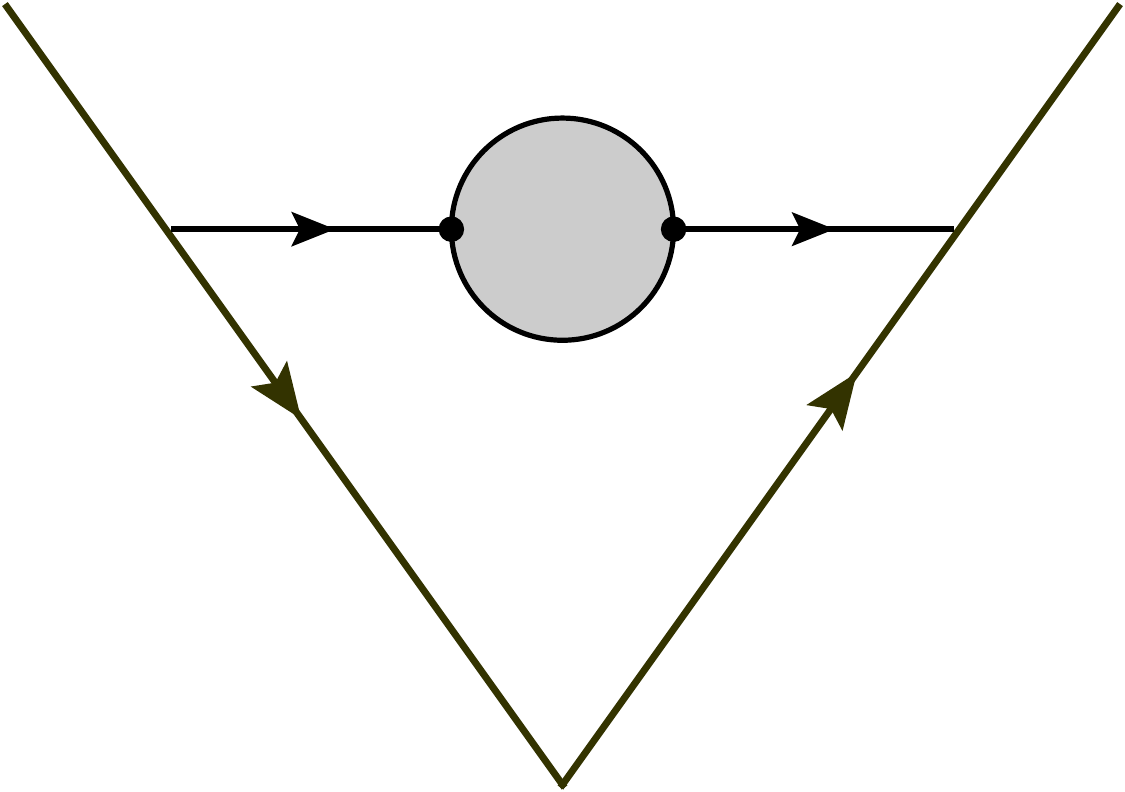}}
\vskip -.3cm
\caption{\label{2loopferm1} One-loop corrected fermions propagators}
\end{wrapfigure}
The simplest fermionic diagram appearing at the second order in perturbation theory
 consists of the 
exchange of the one-loop corrected fermion propagator depicted in fig \ref{2loopferm1}.

The one-loop two-point  function for the spinor fields is  briefly discussed in app. \ref{FRSS}. 
Remarkably  it again displays the  four dimensional behaviour already encountered in the bosonic
case.  Its form, in the DRED scheme, is 
\be
\left\langle
(\psi_{I})_{\hat i}^{\  j}(x)(\bar\psi^{J})_{ k}^{\ \hat l}(y)\right\rangle^{1~\rm \ell oop}_{0}=
-i\zet\left(\frac{2\pi}{\kappa}\right)\delta_{\hat i }^{\\\hat l}\delta^{j}_{k}(N-M)\frac{\Gamma ^{2}\left(\frac{1}{2}-\epsilon \right)}{16 \pi ^{3-2 \epsilon} }\frac{1}{((x-y)^{2})^{1-2\epsilon}}.
\ee
The contribution to the upper block of the Wilson loop takes the following form
\be
\begin{split}
\zet\left(\frac{2\pi\mu^{2\epsilon}}{\kappa}\right)^{2} i MN(N-M) \frac{\Gamma ^{2}\left(\frac{1}{2}-\epsilon \right)}{16 \pi ^{3-2 \epsilon} }\int_{\Gamma}d\tau_{\mbox{\tiny $\displaystyle1\!\!>\!\! 2$}}  \frac{(\eta_{I1}\bar\eta^{I}_{2})}{((x_{1} -x_{2})^{2})^{1-2\epsilon}}
\end{split}
\ee
\subsubsection{Double Exchanges}
\label{DED}
We come now to discuss a more subtle  group of  diagrams, namely those involving two $\langle\psi\bar\psi\rangle$ propagators. They arise when we evaluate  the contribution of the fermionic quadrilinear in \eqref{expaloop}. At this order  its expansion yields  only two sets of  non-vanishing  Wick-contractions, weighted  by different group factor, and thus we arrive at the following integral
\be
\label{DoubleExchange}
\begin{split}
&-4\left(\frac{2\pi\mu^{2\epsilon}}{\kappa}\right)^{\!\!2}\!\int_{\Gamma}\!\!d\tau_{\mbox{\tiny $\displaystyle1\!\!>\!\! 2\!\!>\!\!
3 \!\!>\!\! 4$}}[M^{2} N \underset{(A1)}{S(x_{2}-x_{1}) S(x_{4}-x_{3})}-N^{2} M \underset{(B1)}{S(x_{2}-x_{3}) S(x_{4}-x_{1})}].
\end{split}
\ee
Here the function $S(x_{i}-x_{j})$ is proportional to the two-point fermion correlator already encountered in \eqref{contra} and it can be conveniently written as
\be
S(x_{i}-x_{j})=\frac{(n_{i}\cdot n_{j})}{(\eta_{j}\bar\eta_{i})}(\partial_{\tau_{i}}-\partial_{\tau_{j}})D(x_{i}-x_{j}),
\ee
 \begin{wrapfigure}[10]{l}{92mm}
\centering{
    \includegraphics[width=.54\textwidth, height=.24\textwidth]{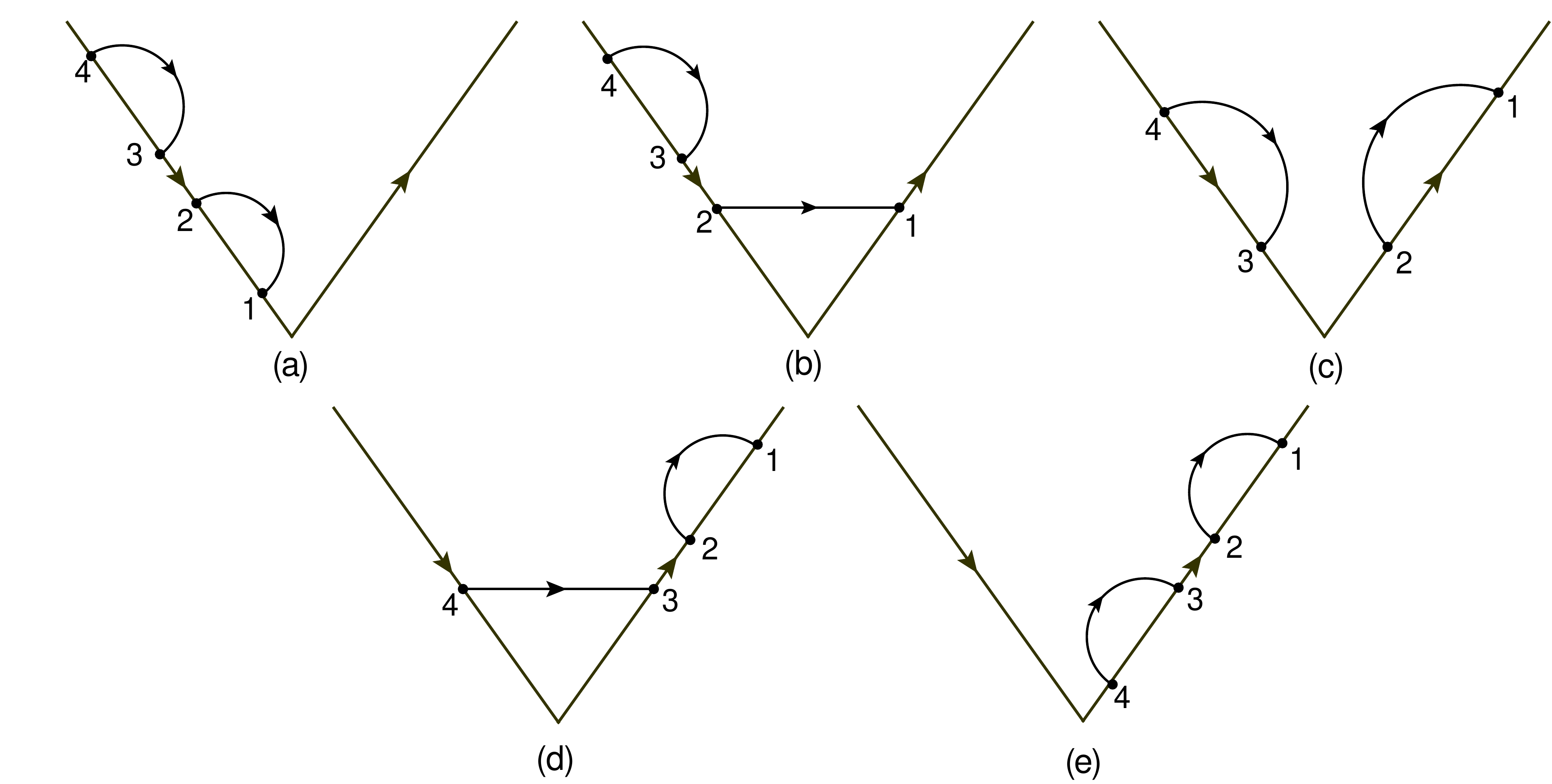}}
\vskip -.3cm
\caption{\label{DoubleExc1class} First group of double-exchange diagrams}
\end{wrapfigure}
where $D(x_{i}-x_{j})$ is the free scalar propagator  defined in \eqref{scal1} while the couple of vec\-tors $(n_{iI}, \bar n_{i}^{I})$ and of  spinors $(\eta_{Ii},\bar\eta_{ i}^{I})$ are defined in sec. \ref{subseccoupl}. We shall consider  the two
contributions in  \eqref{DoubleExchange} separately. In order to evaluate the term (A1) we have to 
split the region of integration in five  sectors that correspond to the five different  Feynman diagrams
depicted in fig. \ref{DoubleExc1class}. Luckily we do not have to compute all of them. In fact graphs,
which are related by a reflection with respect to the axis bisecting the cusp, yield the same result\footnote{This equality can be shown by performing the change of variable  $s_{i}\mapsto -s_{5-i}$ $(i=1,\dots 4)$ and subsequently by restoring the integration in the canonical order.}. In other
words, the following equalities hold among the diagrams of fig.\ref{DoubleExc1class}:  \ref{DoubleExc1class}.(a)=\ref{DoubleExc1class}.(e) and  \ref{DoubleExc1class}.(b)=\ref{DoubleExc1class}.(d). Moreover the graph
\ref{DoubleExc1class}.(c) is simply the square of  \ref{fig:oneloopgraphs}.(b).

To begin with, let us evaluate  the contribution \ref{DoubleExc1class}.(a).  It is given by the following integral
\begin{align}
[\ref{DoubleExc1class}.(a)]_{\rm up}
\!=\!&\left(\frac{2\pi}{\kappa}\right)^{\!\!2}\!\!M^{2}\!N\frac{\Gamma^{2}\left(\frac{3}{2}-\epsilon\right)}{ \pi^{{3}-2\epsilon}}\mu^{4\epsilon}\!\!\int_{-L}^{0}\!\!\!\!\!d\tau_{1}\int_{-L}^{\tau_{1}}\!\!\!\!\!d\tau_{2}\int_{-L}^{\tau_{2}}\!\!\!\!\!d\tau_{3}\int_{-L}^{\tau_{3}}\!\!\!\!\! d\tau_{4}~ (\tau_{1}-\tau_{2})^{2 \epsilon -2} (\tau_{3}-\tau_{4})^{2 \epsilon -2}\!\!=\nonumber\\
   =&
   \left(\frac{2\pi}{\kappa}\right)^{\!\!2}\!\!M^{2}\!N
   \frac{\Gamma^2
   \left(\frac{1}{2}-\epsilon \right)}{16 \pi^{3-2\epsilon}}\frac{\sqrt{\pi}}{ 2^{4\epsilon}}\frac{\Gamma(2\epsilon+1)}{\Gamma(2\epsilon+
   \frac{1}{2})}\frac{(\mu L)^{4\epsilon}}{\epsilon^{2}}.
 \end{align}

The diagram \ref{DoubleExc1class}.(b) instead leads to a different computation
\be
\label{archetti1}
\begin{split}
\!\!\!\!
[\ref{DoubleExc1class}.(b)]_{\rm up}
\!=\!-\!\left(\frac{2\pi}{\kappa}\right)^{\!\!2}\!\!M^{2}N\frac{ \Gamma^{2} \left(\frac{1}{2}-\epsilon \right)
  }{16 \pi ^{{3}-2\epsilon} }\frac{\mu^{4\epsilon}}{\epsilon} \frac{\cos\frac{\theta}{2}}{\cos\frac{\varphi}{2}}
\int_{0}^{L}\!\!\!\!\!d\tau_{1}\!\!\int_{-L}^{0}\!\!\!\!\!d\tau_{2} (L+\tau_{2})^{2 \epsilon }(\partial_{\tau_{2}}\!-\partial_{\tau_{1}})
H(\tau_{1},\tau_{2}),
\end{split}
\ee
where $H(\tau_{1},\tau_{2})=(\tau_{1}^{2}+\tau_{2}^{2}-2\tau_{1}\tau_{2}\cos\varphi)^{-\frac{1}{2}+\epsilon}$. We have performed the two trivial integrations over $\tau_{3}$ and $\tau_{4}$ since they involve a propagator whose endpoints 
belongs to the same edge. To extract the result we are interested in, we do not need the exact
value of the remaining integral, but only its $\epsilon-$expansion up to finite terms discussed in app. \ref{perturbativeintegrals}. We get
\be
[\ref{DoubleExc1class}.(b)]_{\rm up}\!=\!-\left(\frac{2\pi}{\kappa}\right)^{\!\!2}M^{2} N\frac{\cos\frac{\theta}{2}}{\cos\frac{\varphi}{2}}\frac{\Gamma^{2}(\frac{1}{2}-\epsilon)}{16\pi^{3-2\epsilon}}(L\mu)^{4\epsilon}\left[\frac{1}{ \epsilon^{2} }-\frac{2}{\epsilon}\log \left(1+\sec \frac{\varphi }{2}\right)+O(1)\right].
\ee
 \begin{wrapfigure}[10]{l}{85mm}
\centering{
    \includegraphics[width=.55\textwidth, height=.25\textwidth]{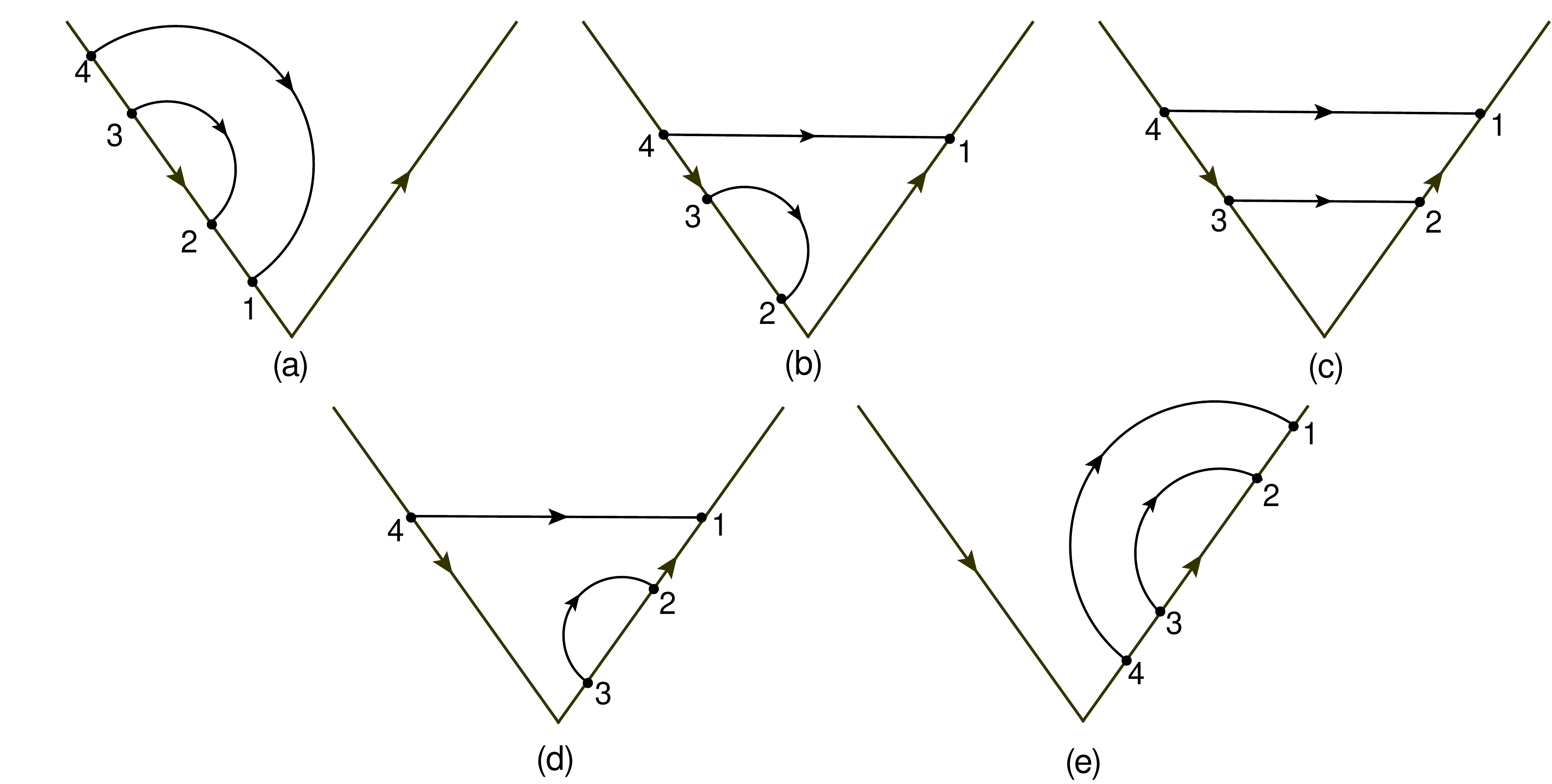}}
\vskip -.3cm
\caption{\label{DoubleExc2class} Second group of double-exchange diagrams}
\end{wrapfigure}
The next step is to consider the term (B1) in \eqref{DoubleExchange}: again we have to separate the region  of
 integration in five sub-sectors and this yields the diagrams in fig. \ref{DoubleExc2class}.
However the same reflection symmetry considered in the case of the term (A1) implies that we have just to compute 
\ref{DoubleExc2class}.(a), \ref{DoubleExc2class}.(b) and \ref{DoubleExc2class}.(c).  The first one can be easily computed in closed form and it gives
\be
[\ref{DoubleExc2class}.(a)]_{\rm up}=\left(\frac{2\pi}{\kappa}\right)^{\!\!2}\!\!M\!N^{\!2}\frac{\Gamma
   \left(\frac{1}{2}-\epsilon \right)^2}{16 \pi ^{3-2 \epsilon}}\frac{ (2 \epsilon -1) }{2 (4 \epsilon -1)}\frac{(\mu L)^{4\epsilon}}{\epsilon^{2}}.
\ee
Concerning the second diagram, we can trivially perform the integration over $\tau_{2}$ and $\tau_{3}$, obtaining
\be
\label{archetti2}
\begin{split}
[\ref{DoubleExc2class}.(b)]_{\rm up}=- \left(\frac{2\pi}{\kappa}\right)^{\!\!2}\!\!N^{2} M \frac{\Gamma^{2}(\frac{1}{2}-\epsilon)}{16\pi^{3-2\epsilon}}\frac{\cos\frac{\theta}{2}}{\cos\frac{\varphi}{2}}\frac{\mu^{4\epsilon}}{\epsilon}
\!\int_{0}^{L}\!\!\!\!d\tau_{1}\int_{-L}^{0}\!\! \!\!d\tau_{4}~{(-\tau_{4})^{2 \epsilon }}
(\partial_{\tau_{4}}-\partial_{\tau_{1}})H(\tau_{1},\tau_{4}).
\end{split}
\ee
With the help of  the results of app. \ref{perturbativeintegrals}, we can then write the following $\epsilon-$expansion
\be
[\ref{DoubleExc2class}.(b)]_{\rm up}=- \left(\frac{2\pi}{\kappa}\right)^{\!\!2}\!\!N^{2} M \frac{\Gamma^{2}(\frac{1}{2}-\epsilon)}{16\pi^{3-2\epsilon}}\frac{\cos\frac{\theta}{2}}{\cos\frac{\varphi}{2}}{(L\mu)^{4\epsilon}}\left[\frac{1}{2\epsilon^{2}}+\frac{1}{\epsilon}\log \left(\frac{1}{4} \cos \frac{\varphi }{2}\sec
   ^4\frac{\varphi }{4}\right) +O(1)\right].
\ee
For the graph \ref{DoubleExc2class}.(c) we shall adopt a different procedure since both propagators connect different edges.  First we rearrange its integral expression as follows
\be
\label{crossed}
\begin{split}
[\ref{DoubleExc2class}.(c)]_{\rm up}=&-2
\left(\frac{2\pi}{\kappa}\right)^{\!\!2}\! N^{2} M\left(\mu^{2\epsilon}\int_{0}^{L}\!\!\!\!\!\!d\tau_{1}\int_{-L}^{0}\!\!\!\!\!\!d\tau_{4} S(x_{4}-x_{1})\right)^{2}
-\\
&\ \  \ \ \ \  -4\left(\frac{2\pi}{\kappa}\right)^{\!\!2}\!M N^{2}  \mu^{4\epsilon}\int_{0}^{L}\!\!\!\!\!d\tau_{1}\int_{0}^{\tau_{1}}\!\!\!\!\!d\tau_{2}\int_{-L}^{0}\!\!\!\!\!d
\tau_{3}\int_{-L}^{\tau_{3}}\!\!\!\!\! d\tau_{4}~ S(x_{2}-x_{4}) S(x_{3}-x_{1}),
\end{split}
\ee
where we  have separated the ``abelian" and ``non-abelian" part of the diagram. The former is given by the first term, which is proportional to the square of \ref{fig:oneloopgraphs}.(e),
 while  the latter is identified 
with the  second term in \eqref{crossed}. This decomposition also has advantage that the leading divergence $1/\epsilon^{2}$ is only present in the first term.

With the help of  the results of app. \ref{perturbativeintegrals}, we obtain  the following $\epsilon-$expansion
for this diagram
\be
\label{crossedR}
\begin{split}
[\ref{DoubleExc2class}.(c)]_{\rm up}=&\frac{1}{2}\left(\frac{2\pi}{\kappa}\right)^{\!\!2} N^{2} M\frac{\Gamma^{2}(\frac{1}{2}-\epsilon)}{16\pi^{3-2\epsilon}}
 (L\mu)^{4\epsilon}    \left[\frac{\cos\frac{\theta}{2}}{\cos\frac{\varphi}{2}}\left(\frac{1}{\epsilon}-2\log\left(1+\sec\frac{\varphi}{2}\right)\right)\right]^{2}
-\\
&\ \  \ \ \ \  -\left(\frac{2\pi}{\kappa}\right)^{\!\!2} N^{2} M\left(\frac{\cos\frac{\theta}{2}}{\cos\frac{\varphi}{2}}\right)^{2}
\frac{\Gamma^{2}(\frac{1}{2}-\epsilon)}{16\pi^{3-2\epsilon}}
\frac{(L\mu)^{4\epsilon}}{\epsilon}\cos^{2}\frac{\varphi}{2}\frac{\varphi}{\sin\varphi}+O(1).
\end{split}
\ee

\subsubsection{Vertex Diagrams} 
The final  group of  fermionic diagrams, which are relevant for our calculation, arises when we expand in perturbation  theory  the term
\be
\label{vertex}
-\frac{2\pi i}{\kappa}\int_{\Gamma}d\tau_{\mbox{\tiny $\displaystyle1\!\!>\!\! 2\!\!>\!\!3$}}\left\langle(\underset{(\mathrm{{\bf A}_2})}{\eta\bar{\psi})_1(\psi\bar{\eta})_2{\cal A}_3 }
+\underset{(\mathrm{{\bf B}_2})}{{\cal A}_1(\eta\bar{\psi})_2(\psi\bar{\eta})_3}+\underset{(\mathrm{{\bf C}_2})}{(\eta\bar{\psi})_1\hat{\cal A}_2 (\psi\bar{\eta})_3}\right\rangle,
\ee
appearing  in the upper block \eqref{expaloop}. At this  order the expectation value in \eqref{vertex} is evaluated by just considering the Wick-contractions of the monomials $({\rm {\bf A}_2})$, $({\rm \mathrm{{\bf B}_2}})$ and $({\rm \mathrm{{\bf C}_2}})$  with the tree-level gauge-fermion vertices present in the Lagrangian \eqref{Lagra}. Then the three different contributions
can be rewritten as follows
  \begin{subequations}
 \label{Vertex1}
\begin{align}
\!\!\!\!\! (\mathrm{{\bf A}_2})=& -\left(\frac{2\pi}{\kappa}\right)^{2}N^{2}M 
 ~\int_{\Gamma}d\tau_{\mbox{\tiny $\displaystyle1\!\!>\!\! 2\!\!>\!\!3$}}~\eta_{1L}\gamma_{\nu}\gamma^{\mu}\gamma_{\lambda}\bar\eta_{2}^{L} \epsilon_{\mu\rho\sigma}
\dot{x}_{3}^{\rho}~
\Gamma^{\nu\lambda\sigma}(x_{1},x_{2},x_{3}),\\
\!\!\!\!\!(\mathrm{\mathrm{{\bf B}_2}})=&  -\left(\frac{2\pi }{\kappa}\right)^{2} N^{2}M~\int_{\Gamma}d\tau_{\mbox{\tiny $\displaystyle1\!\!>\!\! 2\!\!>\!\!3$}}~\eta_{2L}\gamma_{\lambda}\gamma^{\mu}\gamma_{\nu}\bar\eta^{L}_{3}
\epsilon_{\mu\rho\sigma}\dot{x}_{1}^{\rho}~\Gamma^{\sigma\lambda\nu}(x_{1}, x_{2}, x_{3}),
\\
\!\!\!\!\! (\mathrm{\mathrm{{\bf C}_2}}) =&- \left(\frac{2\pi}{\kappa}\right)^{2}  N M^{2}\int_{\Gamma}d\tau_{\mbox{\tiny $\displaystyle1\!\!>\!\! 2\!\!>\!\!3$}}~ \eta_{1L}
\gamma_{\lambda}\gamma^{\mu}\gamma_{\nu}
 \bar\eta_{3}^{\ L}\epsilon_{\mu\rho\sigma}\dot{x}_{2}^{\rho}
~ \Gamma^{\lambda\sigma\nu}(x_{1},x_{2},x_{3}),
\end{align}
\end{subequations}
where $\Gamma^{\lambda\mu\nu}(x_{1},x_{2}, x_{3})$
is a short-hand notation for the three-point function in position space, defined by the integral
\be
\label{threepoint}
\begin{split}
\Gamma^{\lambda\mu\nu}(x_{1},x_{2}, x_{2})=&\left(\frac{\Gamma(\frac{1}{2}-\epsilon)}{4\pi^{3/2-\epsilon}}\right)^{3}\partial_{x_{1}^{\lambda}}\partial_{x_{2}^{\mu}}\partial_{x_{3}^{\nu}}
\int 
\frac{d^{3-2\epsilon}w}{(x_{1w}^{2})^{1/2-\epsilon}(x_{2w}^{2})^{1/2-\epsilon}(x_{3w}^{2})^{1/2-\epsilon}}.
\end{split}
\ee
\begin{wrapfigure}[7]{l}{89mm}
\vskip-.6cm
\centering{
    \includegraphics[width=.50\textwidth, height=.25\textwidth]{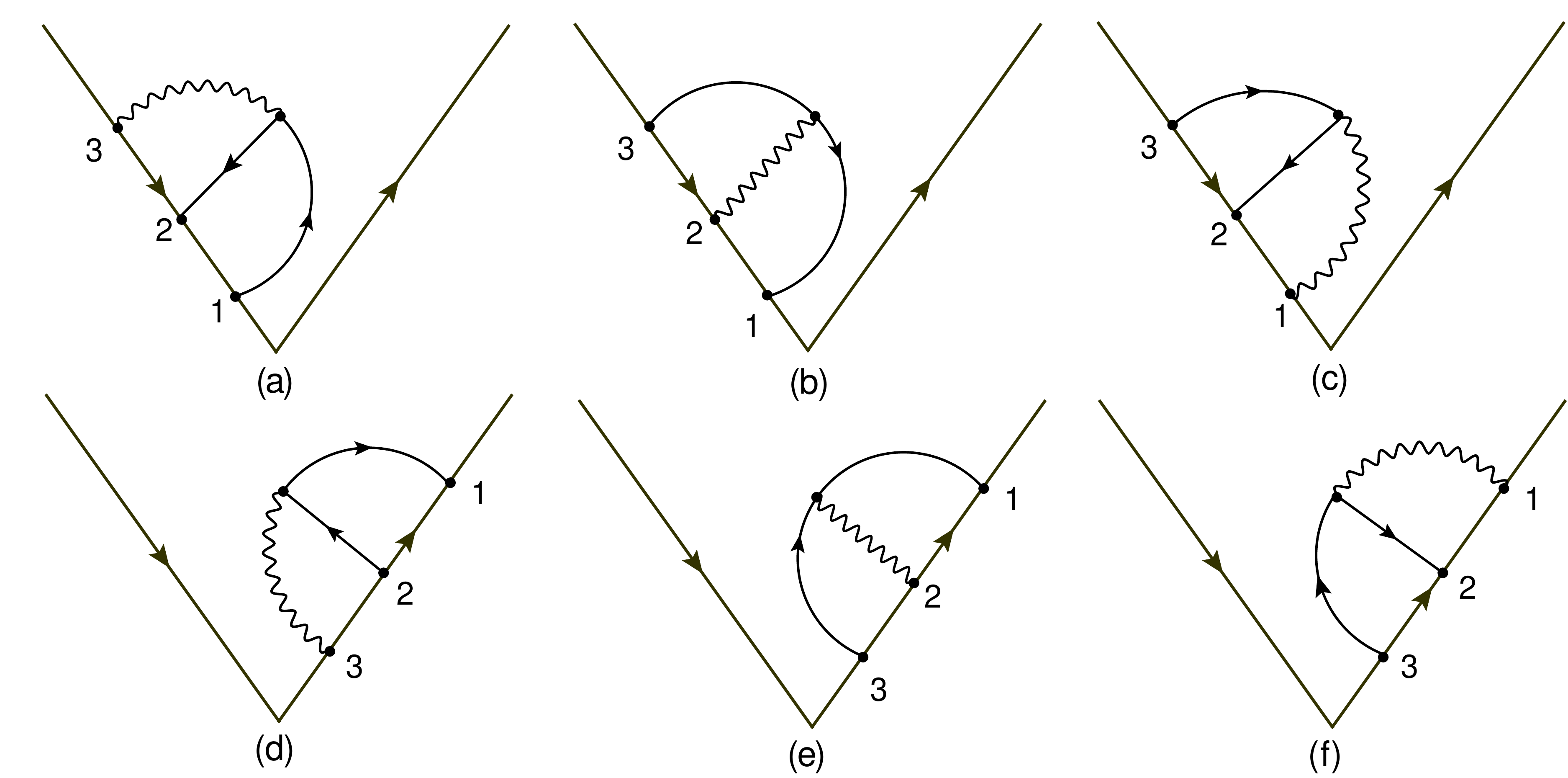}}
\vskip -.3cm
\caption{\label{Vertexa} Vertex diagrams where the fermion propagators are attached to the same edge of the gluon propagator.}
\end{wrapfigure}
Diagrammatically the three contributions \eqref{Vertex1} will lead to graphs which differ for 
the position of the  gauge field along the contour: $(\mathrm{{\bf A}_2})$ and $(\mathrm{{\bf B}_2})$ only  yield diagrams 
where the gluon is respectively the first or the last field we encounter when  $\tau$ runs from 
$-L$ to $L$; $(\mathrm{{\bf C}_2})$  instead corresponds to diagrams where the gauge field is always located between
the two fermionic lines.  For instance, 
\ref{Vertexa}.(a) and \ref{Vertexa}.(d) originate from $(\mathrm{{\bf A}_2})$, \ref{Vertexa}.(b) 
and \ref{Vertexa}.(e) from $(\mathrm{{\bf C}_2})$ and \ref{Vertexa}.($ $c) and \ref{Vertexa}.(f) from $(\mathrm{{\bf B}_2})$.

If we now  expand the spinor bilinears in \eqref{Vertex1} in terms of  the circuit tangent vectors  $\dot{x}_{i}$ and  of the scalar contraction $(\eta_{i}\bar\eta_{j})$, the three  contributions  $(\mathrm{{\bf A}_2})$, $(\mathrm{{\bf B}_2})$ and $(\mathrm{{\bf C}_2})$  can be rewritten as follows
\begin{subequations}
\label{Vertex2}
\begin{align}
\label{Vertex2a}
(\mathrm{{\bf A}_2})=
 &-N^{2}M \left(\frac{2\pi}{\kappa}\right)^{2}i(n_{1}\cdot n_{2})\zet\oint_{\tau_{1}>\tau_{2}>\tau_{3}}
\biggl[\frac{2}{(\eta_{2}\bar\eta_{1})}
\biggl(((\dot{x}_{2}\cdot\dot{x}_{3})\dot{x}_{1\nu}-(\dot{x}_{1}\cdot\dot{x}_{3})\dot{x}_{2\nu})[\Gamma^{\nu\tau\tau}+\nonumber\\
& \phantom{\Biggl [}+\Gamma^{\tau\nu\tau}
-\Gamma^{\tau\tau\nu}] +\dot{x}_{1\sigma} \dot{x}_{2\nu} \dot{x}_{3\lambda}\Gamma^{\nu\lambda\sigma}-\dot{x}_{1\nu}\dot{x}_{2\sigma}\dot{x}_{3\lambda}\Gamma^{\nu\lambda\sigma}+
\dot{x}_{1\sigma} \dot{x}_{2\lambda} \dot{x}_{3\nu}\Gamma^{\nu\lambda\sigma}-\nonumber\\
&-\dot{x}_{1\lambda}\dot{x}_{2\sigma}\dot{x}_{3\nu}\Gamma^{\nu\lambda\sigma}\biggr)+{(\eta_{1}\bar\eta_{2}) \dot{x}_{3\lambda}(\Gamma^{\tau\lambda\tau}-\Gamma^{\lambda\tau\tau})}\Biggr],\\
\label{Vertex2b}
(\mathrm{{\bf B}_2})= &  -N^{2}M\left(\frac{2\pi }{\kappa}\right)^{2} i(n_{3}\cdot n_{2})\zet\oint_{\tau_{1}>\tau_{2}>\tau_{3}}~\Biggl[\frac{2}{(\eta_{3}\bar\eta_{2})}
\biggl([(\dot{x}_{1}\cdot\dot{x}_{3})\dot{x}_{2\nu}-(\dot{x}_{2}\cdot\dot{x}_{1})x_{3\nu}][\Gamma^{\tau\tau\nu}+\nonumber\\
& \phantom{\Biggl [}+\Gamma^{\tau\nu\tau}-\Gamma^{\nu\tau\tau}]+\dot{x}_{1\lambda}\dot{x}_{2\nu} \dot{x}_{3\sigma} \Gamma^{\nu\lambda\sigma}-\dot{x}_{1\lambda}\dot{x}_{2\sigma}\dot{x}_{3\nu}\Gamma^{\nu\lambda\sigma}+\dot{x}_{1\sigma}\dot{x}_{2\nu} \dot{x}_{3\lambda} \Gamma^{\nu\lambda\sigma}-\nonumber\\
&-\dot{x}_{1\sigma}\dot{x}_{2\lambda}\dot{x}_{3\nu}\Gamma^{\nu\lambda\sigma}\biggr)+{(\eta_{2}\bar\eta_{3})\dot{x}_{1\nu}(\Gamma^{\tau\tau\nu}-\Gamma^{\tau\nu\tau})}\Biggr],\displaybreak[2]
\\
\label{Vertex2c}
(\mathrm{{\bf C}_2})=&-N M^{2} \left(\frac{2\pi}{\kappa}\right)^{2}  i(n_{1}\cdot n_{3})\zet\oint_{\tau_{1}>\tau_{2}>\tau_{3}}\Biggl[\frac{2}{(\eta_{3}\bar\eta_{1})}
\biggl(((\dot{x}_{2}\cdot\dot{x}_{3})\dot{x}_{1\nu}-(\dot{x}_{1}\cdot\dot{x}_{2})\dot x_{3\nu})[\Gamma^{\tau\tau\nu}+\nonumber\\
& \phantom{\Biggl [}+\Gamma^{\nu\tau\tau}-\Gamma^{\tau\nu\tau}]+\dot{x}_{1\sigma} \dot{x}_{3\nu} \dot{x}_{2\lambda}\Gamma^{\lambda\sigma\nu}-\dot{x}_{1\nu}\dot{x}_{3\sigma}\dot{x}_{2\lambda}\Gamma^{\lambda\sigma\nu}+\dot{x}_{1\sigma} \dot{x}_{3\lambda} \dot{x}_{2\nu}\Gamma^{\lambda\sigma\nu}-\nonumber\\
&-\dot{x}_{1\lambda}\dot{x}_{3\sigma}\dot{x}_{2\nu}\Gamma^{\lambda\sigma\nu}\biggr)+(\eta_{1}\bar\eta_{3})\dot{x}_{2\nu}(\Gamma^{\tau\tau\nu}-\Gamma^{\nu\tau\tau})\Biggr],
\end{align}
\end{subequations}
where we have dropped all the terms which vanish for planar contours.  To begin with, we   
 shall consider 
 the family of diagrams  of fig.~\ref{Vertexa}, whe\-re all the bosonic and fermionic lines terminate
on the same edge of the  cusp. In this case all the  terms    proportional to the factor
$2/(\eta_{i}\bar\eta_{j})$ in \eqref{Vertex2} drop out  because the tangent vectors obey the relation
\be
\label{eq1}
\dot{x}_{1}=\dot{x}_{2}=\dot{x}_{3},
\ee
for each diagram in fig.~\ref{Vertexa}. Only the last  terms in  \eqref{Vertex2a}, \eqref{Vertex2b} and  \eqref{Vertex2c}    that  are proportional to the bilinear $(\eta_{i}\bar\eta_{j})$ are  different from zero and we are left with
\begin{subequations}
\label{Vertex3}
\begin{align}
\label{Vertex3a}
(\mathrm{{\bf A}_2})=
 &2 N^{2}M \left(\frac{2\pi}{\kappa}\right)^{2}\zet\oint_{\tau_{1}>\tau_{2}>\tau_{3}}
\dot{x}_{3\lambda}(\Gamma^{\tau\lambda\tau}-\Gamma^{\lambda\tau\tau})\equiv-\left(\frac{2\pi}{\kappa}\right)^{2} N^{2}M (\mathfrak{a})\\
(\mathrm{{\bf B}_2})= &  2 N^{2}M\left(\frac{2\pi }{\kappa}\right)^{2} \zet\oint_{\tau_{1}>\tau_{2}>\tau_{3}}~\dot{x}_{1\nu}(\Gamma^{\tau\tau\nu}-\Gamma^{\tau\nu\tau})\equiv-\left(\frac{2\pi}{\kappa}\right)^{2} N^{2}M (\mathfrak{b})\displaybreak[2]
\\
(\mathrm{{\bf C}_2})=&2N M^{2} \left(\frac{2\pi}{\kappa}\right)^{2} \zet\oint_{\tau_{1}>\tau_{2}>\tau_{3}}\dot{x}_{2\nu}(\Gamma^{\tau\tau\nu}-\Gamma^{\nu\tau\tau})\equiv-\left(\frac{2\pi}{\kappa}\right)^{2} N M^{2} (\mathfrak{c}),
\end{align}
\end{subequations}
where we used that $\eta_{i}\bar\eta_{j}=2i$ and $(n_{i}\cdot n_{j})=1$. There is a further simplification: in fact  we do not have to compute  all  the diagrams originating from $(\mathrm{{\bf A}_2})$,
$(\mathrm{{\bf B}_2})$ and $(\mathrm{{\bf C}_2})$ and depicted in fig. \ref{Vertexa}. First of all, we can  restrict ourselves to considering only the   diagrams \ref{Vertexa}.(a), \ref{Vertexa}.(b) and \ref{Vertexa}.($ $c).  The other three graphs will simply double the final result. Next, we note that the following identity holds for this subclass of diagrams
\be
(\mathfrak{a})+(\mathfrak{b})=(\mathfrak{c}),
\ee
{\it i.e.} it is sufficient to evaluate only the integral  
\begin{align}
\label{t8}
(\mathfrak{c})=&-2 \zet\int_{-L}^{0}\!\!\!\!d\tau_{1}\int_{-L}^{\tau_{1}}\!\!\!\!d\tau_{2}\int_{-L}^{\tau_{2}}\!\!\!\!d\tau_{3}~{\dot{x}_{2\nu}(\Gamma^{\tau\tau\nu}-\Gamma^{\nu\tau\tau})}
\end{align}
 to reconstruct the result of all the diagrams in fig. \ref{Vertexa}.  Moreover the three-point functions appearing
 in \eqref{t8}  always possess two contracted indices: in this case the integral \eqref{threepoint} can be easily evaluated in terms of product of scalar propagators and one finds
 \be
 \label{525}
 \Gamma^{\tau\tau\nu}=\partial_{x_{3}^{\nu}}\Phi_{3,12},\ \ \ \ 
\Gamma^{\tau\nu\tau}
=
\partial_{x_{2}^{\nu}}\Phi_{2,13},\ \ \ \ 
\Gamma^{\nu\tau\tau}
=\partial_{x_{1}^{\nu}}\Phi_{1,23},
 \ee
 where 
 \be
\begin{split}
\Phi_{i,jk}=& -\frac{\Gamma^{2}(1/2-\epsilon)}{32\pi^{3-2\epsilon}}\!\!
\left[\frac{1}{(x^{2}_{ij})^{\frac{1}{2}-\epsilon}(x^{2}_{ik})^{\frac{1}{2}-\epsilon}}-\frac{1}{(x^{2}_{ij})^{\frac{1}{2}-\epsilon}(x^{2}_{kj})^{\frac{1}{2}-\epsilon}}-\frac{1}{(x^{2}_{ik})^{\frac{1}{2}-\epsilon}(x^{2}_{jk})^{\frac{1}{2}-\epsilon}}\right]\!\!.
\end{split}
 \ee
See  appendix $C$  for more details. With the help of this result, and recalling \eqref{eq1}, we can 
show that the integrand in \eqref{t8} only contains total derivatives and can be easily computed
\begin{align}
(\mathfrak{c})=&-2 \zet\int_{-L}^{0}\!\!\!\!d\tau_{1}\int_{-L}^{\tau_{1}}\!\!\!\!d\tau_{2}\int_{-L}^{\tau_{2}}\!\!\!\!d\tau_{3}~\left(\frac{d}{d\tau_{3}}\Phi_{3,12}-\frac{d}{d\tau_{1}}\Phi_{1,23}\right)
=\nonumber\\
=&-2\zet \int_{-L}^{0} \!\!\!\! d\tau_{1}\int_{-L}^{\tau_{1}}\!\!\!\! d\tau_{2}(\Phi_{2,12}+\Phi_{1,12}-\Phi_{-L,12}-\Phi_{0,12})=\nonumber\\
=& 2\zet  \frac{\Gamma^{2}(1/2-\epsilon)}{32\pi^{3-2\epsilon}} L^{4 \epsilon } \left(\frac{1}{2 \epsilon ^2}+\frac{1}{2 \epsilon }+O\left(1\right)\right).
\end{align}
\begin{wrapfigure}[8]{l}{85mm}
\centering{
    \includegraphics[width=.45\textwidth, height=.17\textwidth]{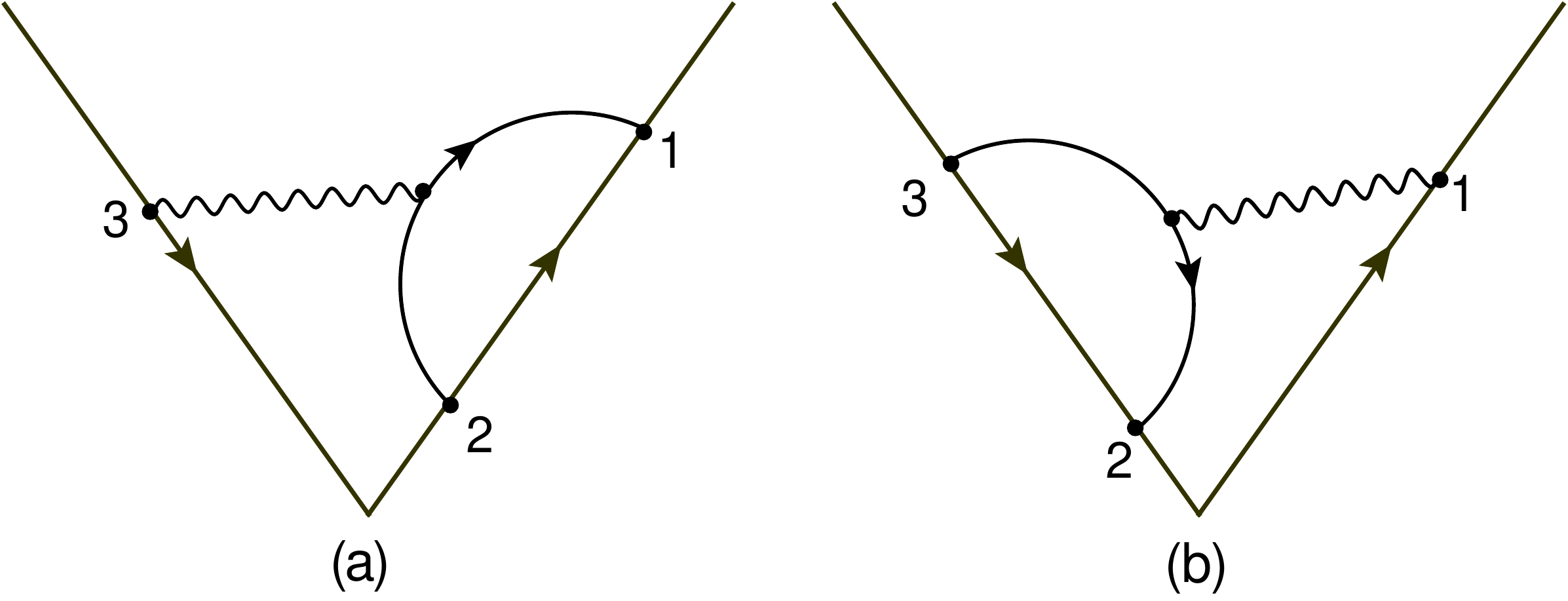}}
\vskip -.3cm
\caption{\label{Vertexb} Vertex diagrams with fermionic propagators attached to the opposite edge of the gluon propagator.}
\end{wrapfigure}
Next we consider  the case where the fermions are both attached to the same line, but the gluon is not. 
We have the two possibilities depicted in fig. \ref{Vertexb}. The diagram \ref{Vertexb}.(a) originates from the contribution
\eqref{Vertex2a} when considering the region of integration $-L\le\tau_{3}\le 0$ and $0\le \tau_{2}\le
\tau_{1}\le L$. 

The diagram \ref{Vertexb}.(b) is instead obtained from \eqref{Vertex2b}, when   $-L\le \tau_{3}\le\tau_{2}\le 0$, $0\le\tau_{1}\le L$ and $\dot{x}_{3}=\dot{x}_{2}$.  No contribution of this kind is instead contained in \eqref{Vertex2c}. 
 Since the two graphs in fig. \ref{Vertexb}  are related by a reflection with respect to the axis bisecting the cusp, they yield the same result and thus we have to  compute only one of them, {\it e.g. } \ref{Vertexb}.(a).
 For this diagram all the terms in \eqref{Vertex2a}, which are not proportional to $(\eta_{1}\bar\eta_{2})$,  will vanish when we use that  $\dot{x}_{1}=\dot{x}_{2}$  and so we get an expression that is similar  to the one considered in 
\eqref{Vertex3a}:
 \be
[\ref{Vertexb}.(a)]_{\rm up}=2 N^{2}M \left(\frac{2\pi}{\kappa}\right)^{2}\zet\int_{0}^{L}\!\!\!\! d\tau_{1}\int_{0}^{\tau_{1}}\!\!\!\!d\tau_{2}\int_{-L}^{0}\!\!\!\!d\tau_{3}~\dot{x}_{3\lambda}(\Gamma^{\tau\lambda\tau}-\Gamma^{\lambda\tau\tau}).
\ee
In order to compute this integral we first observe that the integrand can be rearranged as follows
\be
\label{cache}
\begin{split}
 &\dot x_{3\lambda}
( \Gamma^{\tau\lambda\tau}-\Gamma^{\lambda\tau\tau})=\dot x_{3}\cdot\partial_{x_{2}}\Phi_{2,13}-\dot x_{3}\cdot\partial_{x_{1}}\Phi_{1,23}=\dot x_{3}\cdot\partial_{x_{2}}(\Phi_{2,13}+\Phi_{1,23})+\frac{d}{d\tau_{3}}\Phi_{1,23}\\
&=-\left(\frac{\Gamma(1/2-\epsilon)}{4\pi^{3/2-\epsilon}}\right)^{2}
\frac{1}{(x^{2}_{13})^{1/2-\epsilon}}\frac{d}{d \tau_{3}}
\frac{1}{(x^{2}_{23})^{1/2-\epsilon}}+\frac{d}{d\tau_{3}}\Phi_{1,23}.
\end{split}
\ee
We have two separate contributions, which both appear in the list considered in appendix \ref{perturbativeintegrals}  (see eqs. \eqref{D1} and \eqref{D2}) and thus we can immediately write the final result
\be
\!\!\!
[\ref{Vertexb}.(a)]_{\rm up}\!=
-N^{2}M \left(\frac{2\pi}{\kappa}\right)^{2}
 \zet L^{4\epsilon}\left(\frac{\Gamma(1/2-\epsilon)}{4\pi^{3/2-\epsilon}}\right)^{2}\biggl[\frac{1}{\epsilon} \left({\log \left(\cos \frac{\varphi }{2}\right)}-\frac{1}{2 }\varphi  \cot \varphi \right)\!+O(1)\biggr].
\ee
The final set of diagrams that we have to consider are those where the two fermions end on  different edges of 
the cusp. We have four possible graphs  of this kind, which simply
   \begin{wrapfigure}[12]{l}{85mm}
\centering{
    \includegraphics[width=.40\textwidth, height=.28\textwidth]{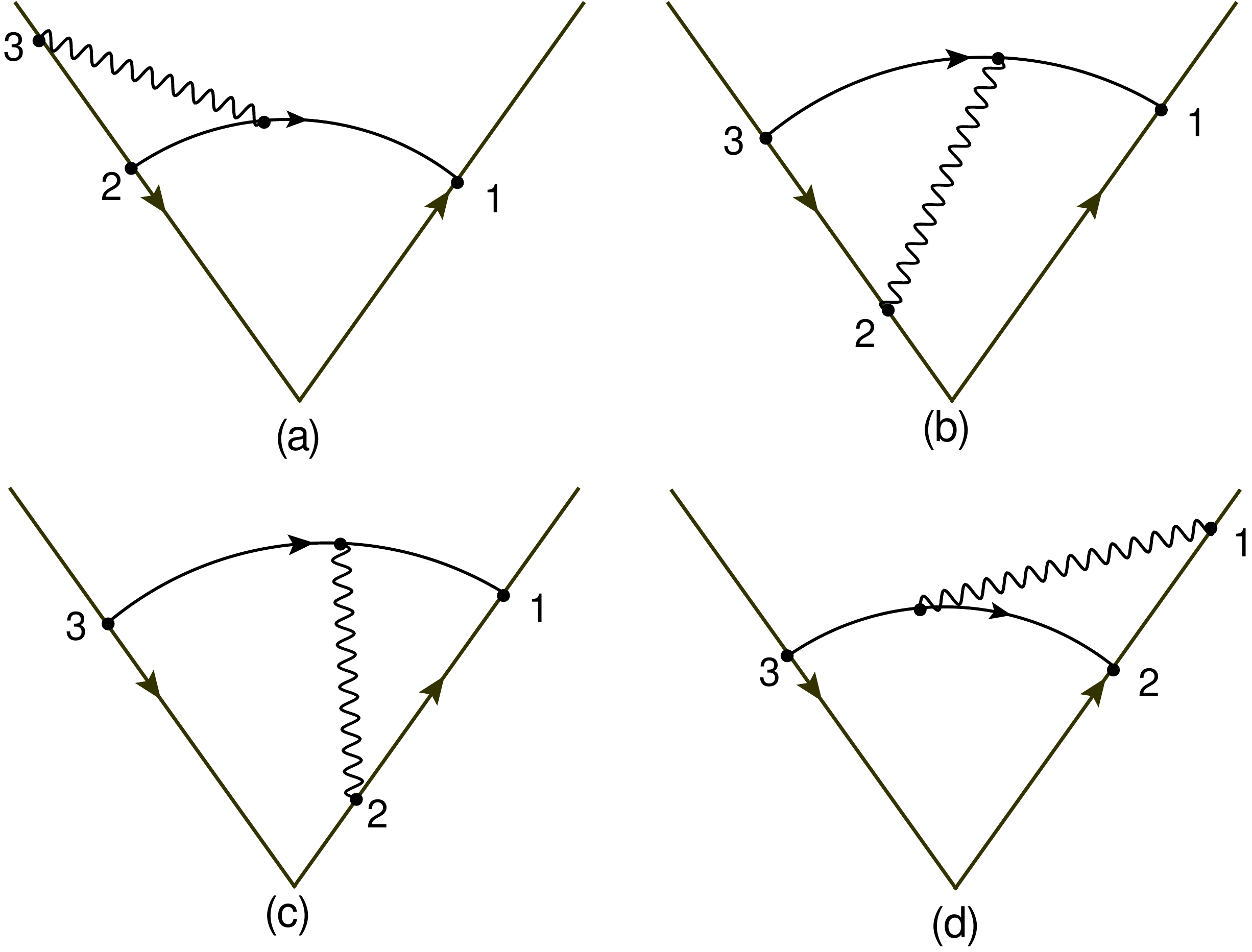}}
\vskip -.3cm
\caption{\label{Vertexc} Vertex diagrams with fermionic propagators are attached to opposite edges.}
\end{wrapfigure}
 differ for the position of the gluon line, and they are displayed in fig.  \ref{Vertexc}. The diagrams  \ref{Vertexc}.(a) and  \ref{Vertexc}.(d) are obtained respectively from \eqref{Vertex2a} and \eqref{Vertex2b}
 when considering the region of integrations  ${\rm (I)}=\{-L\le\tau_{3}\le\tau_{2}\le0$ and $0\le\tau_{1}\le L\}$  and  ${\rm (II)}=\{-L\le\tau_{3}\le 0$ and $0\le\tau_{2}\le\tau_{1}\le L\}$. The diagrams \ref{Vertexc}.(b) and  \ref{Vertexc}.(c) originate instead from \eqref{Vertex2c} when choosing either the  range  (I) or (II) for the parameters $\tau_{i}$. Again graphs, which are related by a reflection with respect the axis bisecting the cusp, produce the same result and we focus our attention only on \ref{Vertexc}.(a)  and  \ref{Vertexc}.(b).

To begin with, we shall  factor out from both diagrams the color and $R-$symmetry dependence  and we shall write
\be
 [\ref{Vertexc}.(a)]_{{\rm up}}\equiv -N^{2}M \left(\frac{2\pi}{\kappa}\right)^{2}\cos\frac{\theta}{2}\zet~ \mathcal{I}_{(a)}\ \ \ \ \  [\ref{Vertexc}.(b)]_{{\rm up}}\equiv -N M^{2} \left(\frac{2\pi}{\kappa}\right)^{2}  \cos\frac{\theta}{2}\zet ~\mathcal{I}_{(b)}.
 \ee
In order to simplify our analysis we shall construct the two independent combinations $\mathcal{I}_{(a)}-\mathcal{I}_{(b)}$ and $\mathcal{I}_{(a)}+\mathcal{I}_{(b)}$. The former is the only combination of the two integrals appearing in the final result when we would take the super-trace  of  the Wilson-loop and it is given by
\begin{align}
\label{supertrace}
\mathcal{I}_{(a)}-\mathcal{I}_{(b)}\!=&\frac{2 i}{(\eta_{2}\bar\eta_{1})}\int_{0}^{L}\!\!\!\!\!d\tau_{1}\!\!\int_{-L}^{0}\!\!\!\!\!d\tau_{2}\!\!\int_{-L}^{\tau_{2}}\!\!\!\!\!d\tau_{3}\!
\left(\!
\dot{x}_{1\nu}-(\dot{x}_{1}\cdot\dot{x}_{3})\dot{x}_{2\nu}-\frac{(\eta_{1}\bar\eta_{2}) (\eta_{2}\bar\eta_{1})}{2}
\dot{x}_{3\nu}\!\right)\!\left(\Gamma^{\tau\tau\nu}-\Gamma^{\tau\nu\tau}\right)\!=\nonumber\\
=&\frac{2 i}{(\eta_{2}\bar\eta_{1})}\int_{0}^{L}\!\!\!\!d\tau_{1}\int_{-L}^{0}\!\!\!\!d\tau_{2}\int_{-L}^{\tau_{2}}\!\!\!\!d\tau_{3}
\left(\dot{x}_{1\nu}+
\dot{x}_{2\nu}\right)\left(\Gamma^{\tau\tau\nu}-\Gamma^{\tau\nu\tau}\right).
\end{align}
In \eqref{supertrace} we were able to get rid of all the terms containing a three-point function contracted with  three $\dot{x}_{i}$, thanks to the identity \eqref{C15a} and to the equality $\dot{x}_{2}=\dot{x}_{3}$, which holds for these diagrams.  We can now use the  relations \eqref{525} and the invariance under translation of the function $\Phi_{i,jk}$ to rewrite the integrand as follows
\be
\label{5.33}
\begin{split}
\left(
\dot{x}_{1\nu}+\dot x_{2\nu}\right)\left(\Gamma^{\tau\tau\nu}-\Gamma^{\tau\nu\tau}\right)
=&\frac{d}{d \tau_{3}} \Phi_{3,12}-\frac{d}{d \tau_{2}} \Phi_{2,13}-\frac{d}{d \tau_{1}} \Phi_{3,12}+\\
&+\left(\frac{\Gamma(1/2-\epsilon)}{4\pi^{3/2-\epsilon}}\right)^{2}
\frac{1}{(x^{2}_{13})^{1/2-\epsilon}}\frac{d}{d \tau_{1}}
\frac{1}{(x^{2}_{12})^{1/2-\epsilon}}.
\end{split}
\ee
The integration over the circuit  can be performed by means of the results given in app. \ref{perturbativeintegrals} and we find
\begin{align}
\label{supertracea}
\mathcal{I}_{(a)}-\mathcal{I}_{(b)}\!=&\frac{L^{4\epsilon}}{2\cos\frac{\varphi}{2}}
\left(\frac{\Gamma(1/2-\epsilon)}{4\pi^{3/2-\epsilon}}\right)^{2}
\left[\frac{1}{2\epsilon}\frac{\varphi}{\sin\varphi}+\frac{1}{\epsilon}\log\left(\cos\frac{\varphi}{2}\right)+\frac{1}{4\epsilon^{2}}-\right.\\
&\left.-\frac{1}{\epsilon} \left(\frac{\varphi }{2 } \cot \varphi -{\log \left(\cos \frac{\varphi }{2}\right)}\right)+\frac{1}{4\epsilon^{2}}+O(1)\right]=\displaybreak[2]\\
=&\left(\frac{\Gamma(1/2-\epsilon)}{4\pi^{3/2-\epsilon}}\right)^{2}\frac{{L^{4\epsilon}}}{4\epsilon}\frac{\varphi}{\sin\frac{\varphi}{2}}+O(1).
\end{align}
The sum  $\mathcal{I}_{(a)}+\mathcal{I}_{(b)}$ is instead the only  combination appearing in the final  result if we  take the trace of the loop operator.  Its expression is less  elegant than the one for the difference and it is given by
\begin{align}
\label{trace}
&\mathcal{I}_{(a)}+\mathcal{I}_{(b)}=
i\int_{0}^{L}\!\!\!\!\!d\tau_{1}\!\!\int_{-L}^{0}\!\!\!\!\!d\tau_{2}\!\!\int_{-L}^{\tau_{2}}\!\!\!\!\!d\tau_{3}~\Biggl[
\frac{2}{(\eta_{2}\bar\eta_{1})}
\biggl(2(\dot{x}_{1\nu}+\dot x_{3\nu})\Gamma^{\nu\tau\tau}
+\dot{x}_{2\nu} \dot{x}_{1\lambda} \dot{x}_{3\sigma}\Gamma^{\nu\lambda\sigma}+\nonumber\\
&+\dot{x}_{2\lambda}\dot{x}_{1\sigma}\dot{x}_{3\nu}\Gamma^{\nu\lambda\sigma}-2
\dot{x}_{1\nu} \dot{x}_{3\lambda} \dot{x}_{2\sigma}\Gamma^{\nu\lambda\sigma}\biggr)+
(\eta_{1}\bar\eta_{2}) ~\dot{x}_{3\nu}(\Gamma^{\tau\nu\tau}+\Gamma^{\tau\tau\nu})\Biggr].
\end{align}
It is not difficult to realize that the integrand in \eqref{trace} is symmetric when exchanging $\tau_{2}$
with $\tau_{3}$: this allows us to extend the integration over $\tau_{3}$ up to $0$ provided dividing the result by two. We can reorganize \eqref{trace} as follows
\begin{align}
\label{trace1}
\mathcal{I}_{(a)}+\mathcal{I}_{(b)}=&
\frac{i}{2}\int_{0}^{L}\!\!\!\!\!d\tau_{1}\!\!\int_{-L}^{0}\!\!\!\!\!d\tau_{2}\!\!\int_{-L}^{0}\!\!\!\!\!d\tau_{3}~\Biggl[
\frac{2}{(\eta_{2}\bar\eta_{1})}
\biggl(2(\dot{x}_{1\nu}+\dot x_{3\nu})\Gamma^{\nu\tau\tau}
+\dot{x}_{2\nu} \dot{x}_{1\lambda} \dot{x}_{3\sigma}\Gamma^{\nu\lambda\sigma}+\nonumber\\
&+\dot{x}_{2\lambda}\dot{x}_{1\sigma}\dot{x}_{3\nu}\Gamma^{\nu\lambda\sigma}-2
\dot{x}_{1\nu} \dot{x}_{3\lambda} \dot{x}_{2\sigma}\Gamma^{\nu\lambda\sigma}\biggr)+
(\eta_{1}\bar\eta_{2}) ~\dot{x}_{3\nu}(\Gamma^{\tau\nu\tau}+\Gamma^{\tau\tau\nu})\Biggr]=\nonumber\\
=&
{i}\int_{0}^{L}\!\!\!\!\!d\tau_{1}\!\!\int_{-L}^{0}\!\!\!\!\!d\tau_{2}\!\!\int_{-L}^{0}\!\!\!\!\!d\tau_{3}~\Biggl[
\frac{2}{(\eta_{2}\bar\eta_{1})}
\biggl((\dot{x}_{1\nu}+\dot x_{3\nu})\Gamma^{\nu\tau\tau}
+\dot{x}_{2\nu} \dot{x}_{1\lambda} \dot{x}_{3\sigma}\Gamma^{\nu\lambda\sigma}-\nonumber\\
&-\dot{x}_{1\nu} \dot{x}_{3\lambda} \dot{x}_{2\sigma}\Gamma^{\nu\lambda\sigma}\biggr)+
(\eta_{1}\bar\eta_{2}) ~\dot{x}_{3\nu}\Gamma^{\tau\tau\nu}\Biggr].
\end{align}
In the second equality in \eqref{trace1} we have identified all the terms which differ by a permutation of
$\tau_{2}$ with $\tau_{3}$, being trivially equivalent. We can distinguish two types of contributions:
one containing only contracted three-point functions and the other  where the three-point functions are saturated with 
three $\dot x_{i}$. The former can be rewritten in terms of the function $\Phi_{i,jk}$ by means  of the relations \eqref{525}
and we obtain
\begin{align}
\label{5.39}
&
{i}\int_{0}^{L}\!\!\!\!\!d\tau_{1}\!\!\int_{-L}^{0}\!\!\!\!\!d\tau_{2}\!\!\int_{-L}^{0}\!\!\!\!\!d\tau_{3}~\Biggl[
\frac{2}{(\eta_{2}\bar\eta_{1})}
(\dot{x}_{1\nu}+\dot x_{3\nu})\Gamma^{\nu\tau\tau}
+
(\eta_{1}\bar\eta_{2}) ~\dot{x}_{3\nu}\Gamma^{\tau\tau\nu}\Biggr]=\nonumber\\
=&{i}\int_{0}^{L}\!\!\!\!\!d\tau_{1}\!\!\int_{-L}^{0}\!\!\!\!\!d\tau_{2}\!\!\int_{-L}^{0}\!\!\!\!\!d\tau_{3}~\Biggl[
\frac{2}{(\eta_{2}\bar\eta_{1})}\biggl(
\frac{d}{d\tau_{1}} \Phi_{1,23}-\frac{d}{d\tau_{2}} \Phi_{1,23}-
\frac{d}{d\tau_{3}}\Phi_{1,23}\biggr)+(\eta_{1}\bar\eta_{2})\frac{d}{d\tau_{3}} \Phi_{3,12}\Biggr].
\end{align}
The divergent part of these integrals can be extracted from the table of integrals presented  in app.  \ref{perturbativeintegrals} and we find
\be
\label{trace3}
\frac{\Gamma^{2}(1/2-\epsilon)}{16\pi^{3-2\epsilon}}\left[\frac{L^{4\epsilon}}{\cos\frac{\varphi}{2}}\left[\frac{2}{\epsilon}\log \left(\sec \left(\frac{\varphi }{2}\right)+1\right)-\frac{1}{2\epsilon^{2}}\right]+\frac{L^{4\epsilon}}{\epsilon}\cos\frac{\varphi}{2}\log\left(\cos\frac{\varphi}{2}\right)+O(1)\right].
\ee
The procedure for determining the divergences of the latter contribution in \eqref{trace1} is more delicate, since we have to deal with the untraced three-point function. 
After a careful analysis, one gets
\begin{align}
\label{trace4}
\frac{2i}{(\eta_{2}\bar\eta_{1})}\int_{0}^{L}\!\!\!\!\!d\tau_{1}\!\!\int_{-L}^{0}\!\!\!\!\!d\tau_{2}\!\!\int_{-L}^{0}\!\!\!\!\!d\tau_{3}~
\biggl(&
\dot{x}_{2\nu} \dot{x}_{1\lambda} \dot{x}_{3\sigma}\Gamma^{\nu\lambda\sigma}-\dot{x}_{1\nu} \dot{x}_{3\lambda} \dot{x}_{2\sigma}\Gamma^{\nu\lambda\sigma}\biggr)=\nonumber\\
&=-\frac{\Gamma^{2}(1/2-\epsilon)}{16\pi^{3-2\epsilon}}\frac{{L^{4\epsilon}}}{\epsilon}\cos \frac{\varphi }{2}\log \left(\cos \frac{\varphi }{2}\right)+O(1).
\end{align}
If we sum \eqref{trace3} and \eqref{trace4}, we can  finally write down the result for $\mathcal{I}_{(a)}+\mathcal{I}_{(b)}$
\be
\mathcal{I}_{(a)}+\mathcal{I}_{(b)}=\frac{\Gamma^{2}(1/2-\epsilon)}{16\pi^{3-2\epsilon}}\frac{L^{4\epsilon}}{\cos\frac{\varphi}{2}}\left[\frac{2}{\epsilon}\log \left(\sec \left(\frac{\varphi }{2}\right)+1\right)-\frac{1}{2\epsilon^{2}}\right]+O(1).
\ee
This completes the evaluation of the divergent part of all diagrams at two loops.
\section{The final result: summing and renormalizing}
In this section we shall add up the different diagrams which appear at two loops.  Because we are actually working with an open contour, we have in principle two possibilities  to perform this  sum: we can take the trace [$\mathcal{W}_{+}$ in \eqref{Wpm}] or the super-trace [$\mathcal{W}_{-}$ in \eqref{Wpm}].  As we shall see, the first choice, that is the correct one for closed contours, appears to be consistent with an exponentiated form. We also discuss the renormalization of our result, paying particular attention to the peculiarities arising in three dimensions and in the presence of the exotic fermionic couplings.

\subsection{Taking the trace}
Let us consider the case of the trace. The bosonic bubble diagrams yield  a  four-dimensional-like  contribution given by
\be
\mathds{B}=-g(\epsilon)\left(\cos\varphi-\cos^{2}\frac{\theta}{2}\right )\frac{1}{\epsilon}\frac{\varphi}{\sin\varphi},
\ee
where we have introduced the short-hand notation 
\be
g(\epsilon)= M N\left(\frac{2\pi}{\kappa}\right)^{\!\!2}\!\frac{\Gamma^{2}\!\left(\frac{1}{2}-\epsilon\right)}{16 \pi^{3-2\epsilon}}(\mu L)^{4\epsilon}
\ee
 for future convenience.
The fermionic bubble instead cancels when we take the trace, since it is odd in the exchange $N\leftrightarrow M$.  The total  result for the complete set of 
double-exchange diagrams is more elaborate and it can be usefully cast in the form
\be
\begin{split}
\mathds{D}=&2 g(\epsilon)
\Biggl[\frac{1}{\epsilon ^2}\left[2-\frac{3}{2} \frac{\cos \frac{\theta }{2}}{ \cos \frac{\varphi }{2}}\right]+\frac{1}{\epsilon }\left[ \frac{\cos \frac{\theta }{2}}{ \cos \frac{\varphi }{2}}\left(4 \log \left(\sec \frac{\varphi }{2}+1\right)+\log \left(\cos \frac{\varphi
   }{2}\right)\right)+1\right]+\\
   &+\frac{1}{4}   \left[\frac{\cos\frac{\theta}{2}}{\cos\frac{\varphi}{2}}\left(\frac{1}{\epsilon}-2\log\left(1+\sec\frac{\varphi}{2}\right)\right)\right]^{2}
 -\frac{1}{2\epsilon}\cos^{2}\frac{\theta}{2}
\frac{\varphi}{\sin\varphi}+O(1)\Biggr].
\end{split}
\ee
The  diagrams which contain the gauge-fermion interaction  yield  instead the following result
\be
\mathds{V}\!=\!
\frac{g(\epsilon)}{\epsilon ^2}\!\left[{\frac{\cos\frac{\theta}{2}}{\cos\frac{\varphi}{2}}-2}\right]+\frac{g(\epsilon)}{\epsilon}\!\left[{\varphi  \cot \varphi -2\! \left(\!2 \frac{\cos\frac{\theta}{2}}{\cos\frac{\varphi}{2}}\log \left(\sec \frac{\varphi }{2}+1\right)+\log \left(\cos
   \frac{\varphi }{2}\right)\!+\!1\!\right)}\right]\!+\!O\left(1\right)\!.
\ee
We shall now sum these three contributions in order to obtain the unrenormalized value of $\mathcal{W}_{+}$ in \eqref{Wpm} at two loops
\be
\begin{split}
[\mathcal{W}_{+}]_{\rm 2-loop}=
&\frac{g(\epsilon)}{2\epsilon^{2}}\left(\frac{\cos\frac{\theta}{2}}{\cos\frac{\varphi}{2}}-2\right)^{2}+\frac{2g(\epsilon)}{\epsilon}\frac{\cos\frac{\theta}{2}}{\cos\frac{\varphi}{2}}\left(2-\frac{\cos\frac{\theta}{2}}{\cos\frac{\varphi}{2}}\right)\log\left(\sec \frac{\varphi }{2}+1\right)+\\
&\ \ \ \ \ \ \ \ \ \ \ \ \ \ \ +\frac{2g(\epsilon)}{\epsilon}\log\left(\cos\frac{\varphi}{2}\right)\left(\frac{\cos\frac{\theta}{2}}{\cos\frac{\varphi}{2}}-1\right)+O(1).
\end{split}
\ee 
In this expression  the structure of the  generalized potential is not manifest. Crucially we observe the presence of double-poles that are not expected to appear in the final expression of the generalized potential. In conventional Wilson loops, where only bosonic couplings are concerned, double-poles at two loops are simply understood as coming from the square of the one-loop result, by virtue of the non-abelian exponentiation theorem \cite{Exp} (that holds even at renormalized level). The non-trivial contribution at second order in perturbation theory comes from the so-called maximally non-abelian part and in ${\cal N}=4$ SYM, for example, involves crossed non-planar bosonic exchanges and interacting diagrams, stretching between the two lines. In our case, due to the presence of the fermionic couplings, we do not have an 
established  exponentiation theorem at hand and we were forced to compute the full two-loop  contribution to the quantum average.  Incidentally, for our loops, double-poles appear both from exchange and interacting diagrams at variance with ${\cal N}=4$ SYM, where non-abelian exponentiation forbids the presence of $1/\epsilon^2$ in vertex or bubble graphs. In order to proceed and extract a generalized potential, taking properly into account the one-loop and two-loop results, we need an exponentiation ansatz: we propose the following form for the unrenormalized loop
\be
\mathcal{W}_{+}=\frac{M \exp(V_{N})+ N \exp(V_{M})}{N+M}. 
\ee
It is not difficult to check that our results are compatible with this  double-exponentiation where 
\begin{align}
\label{unrenpot}
V_{N}=&\left(\frac{2\pi}{\kappa}\right) N\left(\frac{ \Gamma(1/2-\epsilon)}{4\pi^{3/2-\epsilon}}\right)
(\mu L)^{2\epsilon}\left[\frac{1}{\epsilon}\left(\frac{\cos\frac{\theta}{2}}{\cos\frac{\varphi}{2}}-2\right)-2\frac{\cos\frac{\theta}{2}}{\cos\frac{\varphi}{2}}\log \left(\sec \left(\frac{\varphi }{2}\right)+1\right)\right]+\nonumber\\
&+N^2\left(\frac{2\pi}{\kappa}\right)^{\!\!2}\!\frac{\Gamma^{2}\!\left(\frac{1}{2}-\epsilon\right)}{16 \pi^{3-2\epsilon}}(\mu L)^{4\epsilon}\left[\frac{1}{\epsilon} \log\left(\cos^{2}\frac{\varphi}{2}\right)\left(\frac{\cos\frac{\theta}{2}}{\cos\frac{\varphi}{2}}-1\right)+O(1) \right]
\end{align}
The generalized potential $V_{M}$ is of course obtained by exchanging $M$ with $N$ in the above formula. We remark that the actual exponentiation of the one-loop term is a non-trivial support of our assumption and of the correctness of our computations, involving a delicate balance between exchanging and interacting contributions. From the physical point of view we could also justify the presence of two generalized potentials, simply recalling that we have two different test particles running in our contour. Following \cite{Lee:2010hk} it is straightforward to show that in $U(N)\times U(M)$ ${\cal N}=6$ theories two kinds of particles arise from the relevant higgsing procedure and which transform respectively in the $({\bf N},{\bf 1})$ and $({\bf 1},{\bf M})$ representations and their conjugate, that we call $W_N$ and $W_M$ bosons. It is clear that a  pair of  $W_N$ and $W_M$ cannot form a singlet of the color indices and there is no generalization of the quark-antiquark potential in this case. On the other hand a pair of $W_N\bar{W}_N$ or $W_M\bar{W}_M$ do form color singlets, hence there are two potentials in this theory.
\subsection{The renormalized generalized potentials}
\label{outcome}
 The outcome  of our extensive two-loop computation  contains some  puzzling unexpected features which deserve a more detailed analysis. To begin with,  let us consider  the one-loop contribution in \eqref{unrenpot}. 
When $\theta=\varphi=0$ our cusp degenerates into a segment of length $2L$  with the couplings of the $1/2$ BPS straight line and its ({\it unrenormalized}) value is given at the first non-trivial order by 
 \begin{align}
 \label{Ren1}
{W}^{(1)}_{\rm line}=&-\left(\frac{2\pi}{\kappa}\right)\frac{M N}{N+M}\left(\frac{ \Gamma(1/2-\epsilon)}{4\pi^{3/2-\epsilon}}\right)
(\mu L)^{2\epsilon}\left[\frac{1}{\epsilon}+2\log \left(2\right)+O(\epsilon)\right]=\nn \\
=&-\left(\frac{2\pi}{\kappa}\right)\frac{M N}{N+M}\left(\frac{ \Gamma(1/2-\epsilon)}{4\pi^{3/2-\epsilon}}\right)
\frac{(2 L\mu)^{2\epsilon}}{\epsilon}+O(\epsilon).
\end{align}
This divergent result appears to contradict the  expectation  that the  $1/2-$BPS straight-line  is trivial ({\it i.e.} equal to $1$) as  occurs in $\mathcal{N}=4$ SYM.  In that case an analogous computation  for a segment 
of length $2L$ in Feynman gauge  would have led to an exact cancellation between the gauge and the scalar contribution yielding as final result  $W^{(1)}_{\rm line}=0$. We remark, however, that this manifest zero in $\mathcal{N}=4$ SYM is peculiar  of the Feynman gauge. In a generic 
$\alpha-$gauge  the cancellation is only partial and a divergent term similar to  \eqref{Ren1} survives,
\begin{equation}
\label{N=4div}
W^{(1)}_{\rm line}=g^2N(1-\alpha) \frac{\Gamma(1-\epsilon)}{16\pi^{2-\epsilon}}\frac{(2  L\mu)^\epsilon}{\epsilon}. 
\end{equation}
The $\alpha-$dependence in \eqref{N=4div} signals  that we are dealing with a gauge-dependent divergence\footnote{The gauge origin of these additional divergences  is even more transparent when  we consider a circular sector of aperture $2\pi -\theta$ in $\mathcal{N}=4$.  There, next to  the expected result in Feynman gauge, there is  a divergent term given by  \[g^2N(1-\alpha) \frac{\Gamma(1-\epsilon)}{16\pi^{2-\epsilon}} \frac{(4\sin^{2}\frac{\theta}{2})^\epsilon}{\epsilon}.\]
When we close the circle ($\theta=0$),  thus recovering the gauge invariant operator, the coefficient of the  divergence simply vanishes.}, but this is not surprising. In fact the result \eqref{N=4div}  is the expectation  value for a segment of length $2L$, which  does not define a gauge invariant operator unless $L=\infty$.  However the limit $L\to \infty$ is delicate and it cannot be taken before  renormalizing the finite length operator.

The systematic renormalization of Wilson operator on open contours is a subject widely discussed  in the literature
\cite{Dorn1,Dorn2,Aoyama:1981ev,Knauss:1984rx,Dorn:1986dt}
and an exhaustive presentation of the topic is beyond the goal  of this paper. Below we shall recall some general facts  using  YM  or $\mathcal{N}=4$ SYM  as our  pedagogical examples.  The case of ABJM will be considered later.

An efficient frame-work  for discussing the renormalization of path ordered phase factors was introduced 
by \cite{Arefeva:1980zd,Gervais:1979fv}. In this approach these non-local operators  are represented as 
the  two point function
$\left\langle\psi(-L)\bar\psi(L)\right\rangle_{0}$ of the  one-dimensional fermionic  {\it bare} action  \footnote{For open loops the action must also contains boundary terms (see  {\it e.g.} \cite{Dorn:1986dt}) but for simplicity we shall neglect them.}

\be
\label{Spsi}
S=\int_{-L}^{L}\!\! dt ~\bar\psi(i\partial_{t}+  g \mathcal{A}_{\mu}\dot{x}^{\mu} )\psi,
\ee
where $\mathcal{A}_{\mu}$ stands for the connection of which we are computing the quantum holonomy.  
The familiar techniques of renormalization for local Green function can be therefore applied. 

In  this language  the  divergence \eqref{N=4div}  is responsible  for the familiar  wave-function renormalization of the field $\psi$ and it can be in fact eliminated by introducing $\psi_{R}= Z^{-1/2}_{\psi}\psi$.  This interpretation is also consistent with the fact  that its value is gauge-dependent. The usual 
perimeter divergence, present in a cut-off regularization,  appears as a mass counter-term for the spinor $\psi$
 in the renormalized action. 

According to the previous discussion, the renormalized  operator  for an open smooth contour $C$ is obtained as \cite{Dorn1,Dorn2,Aoyama:1981ev,Knauss:1984rx,Dorn:1986dt}
\be
\label{ren-open}
\begin{split}
{\cal W}_{\rm ren.}=&\left\langle\psi_{R}(-L)\bar\psi_{R}(L)\right\rangle_{0}=\\
=&Z^{-1}_{\psi}
\left\langle\psi(-L)\bar\psi(L)\right\rangle_{0}=Z^{-1}_{\psi} e^{-\ell \delta m}\left\langle
P\exp\left(i g_{\rm ren.} \frac{Z_{1}}{ Z_{3}}\int_{C} dx^{\mu} \mathcal{A}^{\rm ren.}_{\mu}\right)\right\rangle_{0},
\end{split}
\ee
where $Z_{1}$ and $Z_{3}$ are the usual renormalization for the gauge coupling constant and the wave-function renormalization for $\mathcal{A}_{\mu}$. Moreover $\ell$ is the perimeter of the smooth open contour $C$; the mass renormalization $\delta m$ is zero when dimensional regularization is used since it corresponds to a power-like divergence.

An important remark is now in order.  In dimensional regularization the new renormalization constant $Z_{\psi}$ can be shown to be  independent of the shape of the  smooth contour \cite{Dorn1,Dorn2,Aoyama:1981ev,Knauss:1984rx,Dorn:1986dt}  (up to  a redefinition of the renormalization scale). Accordingly its value can be computed for a  finite segment and then used for other smooth  contours.

When we close the circuit, thus considering a Wilson loop,   a new divergence appears  \cite{Dorn1,Dorn2,Aoyama:1981ev,Knauss:1984rx,Dorn:1986dt}, since the two fields  in $\left\langle\psi(-L)\bar\psi(L)\right\rangle_{0}$ are now located at the same point.  More correctly  the closed loop does not define a two-point function, but the expectation value of a composite operator: this explains the need of a further renormalization.   However the effect of this additional ingredient  is to exactly cancel  the factor 
$Z^{-1}_{\psi}$ \cite{Dorn1,Dorn2,Aoyama:1981ev,Knauss:1984rx,Dorn:1986dt} and one recovers the 
familiar  and simple result\footnote{This result was first shown in \cite{Exp}  using combinatorial techniques.}
\be
\begin{split}
\label{ren-closed}
{\cal W}^{\rm clos.~ loop}_{\rm ren.}= e^{-\ell \delta m}\left\langle
P\exp\left(i g_{\rm ren.} \frac{Z_{1}}{ Z_{3}}\oint_{C} dx^{\mu} \mathcal{A}^{\rm ren.}_{\mu}\right)\right\rangle_{0},
\end{split}
\ee
{\it i.e.}  a smooth Wilson loop does not contain any new divergence with respect to  those of the gauge theory, apart from the one proportional to the perimeter of the contour.  
Since $Z_{\psi}$  is only present when dealing with open circuits, but disappears for closed loops,  it  is also named  $Z_{\rm open}$. 

Let us remark that  the final equalities in \eqref{ren-open} and \eqref{ren-closed}  define a procedure for  renormalizing  smooth  path ordered phase factors independently of the fermionic representation used to prove them.

We come back  to the example of the segment in $\mathcal{N}=4$ SYM. If we introduce the wave-function
renormalization 
\be
Z_{\rm open}=1+g^2N(1-\alpha) \frac{\Gamma(1-\epsilon)}{16\pi^{2-\epsilon}} \frac{(2  L\mu)^\epsilon}{\epsilon}. 
\ee
the expectation value of the renormalized operator becomes again trivial as occurs in Feynman gauge. 
In the case of ABJM, the divergence  can be handled in the same way  by introducing
 \begin{align}
Z_{\rm open}
=&1-\left(\frac{2\pi}{\kappa}\right)\frac{M N}{N+M}\left(\frac{ \Gamma(1/2-\epsilon)}{4\pi^{3/2-\epsilon}}\right)
\frac{(2 L\mu)^{2\epsilon}}{\epsilon}+O(\epsilon).
\end{align}
In other words, with respect to the familiar $\mathcal{N}=4$ result ($\alpha=1$), the Landau gauge used to compute \eqref{Ren1} in ABJM theories   does not  enjoy  the simplifying property $Z_{\rm open}=1$.

\noindent
 Let us now turn to piecewise smooth contours \cite{Polyakov:1980ca,Korchemsky:1987wg,Dorn1,Dorn2,Knauss:1984rx,Dorn:1986dt}, namely to contours containing points where the derivative $\dot{x}^{\mu}$ is  discontinuous. If there is  a {\it cusp} at  $t=t_{0}$, {\it i.e.} $\lim_{t\to t_{0}^{+}} \dot{x}^{\mu}(t)\ne \lim_{t\to t_{0}^{-}} \dot{x}^{\mu}(t)$, the renormalization of the action \eqref{Spsi} requires 
 an additional counter-term proportional to  $\bar \psi(t_{0}) \psi(t_{0})$ \cite{Knauss:1984rx,Dorn:1986dt}.   To argue the origin of this new counter-term we observe that a reasonable renormalization procedure  should respect the composition rule for path-ordered  phase factors on smooth contours. Specifically if we split a regular contour $C$ $\{ x(t) | -L\le t\le L\}$ into two  sub-contours $C_{1}$ $\{ x(t) | -L\le t\le t_{0}\}$  and $C_{2}$ $\{ x(t) | t_{0}\le t\le L\}$ 
\be
\mathcal{W}^{\rm ren.} (C_{1})\mathcal{W}^{\rm ren.}(C_{2})= W^{\rm ren.}(C).
\ee
 In terms of the two point function of the one-dimensional  fermion $\psi$  this property reads
 \be
 \langle\psi(-L)\bar\psi(t_{0})\rangle\langle\psi(t_{0})\bar\psi(L)\rangle=
  \langle\psi(-L)(\bar\psi\psi)(t_{0})\bar\psi(L)\rangle=\langle\psi(-L)\bar\psi(L)\rangle.
 \ee
 The intermediate equality implies that  the renormalization factor $Z_{\bar\psi\psi}$ for the composite operator $(\bar\psi\psi)(t_{0})$ is $1$. This is an equivalent manifestation  of the previous statement that $Z_{\rm open}$  drops out  when the two endpoints of a loop are joined  smoothly.  If $t_{0}$ is instead the position of a cusp the factor $Z_{\bar\psi\psi}$  can be in general different from $1$ and it must be included  in the renormalization of the Wilson-operator. Its insertion leads to the following modification of \eqref{ren-open} for open contour with one cusp
 \cite{Dorn1,Dorn2,Knauss:1984rx,Dorn:1986dt}
 \be
 \label{Wcusp}
\begin{split}
{\cal W}_{\rm ren.}=Z^{-1}_{\rm open} Z_{\bar\psi\psi} e^{-\ell \delta m}\left\langle
P\exp\left(i g_{\rm ren.} \frac{Z_{1}}{ Z_{3}}\int_{C} dx^{\mu} \mathcal{A}^{\rm ren.}_{\mu}\right)\right\rangle_{0}.
\end{split}
\ee
In the following we shall replace the symbol $Z_{\bar\psi\psi}$ with the more familiar $Z_{\rm cusp}$.

The renormalization factor $Z_{\rm cusp}$ can be shown to depend only on the angle $\varphi$ of the cusp  and not on the global geometry of the circuit, and to be gauge invariant.  Moreover it must satisfy a simple renormalization condition \cite{Dorn2,Knauss:1984rx,Dorn:1986dt}
\be
\label{rescind}
\left. Z_{\rm cusp}\right|_{\varphi=0}=1,
\ee
since  the cusp disappears for $\varphi=0$ and no new renormalization is needed  apart from $Z_{\rm open}$.  This condition also appears in \cite{Korchemsky:1987wg}  as a   Ward identity  for the vertex  in the one-dimensional field theory. The new  factor will give  origin to the  well-known cusp-anomalous dimension, which is defined through the relation 
\be
\label{r1bps}
\gamma=\mu\frac{d}{d\mu} \log Z_{\rm cusp}.
\ee
We expect that the above renormalization procedure   carries over to the case of the Wilson loop in  ABJ theory  with minor changes.  In fact  the structure of eq. \eqref{Wcusp}  is substantially independent  of the specific form of $\mathcal{A}_{\mu}$ and  of the route used to prove it.  A detailed
proof of the above results, in the case of the phase operator defined by the super-connection \eqref{superconnection}, could be obtained  by using   the supersymmetric quantum 
mechanics  discussed in \cite{Lee:2010hk} as a starting point instead of  \eqref{Spsi}.   An obvious difference with the above discussion arises when considering the  renormalization 
condition. For our operators  eq. \eqref{rescind} must be  replaced by
\be
\label{r1abps}
\left. Z_{\rm cusp}\right|_{\varphi=\theta=0}=1.
\ee
Recall, in fact, that we have also a cusp in the $R-$symmetry directions  governed by the angle $\theta$ next to geometrical one  given by $\varphi$.  In this language  the BPS condition $\theta=\varphi$  should translate into the following 
\be
\label{r2bps}
\left. Z_{\rm cusp}\right|_{\varphi=\theta}=1.
\ee 
Eq. \eqref{r2bps} is not equivalent to \eqref{r1abps}. Thus  the BPS condition still provides a check of the correctness of our computation.

\noindent
Having in mind  the above discussion, it is straightforward  to extract the renormalized generalized potential $V^{\rm Ren.}_{N}$ from \eqref{unrenpot}. We obtain
\begin{align}  
\!\!V^{\rm Ren.}_{N}=&\left(\frac{2\pi}{\kappa}\right) N\left(\frac{ \Gamma(\frac{1}{2}-\epsilon)}{4\pi^{3/2-\epsilon}}\right)\!
(\mu L)^{2\epsilon}
\!\!\left[\frac{1}{\epsilon}\left(\frac{\cos\frac{\theta}{2}}{\cos\frac{\varphi}{2}}-1\right)\!\!-2\frac{\cos\frac{\theta}{2}}{\cos\frac{\varphi}{2}}\log \left(\sec \left(\frac{\varphi }{2}\right)\!+\!1\right)\!\!+\log 4\right]\!+\nonumber\\
&+\left(\frac{2\pi}{\kappa}\right)^{\!\!2}N^2\!\left(\frac{\Gamma\!\left(\frac{1}{2}-\epsilon\right)}{4 \pi^{3/2-\epsilon}}\right)^2(\mu L)^{4\epsilon}\left[\frac{1}{\epsilon}\log\left(\cos\frac{\varphi}{2}\right)^2\left(\frac{\cos\frac{\theta}{2}}{\cos\frac{\varphi}{2}}-1\right)+O(1)\right],
\end{align}
where we have included the finite terms for completeness. The terms proportional to $1/\epsilon$ give the logarithm of the celebrated  $Z_{\rm cusp}$.  It is trivial to check that $\left. Z_{\rm cusp}\right|_{\varphi=\theta}=1$.

The quark-antiquark potential is recovered by taking the limit $\varphi\to \pi$ and following the prescription of \cite{Drukker:2011za}
\begin{equation}
V^{(s)}_{N}(R)=\frac{N}{k}\frac{1}{R}-\left(\frac{N}{k}\right)^2\frac{1}{R}\log\left(\frac{T}{R}\right).
\end{equation}

We observe a logarithmic, non-analytic term in $T/R$ at the second non-trivial order that, as in four dimensions, is expected to disappear when resummation of the perturbative series is performed. We can also perform the opposite limit, taking large imaginary $\varphi$, and we recover the universal cusp anomaly 
\be
\gamma_{cusp}=\frac{N^2}{k^2},
\ee
that is the result obtained directly from the light-like cusp \cite{Henn:2010ps}.
\subsection{$1/2-$BPS line versus $1/6-$BPS line}
\label{versus}
In the previous subsection we have discussed the appearence of spurious divergences in the quantum computation of our cusped Wilson loops and explained their subtraction procedure: we have also remarked that these divergences obstinately persits in the case of $1/2-$BPS straight-lines, although not contradicting their triviality. However there is still an additional feature 
that may appear puzzling. In \cite{Drukker:2009hy} it was pointed out that the $1/2-$BPS straight-line is cohomologically equivalent to its $1/6-$BPS counterpart, defined in \cite{Drukker:2008zx}.  One can easily show that, at least at one loop, the  expectation value of the latter is trivial without requiring any renormalization, exactly as in ${\cal N}=4$: encountering  divergences in  the evaluation of $1/2-$BPS straight-line seems therefore to contradict the cohomological equivalence. 

The key point of \cite{Drukker:2009hy}, in order  to establish the equivalence of the two observables, was to observe that the difference between 
$\mathcal{W}^{1/2}_{\rm line}$ and $\mathcal{W}^{1/6}_{\rm line}$ can be cast into a $Q-$exact term
\be
\label{pip}
\mathcal{W}^{1/2}_{\rm line}-\mathcal{W}^{1/6}_{\rm line}=Q V,
\ee 
where  the supercharge  $Q$ is that  generated by the spinor  $ \bar\theta^{IJ\beta}= (\bar n^{I} \bar w^{J}-\bar n^{J} \bar w^{I})\bar\eta^{\beta} -i\epsilon^{IJKL} n_{K} w_{L}\eta^{\beta}$, while  the scalar couplings in  $\mathcal{W}^{1/6}_{\rm line}$ are governed  by  the matrices $M_{J}^{\ \ I}=
\widehat M_{J}^{\ \ I}= \delta^{I}_{J}-2  n_{J} \bar n^{I}-2  w_{J} \bar w^{I}.$  A complete expression for $V$ has been presented in  \cite{Drukker:2009hy}, but we shall not report it  here.   To understand why the above identity fails, it will suffice to consider its lowest non trivial order  in $1/k$: using the notation of \cite{Drukker:2009hy} we explicitly obtain
\be
\label{pip1}
{\small
\begin{split}
-i\int_{-\infty}^{\infty} \!\!\!\! \!\!d t~ \tilde{\mathcal{L}}_{B}-\int_{-\infty}^{\infty} \!\!\!\! \!\!dt_{1}\int_{-\infty}^{t_{1}} \!\!\!\! \!\!d t_{2}
\mathcal{L}_{F}(t_{1})\mathcal{L}_{F}(t_{2})=-\frac{1}{2} Q\left(\int_{-\infty}^{\infty}\!\!\!\! \!\! dt_{1}\int_{-\infty}^{t_{1}}\!\!\!\! \!\! d t_{2}[ \Lambda(\tau_{1}) \mathcal{L}_{F}(\tau_{2})-\mathcal{L}_{F}(\tau_{1}) \Lambda(\tau_{2})) ]\!
\right)\!\!.
\end{split}}
\ee
The quantities $\tilde{\mathcal{L}}_{B},~\mathcal{L}_{F}$ and $\Lambda $ in \eqref{pip1}  are defined by the following matrices
\be
{\small
\begin{split}
\!\!\!\tilde{\mathcal{L}}_{B}=- \frac{4 \pi i }{k}  |\dot x| \begin{pmatrix}
C_{\bar w}\bar C^{w}
&0\\
0 &
\bar C^{w}C_{\bar w}
\end{pmatrix} 
\  \ 
\mathcal{L}_{F}=-i \sqrt{\frac{2\pi}{k}}  |\dot x| \begin{pmatrix}
0
& \eta\bar\psi\\
   \psi\bar{\eta} &
0
\end{pmatrix} 
\ \ 
 \Lambda=-\frac{1}{2} \sqrt{\frac{2\pi}{k}}|\dot x|\begin{pmatrix} 0 &  i\bar C^{w}\\   C_{\bar w} & 0 \end{pmatrix}
 \end{split}}
\ee
where  the scalars are given by $C_{\bar w}= \bar w^{I} C_{I}$ and $\bar C^{w}= \bar C^{J} w_{J}$, the reduced  spinors  are written as  $\bar\psi=\bar\psi^{I} n_{I}$  and $\psi=\psi_{I}\bar n^{I}$.

When we replace the  infinite straight-line with a segment of length  $2L$ to tame the infrared divergences,
the above equality receives a correction from the  value of the scalar fields on the  boundary. Taking properly into account some total derivatives, usually discarded for infinite lenght, \eqref{pip1} is replaced by
\be
\label{pip3}
{\small
\begin{split}
\int_{-\infty}^{\infty} \!\!\!\! \!\!d t~ &\tilde{\mathcal{L}}_{B}+\int_{-\infty}^{\infty} \!\!\!\! \!\!dt_{1}\int_{-\infty}^{t_{1}} \!\!\!\! \!\!d t_{2}
\mathcal{L}_{F}(t_{1})\mathcal{L}_{F}(t_{2})+4 i \int_{-L}^{L}\!\!\!\!d t \left(\frac{\Lambda(L) \Lambda(t)}{|\dot{x}_{L}|}+\frac{\Lambda(t) \Lambda(-L)}{|\dot{x}_{-L}|}\right)=\\
&=-\frac{1}{2} Q\left(\int_{-L}^{L}\!\!\!\! \!\! dt_{1}\int_{-L}^{t_{1}}\!\!\!\! \!\! d t_{2}[ \Lambda(\tau_{1}) \mathcal{L}_{F}(\tau_{2})-\mathcal{L}_{F}(\tau_{1}) \Lambda(\tau_{2})) ]\!
\right).
\end{split}}
\ee
In other words, if defined on a segment the two Wilson operator are not cohomologically equivalent! Actually we can go further and observe that the divergence of the $1/2-$BPS line comes entirely from these {\it boundary terms}, when evaluated at quantum level. For instance it is easy to check that the new term 
in \eqref{pip3} is accountable for  the  result \eqref{Ren1}. The renormalization procedure described in the 
previous subsection is built to subtract exactly these spurious contributions.


\section{Conclusions and outlook}
In this paper we have studied a family of cusped Wilson loops in ABJ(M) super Chern-Simons theory, constructed from two 1/2 BPS lines implying the presence of peculiar fermionic  couplings \cite{Drukker:2009hy}. They depend on two parameters, $\varphi$ and $\theta$, that describe the geometrical and $R$-symmetry angles, respectively, between the two rays. We have studied the supersymmetric properties of these configurations and their relation with closed contours, obtained through conformal transformations. Different limits on the parameters allow to reach interesting observables, as the analogous of the quark-antiquark potential or the universal cusp anomalous dimension. We have performed an explicit two-loop computation in dimensional regularization and we have obtained the divergent part of these contour operators. Our results suggest the existence of two generalized potentials in this theory and, after renormalization, we have obtained in the relevant limits the universal cusp anomaly and the $W_{N(M)}\bar{W}_{N(M)}$ binding energy. 

The construction of a generalized potential from a cusped Wilson loop opens many interesting possibilities in ABJ(M) theory: one could try to compute the radiation of a particle moving along an arbitrary smooth path, as done in ${\cal N}=4$ SYM \cite{Correa:2012at}. Further, one could hope to find a three dimensional analogue of the set of TBA integral equations, recently discovered in \cite{Drukker:2012de,Correa:2012hh}, describing non-perturbatively the D=4 generalized cusp (see also \cite{Bykov:2012sc,Henn:2012qz,Gromov:2012eu} for very recent developments). It is also tempting to speculate on the possibility to derive the infamous interpolating function $h(\lambda)$ \cite{H,Grignani:2008is,Nishioka:2008gz,BT1,BT2,MZ,GGY}, by comparing the integrability computations with exact results obtained through localization \cite{Kapustin:2009kz}. An important step in this program would be the derivation of general class of Wilson loops with lower degree of supersymmetry, specifically some analogue of the DGRT loops \cite{Drukker:2007qr} in ${\cal N}=6$ super Chern-Simons theory. A particular case, the wedge on $S^2$, has been discussed here in sec. 3: a general construction of BPS loops on $S^2$, preserving fractions of supersymmetry, will be presented soon \cite{CGMS}. It would be of course important to compute their quantum expectation value at weak coupling, by perturbation theory, and at strong coupling, using string techniques. Hopefully their exact expression could be derived through localization methods.

\section*{Acknowledgements}
This work was supported in part by the MIUR-PRIN contract 2009-KHZKRX. We warmly thank Antonio Bassetto, Valentina Cardinali, Valentina Forini, Valentina Giangreco Marot\-ta Puletti and especially Nadav Drukker for useful discussions.

\newpage

\appendix
 \begin{flushleft}
 {\huge \bf Appendices}\hfill
 \end{flushleft}
\addcontentsline{toc}{section}{\large Appendices}
\renewcommand{\theequation}{\Alph{section}.\arabic{equation}}

\section{Basics of  ABJ(M) action}
\label{ABJ}
Here we will collect some basic facts about the ABJ(M) action in Euclidean space-time. The gauge sector consists of  two  gauge fields $(A_\mu)_{i}^{\ j}$ and
$(\hat{A}_\mu)_{\hat{i}}^{\ \hat{j}}$ belonging respectively  to the adjoint of $U(N)$ and $U(M)$. The
matter sector  instead contains   the complex fields $(C_I)_{i}^{\ \hat{i}}$ and $({\bar C}^{I})_{\hat{i}}^{\  i}$ as well as the fermions $(\psi_I)_{\hat{i} }^{ \ i}$ and $({\bar\psi}^{I})_{i }^{\ \hat{i} }$. The fields $(C,\bar \psi)$ transform in  the $({\bf N},{\bf \bar M})$  of the gauge group $U(N)\times U(M)$ while the couple $(\bar C, \psi)$ lives in the $({\bf \bar N},{\bf M})$. The additional  capitol index 
$I=1,2,3,4$  belongs to the $R-$symmetry group $SU(4)$.  In order to quantize the theory at the perturbative level, we have introduced  the covariant gauge fixing function 
$\partial_\mu A^\mu$ for both gauge fields and two sets of ghosts $(\bar c,c)$ and
$(\bar{\hat c},\hat c)$. Therefore  we work with the following Euclidian space action 
(see \cite{Chen:1992ee,Aharony:2008ug,Benna:2008zy})
\begin{align}
\label{Lagra}
S_\text{CS} & = -i\frac{ k}{4\pi}\,\int d^3x\, \varepsilon^{\mu\nu\rho} \,\Bigl [\,
\mathrm{Tr} (A_\mu\partial_\nu A_\rho+\frac{2}{3}\,A_\mu A_\nu A_\rho)- 
\mathrm{Tr} (\hat{A}_\mu\partial_\nu \hat{A}_\rho+\frac{2}{3}\, \hat{A}_\mu \hat{A}_\nu \hat{A}_\rho)\, \Bigr ] \nonumber\\
S_\text{gf} & = \frac{k}{4\pi}\, \int d^3 x\, \Bigl [\,\frac{1}{\xi}\, \mathrm{Tr}(\partial_\mu A^\mu)^2
+\mathrm{Tr}(\partial_\mu \bar c\, D_\mu c) - \frac{1}{\xi}\, \mathrm{Tr}(\partial_\mu \hat{A}^\mu)^2
+\mathrm{Tr}(\partial_\mu \bar{ \hat c}\, D_\mu \hat c) \, \Bigr ]\nonumber\\
S_\text{Matter} & = \int d^3 x\, \Bigl [\, \mathrm{Tr}(D_\mu\, C_I\, D^\mu {\bar C}^{I}) + i\, \mathrm{Tr}(\bar\psi^I
\, \slsh{D}\, \psi_I)\, \Bigr ] + S_\text{int}
\end{align}
Here $S_\text{int}$ consists of the sextic scalar potential and $\psi^2 C^2$ Yukawa type potentials
spelled out in \cite{Aharony:2008ug}. The matter covariant derivatives are defined as
\be
\begin{aligned}
D_\mu C_I &= \partial_\mu C_I + i (A_\mu\, C_I - C_I\, \hat{A}_\mu) \\ 
D_\mu {\bar C}^{I} &= \partial_\mu {\bar C}^{I} - i ({\bar C}^{I}\,A_\mu -  \hat{A}_\mu\, {\bar C}_{I})  \\ 
D_\mu \psi_I &= \partial_\mu \psi_I + i (\hat{A}_\mu\, \psi_I - \psi_I\, A_\mu)  \\ 
D_\mu {\bar\psi}^{I} &= \partial_\mu {\bar \psi}^{I} - i ({\bar \psi}^{I}\,\hat{A}_\mu -  A_\mu\, 
{\bar \psi}^{I})\, . 
\end{aligned}
\label{cov-deriv}
\ee

\section{Feynman rules, useful perturbative results and  some spinorology} 
\label{FRSS}
\paragraph{\sc Feynman rules:} In the first part of this appendix we shall briefly review the 
Euclidean Feynman rules relevant for our computation and some general conventions.   We use the 
position-space propagators, which  are obtained  from those in momentum space (see {\it e.g.} \cite{Drukker:2008zx}) by means of  the following  master integral
\be
\int \frac{d^{3-2\epsilon} p}{(2\pi)^{3-2\epsilon}} \frac{e^{i p\cdot x}}{(p^{2})^{s}}=\frac{\Gamma\left(\frac{3}{2}-s-\epsilon\right)}{4^{s} \pi^{\frac{3}{2}-\epsilon}\Gamma(s)}\frac{1}{(x^{2})^{\frac{3}{2}-s-\epsilon}}.
\ee
In Landau gauge, for the gauge field propagators   we find
\be
\begin{split}
\langle (A_{\mu})_{i}^{\ j}(x)  (A_{\nu})_{k}^{\  l} (y)\rangle_{0}=&\delta_{i}^{l}\delta_{k}^{j}\left(\frac{2\pi i}{\kappa}\right)\epsilon_{\mu\nu\rho}\partial^{\rho}_{x}\left(\frac{\Gamma\left(\frac{1}{2}-\epsilon\right)}{4 \pi^{\frac{3}{2}-\epsilon}}\frac{1}{((x-y)^{2})^{\frac{1}{2}-\epsilon}}\right),\\
\langle (\widehat A_{\mu})_{\hat i}^{\ \hat j}(x)  (\widehat A_{\nu})_{\hat k}^{\ \hat l} (y)\rangle_{0}=&-\delta_{\hat i}^{\hat l}\delta^{\hat j}_{\hat k}\left(\frac{2\pi i}{\kappa}\right)\epsilon_{\mu\nu\rho}\partial^{\rho}_{x}\left(\frac{\Gamma\left(\frac{1}{2}-\epsilon\right)}{4 \pi^{\frac{3}{2}-\epsilon}}\frac{1}{((x-y)^{2})^{\frac{1}{2}-\epsilon}}\right).\\
\end{split}
\ee
The scalar propagators are instead given by
\be
\label{scal1}
\begin{split}
\langle (C_{I})_{i}^{\ \hat j}(x)  (\bar C^{J})_{\hat k}^{\  l} (y)\rangle_{0}=&\delta^{J}_{I}~\delta_{i}^{l}~\delta_{\hat k}^{\hat j}\frac{\Gamma\left(\frac{1}{2}-\epsilon\right)}{4 \pi^{\frac{3}{2}-\epsilon}}\frac{1}{((x-y)^{2})^{\frac{1}{2}-\epsilon}}\equiv  \delta^{J}_{I}~\delta_{i}^{l}~\delta_{\hat k}^{\hat j} D(x-y).
\end{split}
\ee
Finally we shall consider  the  case of the tree level  fermionic two-point function
\be
\label{fermionpropagator}
\left\langle
(\psi_{I})_{\hat i}^{\  j}(x)(\bar\psi^{J})_{ k}^{\ \hat l}(y)\right\rangle_{0}=\delta^{J}_{I}~\delta_{\hat i}^{\hat l}~\delta_{ k}^{ j} i\gamma^{\mu}\partial_{\mu}\left(\frac{\Gamma\left(\frac{1}{2}-\epsilon\right)}{4 \pi^{\frac{3}{2}-\epsilon}}\frac{1}{((x-y)^{2})^{\frac{1}{2}-\epsilon}}\right).
\ee
In our computation the interaction vertices  will become relevant  when considering either  the one-loop correction to the   propagators or the  graphs  containing the three-point functions.
The one-loop  two-point function for the gauge field  was computed in \cite{Drukker:2008zx} and in momentum space it is given by
\be
\langle (A_{\mu})_{i}^{\ j}(p)  (A_{\nu})_{k}^{\  l} (-p)\rangle^{\rm 1loop}_{0}=-\delta_{i}^{l}\delta_{k}^{j}\left(\frac{2\pi}{\kappa}\right)^{2}  M   \frac{2^{4 \epsilon -1} \pi ^{\epsilon } \sec (\pi  \epsilon )}{\Gamma (1-\epsilon )} (p^{2})^{-\frac{1}{2}-\epsilon }
    \left(\delta_{\mu\nu}-
 \frac{p_{\mu} p_{\nu}}{p^{2}}\right) .
\ee
In  coordinate space it takes the form
\be
\label{oneloopgauge}
\begin{split}
\langle (A_{\mu})_{i}^{\ j}(x)  (A_{\nu})_{k}^{\  l} (y)\rangle^{1\rm loop}_{0}\!\!=&\delta_{i}^{l}\delta_{k}^{j}\left(\frac{2\pi}{\kappa}\right)^{\!\!2}\!\frac{M\Gamma^{2}\left(\frac{1}{2}-\epsilon\right)}{4 \pi^{3-2\epsilon}} \!\! \left(\frac{\delta_{\mu\nu}}{((x-y)^{2})^{{1}-2\epsilon}}
-\partial_{\mu}\partial_{\nu}\!\left(\frac{((x-y)^{2})^{\epsilon}}{4\epsilon(1+2 \epsilon)}\right)\!\!\right)\! .
\end{split}
\ee
The correction to the gauge propagator of $\widehat A$ is very similar to  \eqref{oneloopgauge}: we have simply to replace $M$ with $N$ and  $\delta_{i}^{l}~\delta_{k}^{j}$ with $\delta_{\hat i}^{\hat l}~\delta^{\hat j}_{\hat k}$.

Next one to consider the one-loop  corrections to the fermion propagator. In momentum space it is given by
\be
\label{ferm1}
\left\langle
(\psi_{I})_{\hat i}^{\  j}(p)(\bar\psi^{J})_{ k}^{\ \hat l}(-p)\right\rangle^{1~\rm \ell oop}_{0}=-2 i\zet\delta_{\hat i }^{\\\hat l}\delta^{j}_{k}(N-M) \frac{16^{\epsilon -1} \pi ^{\epsilon } \left(p^2\right)^{-\frac{1}{2}-\epsilon } \sec (\pi  \epsilon )}{\Gamma (1-\epsilon )}.
\ee
Notice that this expression is finite when $\epsilon$ approaches zero. Its expression in coordinate space
is then obtained  by taking the Fourier-transform 
\be
\label{ferm2}
\left\langle
(\psi_{I})_{\hat i}^{\  j}(x)(\bar\psi^{J})_{ k}^{\ \hat l}(y)\right\rangle^{1~\rm \ell oop}_{0}=
-i\zet\delta_{\hat i }^{\\\hat l}\delta^{j}_{k}(N-M)\frac{\Gamma ^{2}\left(\frac{1}{2}-\epsilon \right)}{16 \pi ^{3-2 \epsilon} }\frac{1}{((x-y)^{2})^{1-2\epsilon}}.
\ee
The last ingredient that is necessary for our analysis of  the two-loop behavior of the cusp in 
ABJ(M) is  the integral 
\be
\label{B9}
\begin{split}
\Gamma^{\lambda\mu\nu}(x_{1},x_{2}, x_{2})=&\left(\frac{\Gamma(\frac{1}{2}-\epsilon)}{4\pi^{3/2-\epsilon}}\right)^{3}\partial_{x_{1}^{\lambda}}\partial_{x_{2}^{\mu}}\partial_{x_{3}^{\nu}}
\int 
\frac{d^{3-2\epsilon}w}{(x_{1w}^{2})^{1/2-\epsilon}(x_{2w}^{2})^{1/2-\epsilon}(x_{3w}^{2})^{1/2-\epsilon}}\equiv\\
\equiv&\partial_{x_{1}^{\lambda}}\partial_{x_{2}^{\mu}}\partial_{x_{3}^{\nu}}\Phi,
\end{split}
\ee
which governs all the three point functions appearing in our analysis. Actually, for planar loops  we shall never need  the closed form of \eqref{B9} but only  its value  when two of the indices are contracted
\begin{subequations}
\label{C10}
\begin{align}
\Gamma^{\lambda\lambda\rho}(x_{1},x_{2},x_{3})=&\partial_{x_{3}^{\rho}}(\partial_{x_{1}}\cdot \partial_{x_{2}})\Phi=
\frac{1}{2}
\partial_{x_{3}^{\rho}}[\square_{x_{3}}-\square_{x_{1}}-\square_{x_{2}}]\Phi\equiv
\partial_{x_{3}^{\rho}}\Phi_{3,12},\\
\Gamma^{\lambda\rho\lambda}(x_{1},x_{2},x_{3})
=&\partial_{x_{2}^{\rho}}(\partial_{x_{1}}\cdot \partial_{x_{3}})\Phi=\frac{1}{2}
\partial_{x_{2}^{\rho}}[\square_{x_{2}}-\square_{x_{1}}-\square_{x_{3}}]\Phi\equiv
\partial_{x_{2}^{\rho}}\Phi_{2,13},\\
\Gamma^{\rho\lambda\lambda}(x_{1},x_{2},x_{3})
=&\partial_{x_{1}^{\rho}}(\partial_{x_{2}}\cdot \partial_{x_{3}})\Phi=\frac{1}{2}
\partial_{x_{1}^{\rho}}[\square_{x_{1}}-\square_{x_{2}}-\square_{x_{3}}]\Phi\equiv
\partial_{x_{1}^{\rho}}\Phi_{1,23},
\end{align}
\end{subequations}
where  we took advantage of the invariance of  the scalar function $\Phi$ under translations [$(\partial_{x_{1}^{\lambda}}+\partial_{x_{2}^{\lambda}}+\partial_{x_{3}^{\lambda}})\Phi=0$]  and introduced the short-hand  notation
\be
\label{C11}
\begin{split}
\Phi_{i,jk}=& -\frac{\Gamma^{2}(1/2-\epsilon)}{32\pi^{3-2\epsilon}}\!\!
\left[\frac{1}{(x^{2}_{ij})^{\frac{1}{2}-\epsilon}(x^{2}_{ik})^{\frac{1}{2}-\epsilon}}-\frac{1}{(x^{2}_{ij})^{\frac{1}{2}-\epsilon}(x^{2}_{kj})^{\frac{1}{2}-\epsilon}}-\frac{1}{(x^{2}_{ik})^{\frac{1}{2}-\epsilon}(x^{2}_{jk})^{\frac{1}{2}-\epsilon}}\right]\!\!.
\end{split}
\ee 
In our computation we are also led to consider the value of  $\Phi_{i,jk}$ at coincident points. For $\epsilon>1/2$ they are finite and given by
\be
\Phi_{i,ik}= \frac{1}{2}\left(\frac{\Gamma(1/2-\epsilon)}{4\pi^{3/2-\epsilon}}\right)^{2}
\frac{1}{(x^{2}_{ik})^{1-2\epsilon}},\ \ \ \ \ \ \ \ \ \ \  
\Phi_{i,jj}= -\frac{1}{2}\left(\frac{\Gamma(1/2-\epsilon)}{4\pi^{3/2-\epsilon}}\right)^{2}
\frac{1}{(x^{2}_{ij})^{1-2\epsilon}}.
\ee 
In the spirit of dimensional regularization we extend these result to any value of $\epsilon$\footnote{
This is equivalent to the usual statement that massless tadpoles vanish in dimensional regularization.}.

The three point-function \eqref{B9} for our specific choice of $x_{1}, x_{2}$ and $x_{3}$  obeys a set 
of useful identities. Consider for instance the case when $x_{2}$ and $x_{3}$  belong to the first edge of the cusp,
while $x_{1}$ is located  on the opposite one. Then we can introduce the  three orthonormal vectors
\be
\dot{x}_{2}^{\nu},\ \ \  \ v^{\nu}=\frac{1}{r} (\dot{x}_{1}^{\nu}-(\dot{x}_{1}\cdot \dot x_{2}) \dot x^{\nu}_{2}),\ \ \  \  n^{\nu}=\frac{1}{r}\epsilon^{\nu\alpha\beta} \dot x_{2\alpha} \dot{x}_{1\beta}\ \ \ \  
\left[r\equiv\sqrt{|\dot x_{1}|^{2}|\dot x_{2}|^{2}-(\dot{x}_{1}\cdot \dot x_{2}) ^{2}}~\right].
\ee
and we can write the trivial identity
\be
\label{C14}
 v_{\nu}(\Gamma^{\tau\nu\tau}-\Gamma^{\tau\tau\nu})-\dot{x}_{2\lambda}\dot{x}_{2\rho} v_{\nu}(\Gamma^{\lambda\nu\rho}-\Gamma^{\lambda\rho\nu})=n_{\lambda} n_{\rho} v_{\nu}(\Gamma^{\lambda\nu\rho}-\Gamma^{\lambda\rho\nu})=0,
\ee
which follows from the complexness condition $\delta_{\mu\nu}=\dot{x}_{2\mu}\dot{x}_{2\nu}+n_{\mu}n_{\nu}+v_{\mu} v_{\nu}$. If we use the explicit form  of the vector $v$,  \eqref{C14} takes the form
\be
\label{C15a}
(\dot{x}_{1\nu}-(\dot{x}_{1}\cdot \dot x_{2}) \dot x_{2\nu})(\Gamma^{\tau\nu\tau}-\Gamma^{\tau\tau\nu})=\dot{x}_{2\lambda}\dot{x}_{2\rho} \dot x_{1\nu}(\Gamma^{\lambda\nu\rho}-\Gamma^{\lambda\rho\nu}).
\ee
\paragraph{\sc Fermionic  contractions:} When computing the {\it fermionic diagrams} contributing  to the Wilson loop defined by the super-connection \eqref{superconnection} we often encounter  bilinears constructed with the spinors $\eta$ and $\bar \eta$  defined by the two relations
\be
\label{B13}
{(\dot{x}^{\mu}\gamma_{\mu})_{\alpha}^{\ \ \beta}}=\frac{1}{2 i} |\dot x|(\eta^{\beta} \bar\eta_{\alpha}+
\eta_{\alpha} \bar\eta^{\beta})\ \ \ \ \ \   (\eta^{\beta} \bar\eta_{\alpha}-
\eta_{\alpha} \bar\eta^{\beta})=2i\delta^{\beta}_{\alpha},
\ee
For instance, the most common is
\be
\eta_{1}\gamma^{\mu}\bar\eta_{2},
\ee
where the superscripts $1$ and $2$  denote two different points on the contour.  We can determine its value up to an overall factor  by means of the following corollary of \eqref{B13}
\be
\label{sds}
\bar\eta_{\alpha}\eta^{\beta}=i\left(\mathds{1}+\frac{\dot{x}^{\lambda}\gamma_{\lambda}}{|\dot{x}|}\right)_{\alpha}^{\ \ \beta}.
\ee
Consider in fact the product $(\eta_{1}\bar\eta_{2})(\eta_{1}\gamma^{\mu}\bar\eta_{2})$.
We can rewrite it as 
\be
(\eta_{1}\bar\eta_{2})(\eta_{1}\gamma^{\mu}\bar\eta_{2})=
\bar\eta_{2\alpha}\eta^{\beta}_{2}(\gamma^{\mu})_{\beta}^{\ \ \rho}\bar\eta_{1\rho}\eta^{\alpha}_{1}=-\mathrm{Tr}\left[
\left(1+\frac{\dot{x_{2}}^{\lambda}\gamma_{\lambda}}{|\dot{x}_{2}|}\right)\gamma^{\mu}\left(1+\frac{\dot{x_{1}}^{\nu}\gamma_{\nu}}{|\dot x_{1}|}\right)\right],
\ee
 where we used  \eqref{sds} in order to eliminate the spinors from the expression. Thus
\be
\label{rg6}
\begin{split}
(\eta_{2}\gamma^{\mu}\bar\eta_{1})
=&-\frac{2}{(\eta_{1}\bar\eta_{2})}\left[\frac{\dot{x_{1}}^{\mu}}{|\dot x_{1}|}+\frac{\dot{x_{2}}^{\mu}}{|\dot{x}_{2}|}-i\zet\frac{\dot{x_{2}}^{\lambda}}{|\dot{x}_{2}|}
\frac{\dot{x_{1}}^{\nu}}{|\dot x_{1}|}\epsilon_{\lambda\nu}^{\ \  \ \mu}
\right].
\end{split}
\ee
The only undetermined factor in \eqref{rg6} is the scalar contraction $(\eta_{1}\bar\eta_{2})$. The condition \eqref{sds} however  determines its {\it norm}, {\it i.e}  the product $(\eta_{1}\bar\eta_{2})(\eta_{2}\bar\eta_{1})$
\be
\begin{split}
(\eta_{1}\bar\eta_{2})(\eta_{2}\bar\eta_{1})=&-\mathrm{Tr}\left[\left(1+\frac{\dot{x_{2}}^{\lambda}}{|\dot{x}_{2}|}\gamma_{\lambda}\right)\left(1+\frac{\dot{x_{1}}^{\nu}}{|\dot x_{1}|}\gamma_{\nu}\right)\right]=-2\left[1+\frac{(\dot{x}_{1}\cdot \dot{x}_{2})}{|\dot x_{1}||\dot{x}_{2}| }\right].
\end{split}
\ee
There is a second bilinear that will be relevant, namely the one containing three Dirac matrices 
\be
\eta_{1}\gamma^{\lambda}\gamma^{\mu}\gamma^{\nu}\bar\eta_{2}.
\ee
Its evaluation  reduces to the previous case because of  the following identity
\be
\gamma^{\rho}\gamma^{\mu}\gamma^{\sigma}=
\delta^{\rho\mu}\gamma^{\sigma}+\delta^{\mu\sigma}\gamma^{\rho}-\delta^{\rho\sigma}\gamma^{\mu}+i\zet\epsilon^{\rho\mu\sigma}\mathds{1},
\ee
which holds for three-dimensional Euclidean gamma matrices.

For completeness, we shall also give all the possible scalar contractions for our specific circuit
\begin{align}
&\eta_{1}\bar\eta_{1}=2 i,\ \ \ \eta_{2}\bar\eta_{2}=2i,\ \ \ \eta_{1}\bar\eta_{2}=2 i e^{i\delta}\cos\frac{\varphi}{2},\ \ \  \eta_{2}\bar\eta_{1}=2 i e^{-i\delta}\cos\frac{\varphi}{2},\ \ \ \nonumber\\
&\eta_{2}\eta_{1}=-2 i e^{-i\delta}\sin\frac{\varphi}{2},\ \ \ \bar\eta_{1}\bar\eta_{2}=2 i e^{i\delta}\sin\frac{\varphi}{2}.
\end{align}
Here the indices $1$ and $2$ indicates the two different edges of the cusp.
\section{Perturbative integrals}
\label{perturbativeintegrals}
In this appendix we have collected some results about the  different integrals which we have encountered in our perturbative 
expansion of the cusp.  In particular, below we  show how to extract  their divergent part, which is the relevant quantity in our calculation. We can naturally divide the integrals into two subfamilies:  the double exchange  and the vertex integrals.
\subsection{Double exchange Integrals}
First we consider the integrals of subsec. \eqref{DED}  appearing in the evaluation of  the fermionic double exchange diagrams.  When all  propagators start and end
on the same edge the integrations can be easily performed in terms of  known functions. When at least one of the two propagators  runs between the 
two edges,  the procedure for extracting the divergent part in $1/\epsilon$ is more intricate. We consider first the diagram $[\ref{DoubleExc1class}.(b)]_{\rm up}$, which is 
governed by the contour  integral
\be
\begin{split}
\mathcal{D}_{1}=\int_{0}^{L}\!\!\!\!d\tau_{1}\int_{-L}^{0}\!\!\!\!d\tau_{2}~ (L+\tau_{2})^{2 \epsilon }(\partial_{\tau_{2}}-\partial_{\tau_{1}})
H(\tau_{1},\tau_{2}),
\end{split}
\ee
where  we recall that $H(\tau_{1},\tau_{2})=(\tau_{1}^{2}+\tau_{2}^{2}-2\tau_{1}\tau_{2}\cos\varphi)^{-\frac{1}{2}+\epsilon}$.  Since $\mathcal{D}_{1}$ is
multiplied by $1/\epsilon$ in the expression for $[\ref{DoubleExc1class}.(b)]_{\rm up}$, we are actually interested in its divergent and finite part. In order  to compute them we first   integrate by parts the derivative with respect to $\tau_{2}$ and   perform the integration of the derivative  with respect to $\tau_{1}$.  Then $\mathcal{D}_{1}$ takes the following form
\begin{align}
\label{DY1}
\mathcal{D}_{1}
=&\!\!\int_{0}^{L}\!\!\!\!\!\!d\tau_{1} ~L^{2\epsilon} (H(\tau_{1},0)-H(\tau_{1},-L))-
2\epsilon\int_{0}^{L}\!\!\!\!\!\!d\tau_{1}\int_{-L}^{0}\!\!\!\!\!\!d\tau_{2} ~(L+\tau_{2})^{2 \epsilon -1}
(H(\tau_{1},\tau_{2})-\nonumber\\
&-H(\tau_{1},-L))-\int_{-L}^{0}\!\!\!\!\!\!d\tau_{2} ~(L+\tau_{2})^{2 \epsilon }(H(L,\tau_{2})-H(0,\tau_{2})).
\end{align}
The double integral in \eqref{DY1}, which is multiplied by $\epsilon$, is finite and thus we can drop it since it does not yield  divergent or finite terms. 
For the two single integrals the relevant contribution is easily evaluated and it is given by 
\be
\begin{split}
\mathcal{D}_{1}
=&L^{4\epsilon}\left[\frac{1}{\epsilon }-2\log \left(\sec \left(\frac{\varphi }{2}\right)+1\right)+O(\epsilon)\right]
\end{split}
\ee
Next we examine the contour integral controlling the graph $[\ref{DoubleExc2class}.(b)]_{\rm up}$:
\be
\mathcal{D}_{2}=\!\int_{0}^{L}\!\!\!\!d\tau_{1}\int_{-L}^{0}\!\! \!\!d\tau_{4}~{(-\tau_{4})^{2 \epsilon }}
(\partial_{\tau_{4}}-\partial_{\tau_{1}})H(\tau_{1},\tau_{4}).
\ee
If we  introduce the short-hand notation
$
G(\tau_{4},\tau_{1})=(\partial_{\tau_{4}}-\partial_{\tau_{1}})H(\tau_{1},\tau_{4})
$, 
we can rewrite  the above integral as follows
\begin{align}
\mathcal{D}_{2}=&L^{4\epsilon}\int_{0}^{1}\!\!\!\!d\tau_{1}\int_{0}^{1}\!\!\!\!d\tau_{4}~
{\tau_{4}^{2 \epsilon }}~
G(-\tau_{4},\tau_{1})=
L^{4\epsilon}\int_{0}^{1}\!\!\!\!d\tau_{1}\int_{0}^{\tau_{1}}\!\!\!\!d\tau_{4}~
[{\tau_{4}^{2 \epsilon }}~
G(-\tau_{4},\tau_{1})+{\tau_{1}^{2 \epsilon }}~
G(-\tau_{1},\tau_{4})]=\nonumber\\
&=L^{4\epsilon}\int_{0}^{1}\!\!\!\!d\tau_{1}\int_{0}^{1}\!\!\!\!d\tau_{4}~\tau_{1}
[{(\tau_{1}\tau_{4})^{2 \epsilon }}~
G(-\tau_{4} \tau_{1},\tau_{1})+{\tau_{1}^{2 \epsilon }}~
G(-\tau_{1},\tau_{4}\tau_{1})].
\end{align}
If we now observe that $G(\tau_{1},\tau_{2})$ is an homogeneous function of degree $-2+
2 \epsilon$ in both variables, we can factor out the integration
over $\tau_{1}$
\begin{align}
\mathcal{D}_{2}&=L^{4\epsilon}\int_{0}^{1}\!\!\!\!d\tau_{1}\int_{0}^{1}\!\!\!\!d\tau_{4}~\tau_{1}^{-1+4\epsilon}
[{\tau_{4}^{2 \epsilon }}~
G(-\tau_{4} , {1})+
G(-1,\tau_{4})]
=\frac{L^{4\epsilon}}{4\epsilon}\int_{0}^{1}\!\!\!\!d\tau_{4}~
\left( {\tau_{4}^{2 \epsilon }}+1\right)
G(-1,\tau_{4}).
\end{align}
Subsequently we expand in power of $\epsilon$  the integrand and perform the integration over $\tau_{4}$. We find 
\begin{align}
\mathcal{D}_{2}={L^{4\epsilon}}\left[\frac{1}{2\epsilon}+\log \left(\frac{1}{4} \cos \left(\frac{\varphi }{2}\right) \sec
   ^4\left(\frac{\varphi }{4}\right)\right) +O(\epsilon)\right].
\end{align}
The last non trivial integral is the one appearing in  the diagram $[\ref{DoubleExc2class}.(c)]_{\rm up}$ and it is given by
\be
\label{crossedD}
\begin{split}
\mathcal{D}_{3}= &\int_{0}^{L}\!\!\!\!\!\!d\tau_{1}\int^{\tau_{1}}_{0}\!\!\!\!\!\!d\tau_{2}\int_{-L}^{0}\!\!\!\!\!\!d\tau_{3}\int_{-L}^{\tau_{3}}\!\!\!\!\!\!d\tau_{4}~
(\partial_{\tau_{1}}-\partial_{\tau_{3}}) H(\tau_{1},\tau_{3})
(\partial_{\tau_{4}}-\partial_{\tau_{2}})H(\tau_{2},\tau_{4})=
\\
=&\int_{0}^{L}\!\!\!\!\!\!d\tau_{1}\int^{\tau_{1}}_{0}\!\!\!\!\!\!d\tau_{2}\int_{-L}^{0}\!\!\!\!\!\!d\tau_{3}\int_{-L}^{\tau_{3}}\!\!\!\!\!\!d\tau_{4}~\left[
\partial_{\tau_{1}}H(\tau_{1},\tau_{3})\partial_{\tau_{4}}H(\tau_{2},\tau_{4})+\partial_{\tau_{3}} H(\tau_{1},\tau_{3})
\partial_{\tau_{2}}H(\tau_{2},\tau_{4})-\right.\\
&\ \ \ \ \ \ \ \ \ \ \ \ \ -\left.
\partial_{\tau_{1}}H(\tau_{1},\tau_{3})
\partial_{\tau_{2}}H(\tau_{2},\tau_{4})-\partial_{\tau_{3}} H(\tau_{1},\tau_{3})
\partial_{\tau_{4}}H(\tau_{2},\tau_{4})\right]
\end{split}
\ee
Consider the first two term  in \eqref{crossedD}: we can  perform two of the four integrations and  we can write
\begin{align}
\label{werp}
L^{4\epsilon}\int_{0}^{1}\!\!\!\!\!\!d\tau_{1}\int_{-1}^{0}\!\!\!\!\!\!d\tau_{2}~&\biggl[
[H(1,\tau_{2})-H(\tau_{1},\tau_{2})][H(\tau_{1},\tau_{2})-H(\tau_{1},-1)]
+\nonumber\\
&+[H(\tau_{1},0)-H(\tau_{1},\tau_{2})][H(\tau_{1},\tau_{2})-H(0,\tau_{2})]\biggr].
\end{align}
When  expanding the integrand of \eqref{werp} we encounter three basic divergent integrals, whose $\epsilon-$expansion can be easily determined. 
The first one is very simple and yields
\begin{align}
\!\!
\mathrm{(I)}_{\ref{werp}}=&\int_{0}^{L}\!\!\!\!d\tau_{1}\int_{-L}^{0}\!\!\!\! d\tau_{2}H(\tau_{1},0) H(0,\tau_{2})=
\frac{L^{4\epsilon}}{4\epsilon^{2}},\ 
\end{align}
The other two require a little effort. We have an integral which has the  structure of the single exchange of a scalar in four dimensions  up to the redefinitions
$2\epsilon\to 4\epsilon$
\begin{align}
\label{D11}
\mathrm{(II)}_{\ref{werp}}=&\int_{0}^{L}\!\!\!\! d\tau_{1}\int^{0}_{-L}\!\!\!\! d\tau_{2} H(\tau_{1},\tau_{2})^{2}=\int_{0}^{L}\!\!\!\! d\tau_{1}\int^{0}_{-L}\!\!\!\! d\tau_{2}\frac{1}{(\tau_{1}^{2}+\tau_{2}^{2}-2\tau_{1}\tau_{2}\cos\varphi)^{1-4\epsilon}}=\nonumber\\
=&\frac{L^{4\epsilon}}{4\epsilon}\frac{\varphi}{\sin\varphi}+O(1).
\end{align}
Finally we have to consider 
\begin{align}
&\mathrm{(III)}_{\ref{werp}}=
\int_{0}^{L}\!\!\!\! d\tau_{1}\int_{-L}^{0}\!\!\!\! d\tau_{2}~ H(\tau_{1},\tau_{2})H(\tau_{1},0)=L^{4\epsilon}
\int_{0}^{1}\!\!\!\! d\tau_{1}\int_{0}^{1}\!\!\!\! d\tau_{2}~ H(\tau_{1},-\tau_{2})\tau_{1}^{-1+2\epsilon}=\nonumber\\
&=L^{4\epsilon}\int_{0}^{1} d\tau_{1}\int_{0}^{\tau_{1}}d\tau_{2}~ H(\tau_{1},-\tau_{2})\tau_{1}^{-1+2\epsilon}+L^{4\epsilon}
\int_{0}^{1} d\tau_{2}\int_{0}^{\tau_{2}}d\tau_{1}~ H(\tau_{1},-\tau_{2})\tau_{1}^{-1+2\epsilon}.
\end{align}
If we now use that $H(\tau_{1},\tau_{2})$ it is an homogeneous function of degree $-1+2\epsilon$ we can easily perform one of the two integrations  and we find
\begin{align}
=&\frac{L^{4\epsilon}}{4\epsilon}\int_{0}^{{1}}d\tau_{2}~ H({1},-\tau_{2})+\frac{L^{4\epsilon}}{4\epsilon}
\int_{0}^{1}d\tau_{1}~ [H(\tau_{1},-1)-1]\tau_{1}^{-1+2\epsilon}+\frac{L^{4\epsilon}}{4\epsilon}\int_{0}^{1}d\tau_{1}~ \tau_{1}^{-1+2\epsilon}=\nonumber\\
=&-\frac{L^{4\epsilon}}{2\epsilon}\log\left(\cos\frac{\varphi}{2}\right)+\frac{L^{4\epsilon}}{8\epsilon^{2}}+O(1).
\end{align}
Then the divergent part of the integral \eqref{werp} is obtained by considering the combination
\be
-\mathrm{(I)}_{\ref{werp}}-2 \mathrm{(II)}_{\ref{werp}}+2 \mathrm{(III)}_{\ref{werp}}=-\frac{L^{4\epsilon}}{2\epsilon}\frac{\varphi}{\sin\varphi}
-\frac{L^{4\epsilon}}{\epsilon}\log\left(\cos\frac{\varphi}{2}\right)+O(1).
\ee
The last two terms  in \eqref{crossedD} can be shown to be equal by performing the change of variables $\tau_{1}\leftrightarrow-\tau_{4}$ and $\tau_{2}\leftrightarrow-\tau_{3}$.  Thus we double the first one 
and we  perform  one of the four integration. We obtain
\be
\label{ccc}
\begin{split}
\mathcal{D}_{4}\equiv-2L^{4\epsilon}\int_{0}^{1}\!\!\!\!d\tau_{1}\int_{-1}^{0}\!\!\!\!\!\!d\tau_{3}\int_{-1}^{\tau_{3}}\!\!\!\!\!\!d\tau_{4}~
\partial_{\tau_{1}}H(\tau_{1},\tau_{3})
[H(\tau_{1},\tau_{4})-H(0,\tau_{4})]
\end{split}
\ee
This integral  is quite hard to compute and it will also appear when consider the vertex diagrams.   To evaluate it,  we shall introduce the auxiliary function
$F(\tau_{1},\tau_{3})\equiv H(\tau_{1},\tau_{3})-H(0,\tau_{3})$, which  obeys the following relation
\be
\label{Eu}
\tau_{1}\partial_{\tau_{1}}F(\tau_{1},\tau_{3})+\tau_{3}\partial_{\tau_{3}} F(\tau_{1},\tau_{3})-(2\epsilon-1)F(\tau_{1},\tau_{3})=0,
\ee
since it is a homogeneous of degree $(2\epsilon-1)$. Next we use \eqref{Eu} to eliminate $\partial_{\tau_{1}}H(\tau_{1},\tau_{3})[\equiv \partial_{\tau_{1}}F(\tau_{1},\tau_{3}) ]$ from \eqref{ccc} and we find
\begin{align}
\label{ccc1}
\frac{\mathcal{D}_{4}}{2}&
=-(2\epsilon-1)L^{4\epsilon}\int_{0}^{1}\!\!\frac{d\tau_{1}}{\tau_{1}}\int_{-1}^{0}\!\!\!\!\!\!d\tau_{3}\int_{-1}^{\tau_{3}}\!\!\!\!\!\!d\tau_{4}~
F(\tau_{1},\tau_{3})
F(\tau_{1},\tau_{4})+\nonumber\\
&+L^{4\epsilon}\!\!\int_{0}^{1}\!\!\frac{d\tau_{1}}{\tau_{1}}\!\!\int_{-1}^{0}\!\!\!\!\!\!d\tau_{3}\!\!\int_{-1}^{\tau_{3}}
\!\!\!\!\!\!d\tau_{4}F(\tau_{1},\tau_{4})\tau_{3}\partial_{\tau_{3}} F(\tau_{1},\tau_{3})
=
-\frac{(2\epsilon-1)L^{4\epsilon}}{2}\!\!\int_{0}^{1}\!\!\frac{d\tau_{1}}{\tau_{1}}
\left(\int_{-1}^{0}\!\!\!\!\!\!d\tau_{4}~
F(\tau_{1},\tau_{4})\right)^{2}\!\!\!-\nonumber\\
&-L^{4\epsilon}\int_{0}^{1}\!\!\frac{d\tau_{1}}{\tau_{1}}\int_{-1}^{0}\!\!\!\!d\tau_{4}~\tau_{4}  F(\tau_{1},\tau_{4})
F(\tau_{1},\tau_{4})-
\frac{L^{4\epsilon}}{2}\int_{0}^{1}\!\!\frac{d\tau_{1}}{\tau_{1}}\left(\int_{-1}^{0}\!\!\!\!d\tau_{4}~
F(\tau_{1},\tau_{4})\right)^{2}=\nonumber\\
&=-\epsilon L^{4\epsilon}\!\!\int_{0}^{1}\!\!\frac{d\tau_{1}}{\tau_{1}}
\left(\int_{-1}^{0}\!\!\!\!\!\!d\tau_{4}~
F(\tau_{1},\tau_{4})\right)^{2}-L^{4\epsilon}\!\!\int_{0}^{1}\!\!\frac{d\tau_{1}}{\tau_{1}}\int_{-1}^{0}\!\!\!\!d\tau_{4}~\tau_{4}F^{2} (\tau_{1},\tau_{4}),
\end{align}
where we have performed the integration by parts over $\tau_{3}$. We shall now  consider the two integrals in \eqref{ccc1}: the former,
\be
V\equiv\int_{0}^{1}\!\!\frac{d\tau_{1}}{\tau_{1}}
\left(\int_{-1}^{0}\!\!\!\!\!\!d\tau_{4}~
F(\tau_{1},\tau_{4})\right)^{2},
\ee
can be simplified by recalling the relation \eqref{Eu} obeyed by $F(\tau_{1},\tau_{4})$. In fact
\[
\begin{split}
V=&\frac{1}{2\epsilon-1}\int_{0}^{1}\!\!\frac{d\tau_{1}}{\tau_{1}}
\left(\int_{-1}^{0}\!\!\!\!\!\!d\tau_{4}~
\tau_{1}\partial_{\tau_{1}}F(\tau_{1},\tau_{4})+\tau_{4}\partial_{\tau_{4}} F(\tau_{1},\tau_{4})\right)\left(\int_{-1}^{0}\!\!\!\!\!\!d\tau_{3}~
F(\tau_{1},\tau_{3})\right)=\\
=&\frac{1}{2(2\epsilon-1)}\left[\int_{0}^{1}\!\!{d\tau_{1}}\partial_{\tau_{1}}\left(\int_{-1}^{0}\!\!\!\!\!\!d\tau_{4}~
F(\tau_{1},\tau_{4})\right)^{2}\!\!\!\!+2\!\!
\int_{0}^{1}\!\!\frac{d\tau_{1}}{\tau_{1}}
\int_{-1}^{0}\!\!\!\!\!\!d\tau_{4}~
\tau_{4}\partial_{\tau_{4}} F(\tau_{1},\tau_{4})\!\!\int_{-1}^{0}\!\!\!\!\!\!d\tau_{3}~
F(\tau_{1},\tau_{3})\right]\!\!=\\
=&\frac{1}{2(2\epsilon-1)}\left(\int_{-1}^{0}\!\!\!\!\!\!d\tau_{4}~
F(1,\tau_{4})\right)^{2}+\frac{1}{2\epsilon-1}\int_{0}^{1}\!\!\frac{d\tau_{1}}{\tau_{1}}
~
 F(\tau_{1},-1)\left(\int_{-1}^{0}\!\!\!\!\!\!d\tau_{4}~
F(\tau_{1},\tau_{4})\right)-\frac{V}{2\epsilon-1},\nonumber
\end{split}
\]
where we have performed the integration over $\tau_{1}$ in the first integral and we have integrated the second one by parts 
with respect to $\tau_{4}$.  This is an equation that can be solved for $V$ and we get
\be
\begin{split}
 V=\frac{1}{4\epsilon}\left(\int_{-1}^{0}\!\!\!\!d\tau_{4}~
F(1,\tau_{4})\right)^{2}+\frac{1}{2\epsilon}\int_{0}^{1}\!\!\frac{d\tau_{1}}{\tau_{1}}
~\int_{0}^{1}\!\!\!\! d\tau_{4}~
 F(\tau_{1},-1)
F(\tau_{1},-\tau_{4})
\end{split}
\ee
In the same way we can show that 
\be
W\equiv\int_{0}^{1}\!\!\frac{d\tau_{1}}{\tau_{1}}\int_{-1}^{0}\!\!\!\!d\tau_{4}~\tau_{4}[F(\tau_{1},\tau_{4})]^{2}
\ee
obeys the following relation
\be
W=\frac{1}{4\epsilon}\int_{-1}^{0}\!\!\!\!d\tau_{4}~\tau_{4}[F({1},\tau_{4})]^{2}-\frac{1}{4\epsilon}\int_{0}^{1}\!\!\frac{d\tau_{1}}{\tau_{1}}
~[F(\tau_{1},-1)]^{2}.
\ee
Collecting the two contributions $V$ and $W$, we finally obtain
\be
\label{DFG}
\begin{split}
\frac{\mathcal{D}_{4}}{2}=&- L^{4\epsilon}\left[\frac{1}{4}\left(\int_{-1}^{0}\!\!\!\!d\tau_{4}~
F(1,\tau_{4})\right)^{2}+\frac{1}{2}\int_{0}^{1}\!\!\frac{d\tau_{1}}{\tau_{1}}
~\int_{0}^{1}\!\!\!\! d\tau_{4}~
 F(\tau_{1},-1)
F(\tau_{1},-\tau_{4})-\right.\\
&\left.-\frac{1}{4\epsilon}\int_{0}^{1}\!\!\!\!d\tau_{1}~\tau_{1}[F({1},-\tau_{1})]^{2}-\frac{1}{4\epsilon}\int_{0}^{1}\!\!\frac{d\tau_{1}}{\tau_{1}}
~[F(\tau_{1},-1)]^{2}\right].
\end{split}
\ee
The final step is to evaluate the four integrals in \eqref{DFG}. The first one must be computed up to terms which vanish when $\epsilon\to 0$ and 
it yields
\begin{align}
\int_{-1}^{0}\!\!\!\!d\tau_{4}~
F(1,\tau_{4})=-\frac{1}{2\epsilon}+\log \left(\sec \frac{\varphi }{2}+1\right)+O(\epsilon).
\end{align}
For the remaining three integrals, it is sufficient to determine the divergent part and we have
\begin{subequations}
\begin{align}
&\int_{0}^{1}\!\!\frac{d\tau_{1}}{\tau_{1}}
~\int_{0}^{1}\!\!\!\! d\tau_{4}~
 F(\tau_{1},-1)
F(\tau_{1},-\tau_{4})=\frac{1}{2 \epsilon }\log \left(2 \cos ^2\frac{\varphi }{4}\cos \frac{\varphi }{2}\right)+O(1)\\
&\frac{1}{4\epsilon}\int_{0}^{1}\!\!\!\!d\tau_{1}~\tau_{1}[F({1},-\tau_{1})]^{2}=\frac{1}{16 \epsilon ^2}-\frac{1}{8 \epsilon }\varphi  \cot \varphi -\frac{1}{2 \epsilon }\log
   \left(\sec \left(\frac{\varphi }{2}\right)+1\right)+\\
   &\ \ \ \ \ \ \ \ \ \ \ \ \ \ \ \ \ \  \ \ \  \ \ \ \ \ \ \ \ \ \ \ +\frac{1}{8 \epsilon }\log (2
   \cos (\varphi )+2)+O(1)\nonumber\\
&\frac{1}{4\epsilon}\int_{0}^{1}\!\!\frac{d\tau_{1}}{\tau_{1}}
~[F(\tau_{1},-1)]^{2}=\frac{1}{4\epsilon}\log \left(2 \cos ^4\frac{\varphi }{4} \cos \frac{\varphi }{2}\right)-\frac{1}{8\epsilon}\varphi  \cot \varphi +O(1)
\end{align}
\end{subequations}
When collecting these different contributions, we finally find
\be
\mathcal{D}_{4}=\frac{L^{4\epsilon}}{2\epsilon} \left[{\log \left(\cos \frac{\varphi }{2}\right)}-\frac{1}{2 }\varphi  \cot \varphi \right]+O(1)
\ee
\subsection{Vertex Integrals}
To begin with, we shall determine the divergent part of the integral in \eqref{cache}. We have two contributions: one has the form of the total derivative,
while the second one is similar to the integral $\mathcal{D}_{4}$ encountered in the previous subsection. First we consider 
\begin{align}
\label{D1}
&\int_{0}^{L}\!\!\!\! d\tau_{1}\int_{0}^{\tau_{1}}\!\!\!\!d\tau_{2}\int_{-L}^{0}\!\!\!\!d\tau_{3}~\frac{d}{d\tau_{3}}\Phi_{1,23}=\int_{0}^{L}\!\!\!\! d\tau_{1}\int_{0}^{\tau_{1}}\!\!\!\!d\tau_{2}(\Phi_{1,20}-\Phi_{1,2-L})=\nonumber\\
=&-\frac{\Gamma^{2}(\frac{1}{2}-\epsilon)}{32\pi^{3-2\epsilon}}\int_{0}^{L}\!\!\!\! d\tau_{1}\!\!\int_{0}^{\tau_{1}}\!\!\!\!d\tau_{2}~( (\tau_{1}-\tau_{2})^{-1+2\epsilon} \tau_{1}^{-1+2\epsilon}\!\!-\!\tau^{-1+2\epsilon}_{1} \tau^{-1+2\epsilon}_{2} \!\!-\!(\tau_{1}-\tau_{2})^{-1+2\epsilon} \tau_{2}^{-1+2\epsilon})+\nonumber\\
&+O(1)=\frac{\Gamma^{2}(\frac{1}{2}-\epsilon)}{32\pi^{3-2\epsilon}}\frac{L^{4\epsilon}}{4\epsilon^{2}}+O(1),
\end{align}
where we have neglected $\Phi_{1,2-L}$ since it does not yield divergent contributions. Next we consider the master integral 
 \begin{align}
& \label{D2a}
\int_{0}^{	1} \!\!\!\!d\tau_{1} \int_{0}^{\tau_{1}} \!\!\!\!d\tau_{2}\int_{-1}^{0}\!\!\!\! d\tau_{3}\frac{1}{(x^{2}_{13})^{1/2-\epsilon}}\frac{d}{d \tau_{3}}
\frac{1}{(x^{2}_{23})^{1/2-\epsilon}}.
\end{align}
With  the help of an elementary change of variable, we can relate this integral to $\mathcal{D}_{4}$ and we find
\begin{align}
\label{D2}
\!\frac{\mathcal{D}_{4}}{2}-\!\!\!
\int_{0}^{L}\!\!\!\!\!\!d\tau_{1}
 \int_{-L}^{0}\!\!\!\!\!\!d\tau_{2}\int_{-L}^{\tau_{2}} \!\frac{d\tau_{3}}{(-\tau_{3})^{1-2\epsilon}}
\frac{d}{d \tau_{1}} &H(\tau_{1},\tau_{2})\!=\nonumber\\
=&\biggl[\frac{L^{4\epsilon}}{2\epsilon} \left({\log \left(\cos \frac{\varphi }{2}\right)}-\frac{1}{2 }\varphi  \cot \varphi \right)+\frac{L^{4\epsilon}}{8\epsilon^{2}}+O(1)\biggr]
\end{align}
We come now to discuss the two integrals containing only the traced three point function and appearing
in the sum $\mathcal{I}_{(a)}+\mathcal{I}_{(b)}$ in \eqref{5.39}. We  first evaluate 
\begin{align}
&\int_{0}^{L}\!\!\!\!\!d\tau_{1}\!\!\int_{-L}^{0}\!\!\!\!\!d\tau_{2}\!\!\int_{-L}^{0}\!\!\!\!\!d\tau_{3}~\frac{d}{d\tau_{3}}\Phi_{3,12}=\int_{0}^{L}\!\!\!\!\!d\tau_{1}\!\!\int_{-L}^{0}\!\!\!\!\!d\tau_{2}~(\Phi_{0,12}-\Phi_{-L,12})=\nonumber\\
&=-\frac{\Gamma^{2}(1/2-\epsilon)}{32\pi^{3-2\epsilon}}L^{4\epsilon}\int_{0}^{1} \!\!\!\!d\tau_{1}\int_{0}^{1}\!\!\!\!d\tau_{2}~\left[\frac{1}{\tau_{1}^{1-2\epsilon}\tau_{2}^{1-2\epsilon}}-\frac{2\tau_{1}^{-1+2\epsilon}}{(\tau_{1}^{2}+\tau_{2}^{2}+2\tau_{1}\tau_{2}\cos\varphi)^{1/2-\epsilon}} \right]+O(1)=\nonumber\\
&=-\frac{\Gamma^{2}(1/2-\epsilon)}{32\pi^{3-2\epsilon}}\frac{L^{4\epsilon}}{\epsilon}\log\left(\cos\frac{\varphi}{2}\right)+O(1).
\end{align}
Again the sector of the integrand which is evaluated in   $-L$ ($\Phi_{-L,12}$) yields a finite integral. Subsequently we determine the divergent part of 
\begin{align}
\label{D30}
&\int_{0}^{L} \!\!\!\!d\tau_{1}\int_{-L}^{0}\!\!\!\!d\tau_{2}\int_{-L}^{0}\!\!\!\!d\tau_{3}~\left(\frac{d}{d\tau_{1}} \Phi_{1,23}-\frac{d}{d\tau_{3}} \Phi_{1,23}-\frac{d}{d\tau_{2}} \Phi_{1,23}\right)
=\nonumber\\
=&\int_{0}^{L} \!\!\!\!d\tau_{1}\int_{-L}^{0}\!\!\!\!d\tau_{2}\int_{-L}^{0}\!\!\!\!d\tau_{3}~\left(\frac{d}{d\tau_{1}} \Phi_{1,23}-2\frac{d}{d\tau_{3}} \Phi_{1,23}\right)=\nonumber\\
=&\int_{-L}^{0}\!\!\!\!d\tau_{2}\int_{-L}^{0}\!\!\!\!d\tau_{3}~(\Phi_{L,23}-\Phi_{0,23})-2\int_{0}^{L} \!\!\!\!d\tau_{1}\int_{-L}^{0}\!\!\!\!d\tau_{2}~(\Phi_{1,20}-\Phi_{1,2-L})
\end{align}
We have four different contributions in the above integral and  we shall compute them separately 
\begin{subequations}
\begin{align}
\label{N1}
{\rm (I)}_{\ref{D30}}=&\int_{-L}^{0}\!\!\!\!d\tau_{2}\int_{-L}^{0}\!\!\!\!d\tau_{3}~\Phi_{L,23}=\nonumber\\
&=\frac{\Gamma^{2}(1/2-\epsilon)}{32\pi^{3-2\epsilon}}L^{4\epsilon}\!\!\int_{-1}^{0}\!\!\!\!d\tau_{2}\int_{-1}^{0}\!\!\!\!d\tau_{3}\left[\frac{((\tau_{2}-\tau_{3})^{2})^{-\frac{1}{2}+\epsilon}}{(\tau_{2}^{2}+1-2\tau_{2}\cos\varphi)^{\frac{1}{2}-\epsilon}}+2\leftrightarrow 3\right]+O(1)
=\nonumber\\
&=\frac{ L^{4\epsilon}}{\epsilon}\frac{\Gamma^{2}(1/2-\epsilon)}{16\pi^{3-2\epsilon}}{\log \left(\sec \left(\frac{\varphi }{2}\right)+1\right)}+O(1)
\end{align}
In \eqref{N1} we have first performed the integration over $\tau_{3}$ in order to extract the divergence $1/\epsilon$. The remaining integral over $\tau_{2}$
is finite and it can be computed at $\epsilon=0$. We now come to consider 
\begin{align}
{\rm (II)}_{\ref{D30}}&=\int_{-L}^{0}\!\!\!\!d\tau_{2}\int_{-L}^{0}\!\!\!\!d\tau_{3}~\Phi_{0,23}=\nonumber\\
 &=-\frac{\Gamma^{2}(1/2-\epsilon)}{32\pi^{3-2\epsilon}}\!\!\int_{-L}^{0}\!\!\!\!d\tau_{2}\int_{-L}^{0}\!\!\!\!d\tau_{3}~\left[ \frac{1}{\tau_{2}^{1-2\epsilon}\tau_{3}^{1-2\epsilon}}
 -((\tau_{2}-\tau_{3})^{2})^{-\frac{1}{2}+\epsilon}( \frac{1}{\tau_{2}^{1-2\epsilon}}+\frac{1}{\tau_{3}^{1-2\epsilon}} )\right]=\nonumber\\
&=\frac{\Gamma^{2}(1/2-\epsilon)}{32\pi^{3-2\epsilon}}\frac{L^{4\epsilon}}{2\epsilon^{2}}+O(1)
\end{align}
The third integral is easy to compute since the $\varphi-$dependent  contributions drop out from our calculation. In fact
\begin{align}
&{\rm (III)}_{\ref{D30}}=\int_{0}^{L} \!\!\!\!d\tau_{1}\int_{-L}^{0}\!\!\!\!d\tau_{2}~\Phi_{1,20}=\nonumber\\
&=-\frac{\Gamma^{2}(1/2-\epsilon)}{32\pi^{3-2\epsilon}} L^{4\epsilon}\!\!\int_{0}^{1} \!\!\!\!d\tau_{1}\int_{0}^{1}\!\!\!\!d\tau_{2}\left[\frac{\tau_{1}^{-1+2\epsilon}}{(\tau_{1}^{2}+\tau_{2}^{2}+2\tau_{1}\tau_{2}\cos\varphi)^{\frac{1}{2}-\epsilon}}-(1\leftrightarrow 2)-\tau_{1}^{-1+2\epsilon}\tau_{2}^{-1+2\epsilon} 
\right]=\nonumber\\
&=\frac{\Gamma^{2}(1/2-\epsilon)}{32\pi^{3-2\epsilon}}\frac{L^{4\epsilon}}{4\epsilon^{2}}+O(1)
\end{align}
where the two contributions which are related by the exchange $\tau_{1}\leftrightarrow\tau_{2}$ cancel.
\begin{align}
&{\rm (IV)}_{\ref{D30}}=\int_{0}^{L} \!\!\!\!d\tau_{1}\int_{-L}^{0}\!\!\!\!d\tau_{2}~\Phi_{1,2-L}=\nonumber\\
&=\frac{\Gamma^{2}(1/2-\epsilon)}{32\pi^{3-2\epsilon}}\!\!\int_{0}^{L} \!\!\!\!d\tau_{1}\int_{-L}^{0}\!\!\!\!d\tau_{2}~\left[ (L+\tau_{2})^{-1+2\epsilon} H(\tau_{1},\tau_{2})+H(\tau_{1},-L)(L+\tau_{2})^{-1+2\epsilon} 
\right]+O(1)=\nonumber\\
&=2\frac{\Gamma^{2}(1/2-\epsilon)}{32\pi^{3-2\epsilon}} L^{4\epsilon}\!\!\int_{0}^{1} \!\!\!\!d\tau_{1}\int_{-1}^{0}\!\!\!\!d\tau_{2}~H(\tau_{1},-1)(1+\tau_{2})^{-1+2\epsilon} +O(1)=\nonumber\\
&=\frac{\Gamma^{2}(1/2-\epsilon)}{32\pi^{3-2\epsilon}} \frac{L^{4\epsilon}}{\epsilon}\!\!\log \left(1+\sec \left(\frac{\varphi }{2}\right)\right)+O(1)
\end{align}
\end{subequations}
The integral over $\tau_{1}$ can be computed at $\epsilon=0$ since it is finite.
In the above analysis we have consistently dropped all the integrals which do not produce a divergence. Then the total result for the divergent part is 
\begin{align}
&\int_{0}^{L} \!\!\!\!d\tau_{1}\int_{-L}^{0}\!\!\!\!d\tau_{2}\int_{-L}^{0}\!\!\!\!d\tau_{3}~\left(\frac{d}{d\tau_{1}} \Phi_{1,23}-\frac{d}{d\tau_{3}} \Phi_{1,23}-\frac{d}{d\tau_{2}} \Phi_{1,23}\right)={\rm (I)}_{\ref{D30}}-{\rm (II)}_{\ref{D30}}-\nonumber\\
&-2{\rm (III)}_{\ref{D30}}+2{\rm (IV)}_{\ref{D30}}= \frac{\Gamma^{2}(1/2-\epsilon)}{16\pi^{3-2\epsilon}}L^{4\epsilon}\left[\frac{2}{\epsilon}\log \left(\sec \left(\frac{\varphi }{2}\right)+1\right)-\frac{1}{2\epsilon^{2}}+O(1)\right].
\end{align}
There is one remaining integral in the expansion of $\mathcal{I}_{(a)}+\mathcal{I}_{(b)}$ to be evaluated and it contains the three-point functions saturated
with three $\dot{x}_{i}$. An (almost) straightforward computation leads  to the following result
\begin{align}
&\int_{0}^{L}\!\!\!\!\!d\tau_{1}\!\!\int_{-L}^{0}\!\!\!\!\!d\tau_{2}\!\!\int_{-L}^{0}\!\!\!\!\!d\tau_{3}~\bigl[
\dot{x}_{2\nu} \dot{x}_{1\lambda} \dot{x}_{3\sigma}\Gamma^{\nu\lambda\sigma}-\dot{x}_{1\nu} \dot{x}_{3\lambda} \dot{x}_{2\sigma}\Gamma^{\nu\lambda\sigma}\bigr]=\nonumber\\
&=-\frac{\Gamma^{2}(1/2-\epsilon)}{16\pi^{3-2\epsilon}}\frac{{L^{4\epsilon}}}{\epsilon}\left( \cos ^2\frac{\varphi }{2}\log \left(\cos \frac{\varphi }{2}\right)\right)+O(1).
\end{align}
The difference $\mathcal{I}_{(a)}-\mathcal{I}_{(b)}$ is instead governed by only  one  integral which contains traced
three-point functions, namely
\begin{align}
\label{D8}
&\int_{0}^{L}\!\!\!\!d\tau_{1}\int_{-L}^{0}\!\!\!\!d\tau_{2}\int_{-L}^{\tau_{2}}\!\!\!\!d\tau_{3}
\left(\dot{x}_{1\nu}+
\dot{x}_{2\nu}\right)\left(\Gamma^{\tau\tau\nu}-\Gamma^{\tau\nu\tau}\right)=\!\!
\int_{0}^{L}\!\!\!\!d\tau_{1}\int_{-L}^{0}\!\!\!\!d\tau_{2}\int_{-L}^{\tau_{2}}\!\!\!\!d\tau_{3}\left(\frac{d}{d \tau_{3}} \Phi_{3,12}-\frac{d}{d \tau_{2}} \Phi_{2,13}-\right.\nonumber\\ &\ \  \ -\left.\frac{d}{d \tau_{1}} \Phi_{3,12}\right)
-\left(\frac{\Gamma(1/2-\epsilon)}{4\pi^{3/2-\epsilon}}\right)^{2}\int_{0}^{L}\!\!\!\!d\tau_{1}\int_{-L}^{0}\!\!\!\!d\tau_{2}\int_{-L}^{\tau_{2}}\!\!\!\!d\tau_{3}
\frac{1}{(x^{2}_{13})^{1/2-\epsilon}}\frac{d}{d \tau_{1}}
\frac{1}{(x^{2}_{12})^{1/2-\epsilon}}.
\end{align}
The last integral in \eqref{D8} is equal to minus \eqref{D2a} and thus
\be
 \int_{0}^{L}\!\!\!\!\!\!d\tau_{1}
 \int_{-L}^{0}\!\!\!\!\!\!d\tau_{2}\int_{-L}^{\tau_{2}} \!\!\!\!\!\!d\tau_{3}
\frac{1}{(x^{2}_{13})^{1/2-\epsilon}}\frac{d}{d \tau_{1}}
\frac{1}{(x^{2}_{12})^{1/2-\epsilon}}
\!=\!\biggl[\frac{L^{4\epsilon}}{2\epsilon} \left(\frac{\varphi }{2 } \cot \varphi -{\log \left(\cos \frac{\varphi }{2}\right)}\right)-\frac{L^{4\epsilon}}{8\epsilon^{2}}\!+\!O(1)\biggr]\!.
\ee
The remaining integrals can be then rearranged as follows
\be
 \int_{0}^{L}\!\!\!\!\!\!d\tau_{1}
 \int_{-L}^{0}\!\!\!\!\!\!d\tau_{2}(2 \Phi_{2,12}-\Phi_{-L,12}-\Phi_{0,12})-
 \int_{-L}^{0}\!\!\!\!\!\!d\tau_{2}\int_{-L}^{\tau_{2}} \!\!\!\!\!\!d\tau_{3}(\Phi_{3,L 2}-\Phi_{3,02})
\ee 
Consider first the contributions coming from $\Phi_{-L,12}$: the possible divergence arises when $\tau_{2}$
approaches $-L$.  On the other hand $\Phi_{-L,12}$ is regular when $\tau_{2}\to -L$ and thus the integral of
this quantity is finite.  A similar analysis can be applied to $\Phi_{3,L2}$: the source of divergence is the region $\tau_{3}\to \tau_{2}$, but the function is regular in this limit.  Thus this term will not yield poles in
$\epsilon$. 
It remains to extract the divergent part of 
\be
\label{jhg}
 \int_{0}^{L}\!\!\!\!\!\!d\tau_{1}
 \int_{-L}^{0}\!\!\!\!\!\!d\tau_{2}~(2 \Phi_{2,12}-\Phi_{0,12})+
 \int_{-L}^{0}\!\!\!\!\!\!d\tau_{2}\int_{-L}^{\tau_{2}} \!\!\!\!\!\!d\tau_{3}~\Phi_{3,02},
\ee 
which for each term in \eqref{jhg}  is given by
\begin{subequations}
\begin{align}
{\rm (I)}_{\ref{jhg} }=&2 \int_{0}^{L}\!\!\!\!\!\!d\tau_{1}\!
 \int_{-L}^{0}\!\!\!\!\!\!d\tau_{2}~\Phi_{2,12}=2\left(\!\frac{\Gamma(1/2-\epsilon)}{4\pi^{3/2-\epsilon}}\right)^{2}\!\! \eqref{D11}=\left(\frac{\Gamma(1/2-\epsilon)}{4\pi^{3/2-\epsilon}}\!\right)^{2}\frac{L^{4\epsilon}}{4\epsilon}\frac{\varphi}{\sin\varphi}+O(1),\displaybreak[2]\\
{\rm (II)}_{\ref{jhg} }=& \int_{0}^{L}\!\!\!\!d\tau_{1}
 \int_{-L}^{0}\!\!\!\!d\tau_{2}~\Phi_{0,12}=-\frac{L^{4\epsilon}}{2}\left(\frac{\Gamma(1/2-\epsilon)}{4\pi^{3/2-\epsilon}}\right)^{2}\int_{0}^{1}\!\!\!\!d\tau_{1}
 \int_{0}^{1}\!\!\!\!d\tau_{2}~\Biggl[\tau_{1}^{-1+2\epsilon}\tau_{2}^{-1+2\epsilon}-\nonumber\\
 &-\frac{\tau_{1}^{-1+2\epsilon}}{(\tau_{1}^{2}+\tau_{2}^{2}+2\tau_{1}\tau_{2}\cos\varphi)^{\frac{1}{2}-\epsilon}}-\frac{\tau_{2}^{-1+2\epsilon}}{(\tau_{1}^{2}+\tau_{2}^{2}+2\tau_{1}\tau_{2}\cos\varphi)^{\frac{1}{2}-\epsilon}}\Biggr]=\nonumber\\
 =&-\frac{L^{4\epsilon}}{2}\left(\frac{\Gamma(1/2-\epsilon)}{4\pi^{3/2-\epsilon}}\right)^{2}\frac{1}{\epsilon}\log\left(\cos\frac{\varphi}{2}\right)+O(1),\\
{\rm (III)}_{\ref{jhg} }=&\int_{-L}^{0}\!\!\!\!\!\!d\tau_{2}\int_{-L}^{\tau_{2}} \!\!\!\!\!\!d\tau_{3}~\Phi_{3,02}=\frac{1}{2}\left(\frac{\Gamma(1/2-\epsilon)}{4\pi^{3/2-\epsilon}}\right)^{2}\frac{L^{4\epsilon}}{4\epsilon^{2}}+O(1).
\end{align}
\end{subequations}
If we now collect all the contributions, we obtain a remarkably simple result 
\be
\int_{0}^{L}\!\!\!\!d\tau_{1}\int_{-L}^{0}\!\!\!\!d\tau_{2}\int_{-L}^{\tau_{2}}\!\!\!\!d\tau_{3}
\left(\dot{x}_{1\nu}+
\dot{x}_{2\nu}\right)\left(\Gamma^{\tau\tau\nu}-\Gamma^{\tau\nu\tau}\right)=
\left(\frac{\Gamma(1/2-\epsilon)}{4\pi^{3/2-\epsilon}}\right)^{2}\frac{{L^{4\epsilon}}}{4\epsilon}\varphi\cot\frac{\varphi}{2}+O(1)
\ee

\end{document}